\documentclass[final,5p,times,twocolumn,authoryear]{elsarticle}

\usepackage{amssymb}
\usepackage{lipsum}
\usepackage{graphicx}
\usepackage{upgreek}
\usepackage{derivative}
\usepackage{multirow}
\usepackage{hyperref}
\usepackage{amsmath}
\usepackage{float} 
\usepackage{array}
\usepackage{bm} 

\usepackage{subcaption}

\usepackage[utf8]{inputenc}
\usepackage[T1]{fontenc}

\usepackage{comment}
\usepackage{xspace}

\usepackage{xcolor}
\usepackage[normalem]{ulem}

\usepackage{cuted}

\newcommand{\e}{\ensuremath{\rm e}}
\newcommand{\diff}{\ensuremath{\rm d}}

\newcommand{\ergl}{\ensuremath{\rm erg\,s^{-1}}}
\newcommand{\Gcmc}{\ensuremath{\rm G \, cm^3}}

\newcommand{\add}[1]{\textbf{#1}\xspace}

\renewcommand{\vector}[1]{\ensuremath{\pmb{#1}}}
\newcommand{\Fgeom}{\mathcal{T}}
\newcommand{\Lx}{L_{\rm x}}
\newcommand{\object}[1]{#1}

\defcitealias{Basko-Sunyaev1976}{BS76}
\defcitealias{Abolmasov+2023}{AL23}

\journal{New Astronomy}

\usepackage{amssymb}

\def\muang{\chi} 
\def\Re{R_\mathrm{e}}
\def\Rm{R_\mathrm{m}}
\def\grad{^\mathrm{o}}

\def\ReRns{\xi_{\rm e}}
\def\RRns{\xi}
\def\Rshock{R_\mathrm{shock}}
\def\xishock{\xi_{\rm s}}
\def\ReRa{\xi}
\def\phicenter{\phi_0}

\def\thetaobs{\ensuremath{\thetarot_{\rm obs}}}
\def\thetadisc{\thetamag_{\rm disc}}
\def\Lisoavg{\langle L_{\rm iso} \rangle}

\newcommand{\Rns}{R_\star}  
\newcommand{\Mns}{M_\star}

\newcommand{\thetans}{\thetamag_\star} 
\newcommand{\crossec}{A_\perp}

\newcommand{\thetarot}{\Theta} 

\newcommand{\thetamag}{\theta} 
\newcommand{\phimag}{\phi}

\def\phicenter{\ensuremath{\phimag_0}}

\newcommand{\km}{\ensuremath{\rm km}}
\newcommand{\Msun}{\ensuremath{\rm M_\odot}}

\newcommand{\phiomega}{\varphi}

\begin{document}

\begin{frontmatter}



\title{Pulse profiles of tall accretion columns in X-ray pulsars}


\author[first]{Bagration Megrelishvili}
\author[second]{Pavel Abolmasov}
\affiliation[first]{organization={Moscow State University},
            addressline={Leninskie Gory, 1},
            city={Moscow},
            postcode={119991},
            country={Russia}}
\affiliation[second]{organization={The Raymond and Beverly Sackler School of Physics and Astronomy, Tel Aviv University},
            addressline={Ramat Aviv}, 
            city={Tel Aviv},
            postcode={69978}, 
            country={Israel}}

\begin{abstract}
Accreting X-ray pulsars (XRPs) are strongly magnetized neutron stars (NSs) distinguished by a periodic variability pattern related to the rotation of their magnetospheres (pulsations). Many of these objects show complex asymmetric pulse profiles that are difficult to explain in the framework of existing models. We calculate the pulsations of X-ray radiation from a magnetized NS in the regime of supercritical accretion, when the radiation shock wave is high above the NS surface, and the observed radiation comes from two optically thick accretion columns located downstream of the shocks. 
We show that the accretion column beam pattern calculated from the first principles has a potential of reproducing the variety of XRP pulse shapes. The surfaces of the columns are assumed to radiate locally as blackbodies with the temperature distribution following the solution of \citet{Basko-Sunyaev1976}. The adopted geometry of the field is an inclined dipole. It is taken into account that the two accretion flows above the shocks attenuate and scatter the radiation of the accretion columns. Using the ray-tracing technique, we obtain a vast variety of pulse shapes controlled by the mass accretion rate, magnetic angle, the direction towards the observer, and the parameters that determine the particular set of magnetic field lines being loaded with accreting matter. We propose a universal dimensionless parameter describing the asymmetry of a pulse quantitatively.
The model naturally produces asymmetric pulse profiles observed in many XRPs. Its applicability, however, is limited to the brightest objects like pulsating ultraluminous X-ray sources and strong flares of Galactic XRPs.

\end{abstract}

\begin{keyword}
accretion, accretion discs \sep magnetic fields \sep stars: neutron \sep X-rays: binaries



\end{keyword}

\end{frontmatter}





\section{Introduction}
\label{introduction}

X-ray pulsars (XRPs) are highly magnetized neutron stars (NSs) in binary systems. 
Despite these objects being known for around 50 years  \citep{1971ApJ...167L..67G} and unambiguously identified as accreting NSs in binary systems, many details of their physics remain uncertain \citep[see, for example,][]{Mushtukov-Tsygankov2022}. Accretion occurs either from the disc or the wind of the donor star, with higher accretion rates primarily achieved in the former case.
The presence of magnetic field destroys the inner parts of the hydrodynamic accretion flow (disc or gravitationally captured wind) and creates a magnetosphere with a characteristic radius estimated as the Alfv\'{e}n radius, where magnetic stresses exceed both dynamic and thermal pressure of the accreting matter. 
Within the Alfv\'{e}n radius, the dynamics of the flow is governed by magnetic stresses. 
Consequently, the matter captured by gravitational forces is channelled along the magnetic field lines, converging towards the vicinity of the magnetic poles.

For low mass accretion rates, the observed emission comes from the regions where the mass-loaded field lines cross the NS surface (hot spots within the polar caps, see \citealt{2025ApJ...978...80L} and references within). 
If the energy flux released by accretion exceeds the local Eddington limit, radiation pressure becomes dynamically important and stops the magnetospheric flow above the NS surface in a radiative shock wave (see \citealt{1975PASJ...27..311I} and \citealt{Basko-Sunyaev1976}, the latter hereafter referred to as \citetalias{Basko-Sunyaev1976}). 
The optically thick subsonic sinking flow below the shock is known as the accretion column. 

The presence of coherent pulsations in XRPs suggests that the accretion flows and radiating regions on their surfaces are not axisymmetric with respect to the rotational axis. 
Even in the simplistic dipolar magnetosphere case, the magnetic and rotation axes of the NS are likely misaligned, which makes the accretion flow shape non-axisymmetric~\citep{Bachetti+Romanova2010}. 
This is especially true in the super-Eddington case when the observed radiation comes from multiple radiating surfaces.

Real XRPs exhibit various pulse shapes that change on times longer than the spin period. \citet{2022A&A...662A..62A} provide a simple classification of pulse shapes based on the dominance of the first and second harmonic in the Fourier decomposition of the pulse.
For certain objects, like \object{4U1901+03}, pulse shape types were observed to change. 
Interestingly, \citet{2022A&A...662A..62A} report no association between the pulse shape 
and luminosity or magnetic field strength. 
This suggests the presence of some hidden parameters that are, unlike luminosity or magnetic field, difficult to measure directly, but nonetheless strongly affect the observed pulse shapes. 
It is logical to associate these additional parameters with the geometrical properties of the accretion flows in the magnetosphere of the NS. 
One of them is magnetic angle $\chi$ between the magnetic and the rotation axes. 
But there should be more degrees of freedom related to the shape and the orientation of the accretion flows.

Most of the well-studied XRPs, including those undergoing super-Eddington flares, show noticeably asymmetric pulse profiles. Pulsating Ultra Luminous X-ray Sources (ULXs) also show asymmetries in their pulse profiles~\cite[][]{2017Sci...355..817I, 2024A&A...691A.123P, 2017MNRAS.466L..48I, 2017xru..conf...81F}.
Though these asymmetries are a challenge to some of the models routinely used to interpret the observational data on XRPs, they are scarcely studied and lack a model-independent quantitative measure.

We interpret the observed pulsations as one-dimensional scans through the neutron star’s stable beam pattern. This pattern is traditionally assumed to be axially symmetric around the magnetic axis, as in the Rotating Vector Model (RVM)~\citep{Radhakrishnan-Cooke1969}.
The RVM fails to explain the asymmetric pulse profiles of XRPs like Cen X-3. 
To address this, previous studies have often assumed non-dipolar magnetic fields to break the model's inherent axial symmetry~\citep[e.g.,][]{kraus96,sasaki12}. 
Alternatively, one can relate the asymmetry of the beam pattern with the asymmetry of the accretion flows.

Misalignment of the magnetic and rotational axes creates preferential magnetic longitudes \citep{Kulkarni-Romanova2013,2024ApJ...960L..12D}, leading to lopsided streams even for moderate magnetic inclinations.

In this work, we attempt to predict the observed features of the sources containing high accretion columns. Such columns exhibit a geometry where the beaming pattern becomes strongly misaligned with the magnetic axis \citep{miller1996}. If the column is sufficiently tall, most of its radiation emerges well above the NS surface, allowing us to neglect both relativistic light bending and the extreme variations in scattering cross‑sections expected in strong magnetic fields \citep[e.g.,][]{SokolovaThesis}. We calculate observed pulse profiles for a sample of models to probe the dependence of the pulse fraction, beaming factor, pulse-asymmetry degree (for which we propose a new metric) on the intrinsic model parameters and inclination of the observer. We also discuss whether the features typically observed from luminous XRPs can be naturally obtained in the model of two tall accretion columns.

In Sect.~\ref{s.model}, we describe the model and the numerical algorithm. 
In Sect.~\ref{sec:results}, we present the results of our calculations, and discuss them in Sect.~\ref{sec:discussion}. 
We conclude in Sect.~\ref{sec:conclusions}.

\begin{figure*}[ht]
    \centering
    \begin{subfigure}[t]{0.42\textwidth}
        \centering
        \includegraphics[width=\linewidth]{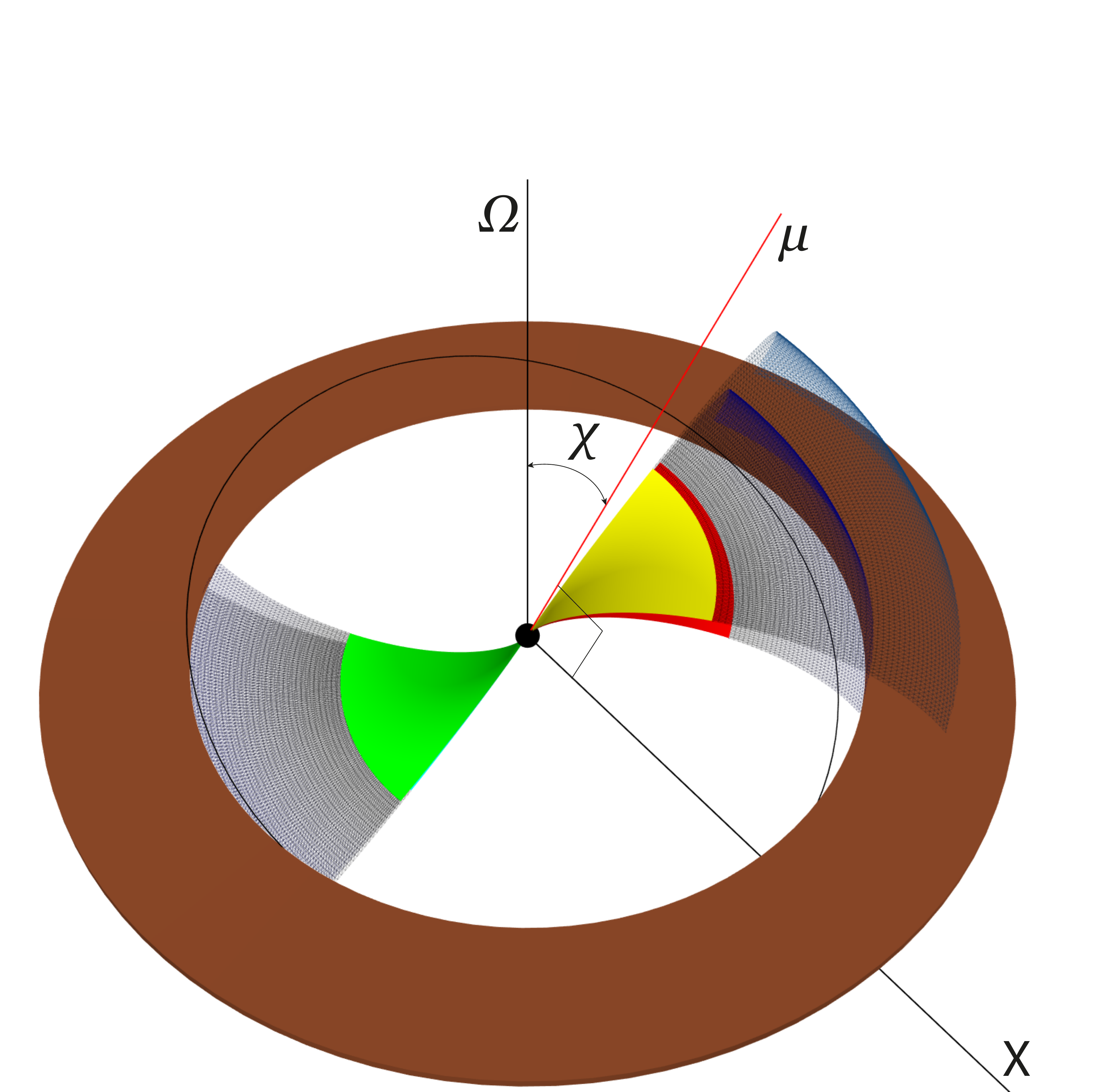}
        \caption{$\thetaobs=40^\circ$}
        \label{fig:Geom_config_a}
    \end{subfigure}\hfill
    ~
    \begin{subfigure}[t]{0.48\textwidth}
        \centering
        \includegraphics[width=\linewidth]{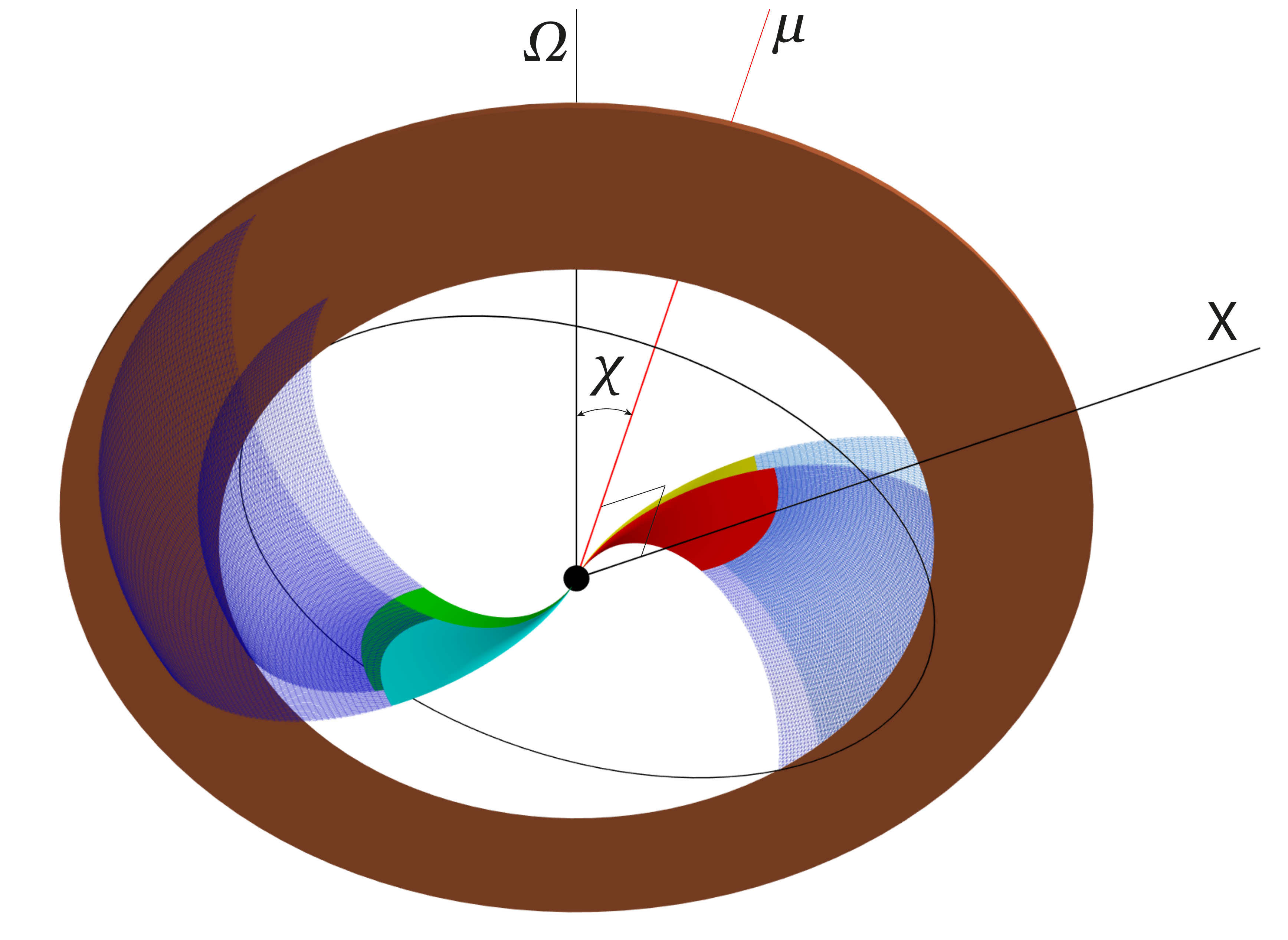}
        \caption{$\thetaobs=140^\circ$}
        \label{fig:Geom_config_c}
    \end{subfigure}
    \caption{Sketches of the adopted geometry of the accretion flows from the points of view of an observer at inclinations $\thetaobs = 40^\circ$ (a) and $\thetaobs = 140^\circ$ (b).
    The geometry of the simulation is set by
    $\muang = 20^\circ$, $\phicenter=60^\circ$, 
    and $a=0.2$. 
    The surfaces of the northern accretion column are shown in red and yellow, of the southern accretion column, in green and cyan. The red and green surfaces are equatorial (facing the magnetic equator of the star), and yellow and cyan are polar (facing the poles). 
    The fine blue mesh shows the funnel flow above the shock, following the dipole field lines. The black line shows the plane of the magnetic equator. The inner region of the disc is shown in brown. 
    }
\label{fig:Geom_config}    
\end{figure*}

\begin{figure}[ht]
    \centering  \includegraphics[width=0.4\textwidth]{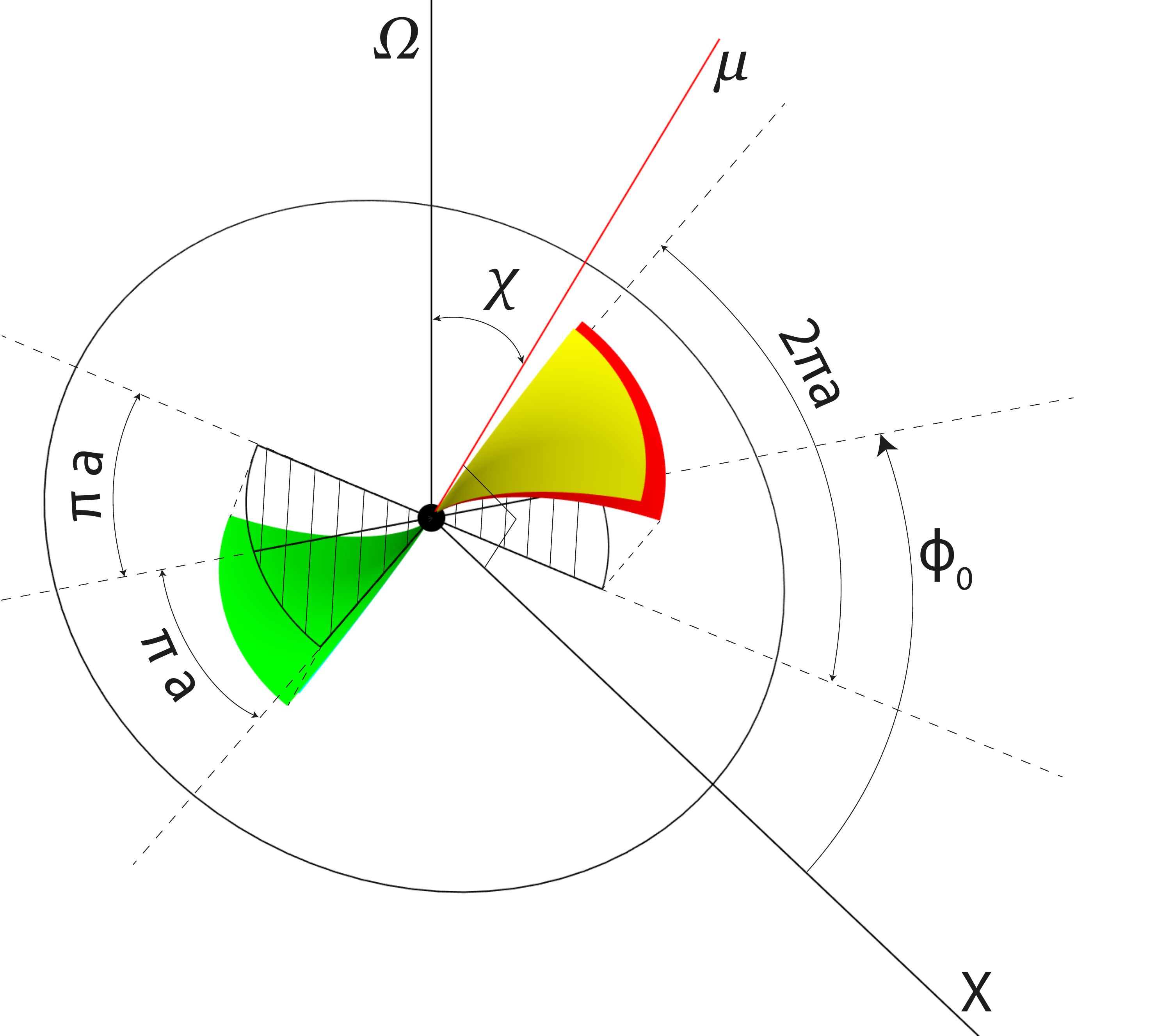}
    \caption{Schematic representation of the columns (without the disc and FF). The projections  of the columns onto the magnetic equator plane (hatched) with the corresponding centre lines are also shown}
    \label{fig:schema}
\end{figure}

\section{Model description}
\label{s.model}

We develop a model calculating pulse profiles produced by two tall accretion columns within an inclined dipole magnetic field geometry. 
The radiation emitted by the columns is assumed to be locally blackbody, and we calculate the bolometric flux using the ray-tracing technique. To calculate the local intensities, we use the analytical solution for a steady-state radiation-pressure supported accretion column obtained by  \citetalias{Basko-Sunyaev1976}.

In the setting of \citetalias{Basko-Sunyaev1976}, the plasma in the magnetosphere is falling free along the field lines until it is stopped in a radiative shock wave above the NS surface.
After passing the shock wave, accreting matter in this model is assumed to sink subsonically towards the surface, losing energy by the sideways diffusion of radiation.

The analytic solution of \citetalias{Basko-Sunyaev1976} is  well reproduced by time-dependent numerical simulations by \citet{Abolmasov+2023} (hereafter \citetalias{Abolmasov+2023}) until the height of the column becomes comparable to the size of the magnetosphere or the flow becomes highly super-Eddington. 
We follow the formalism used by \citetalias{Abolmasov+2023} for dipolar geometry and the shock location and effective temperature distributions from \citetalias{Basko-Sunyaev1976}.
If the analytical prediction for the location of the shock exceeds the inner radius of the disc, the shock position is assumed to coincide with the latter. 

\subsection{Geometry of the flow}\label{sec:geometry cols}

Figure~\ref{fig:Geom_config} illustrates the adopted configuration of the system. 
The rotation axis of the NS (set by the angular frequency vector $\bm\Omega$) coincides with the rotation axis of the disc. 
The magnetic moment $\bm\mu$ is inclined at the magnetic angle $\muang$ with respect to $\vector{\Omega}$.

Two accretion columns are symmetrically positioned relative to the NS centre.
They  are anchored onto the NS surface and curve along the dipole field lines. %
Each accretion column (shown in solid colours in Fig.~\ref{fig:Geom_config}) has two radiating surfaces: the polar one (yellow for the northern column, cyan for the southern) facing the closer of the magnetic poles, and the equatorial one (green for the northern column, red for the southern) facing the magnetic equator. 
Both columns start at the surface of the NS and extend to the shock front, following the dipolar field lines. 

Above the shock, accreting matter flows along the magnetic field lines with free-fall velocity, forming a  funnel flow (hereafter FF). The matter in this region can scatter and obscure radiation emitted by the accretion columns. FF is shown with a blue mesh in Fig.~\ref{fig:Geom_config}.

For all the calculations, we use the reference frame co-rotating with the NS and a spherical coordinate system ($R$, $\thetamag$, $\phimag$) with the polar angle $\thetamag$ counted from $\bm\mu$, azimuthal angle $\phimag$, and radius $R$ (see~\ref{s.app_dl_dS}).
Azimuth  $\phimag$ is counted counter-clockwise from the axis $X$, set by the intersection of the magnetic equatorial plane and the plane set by $\vector{\mu}$ and $\vector{\Omega}$. 
Positive $X$ corresponds to the direction from  $\bm\Omega$ to $\bm\mu$ and may be set by the unit vector
\begin{equation}
    \vector{e}_X = \frac{1}{\sin\chi} \,\vector{e}_\mu \times (\vector{e}_\mu \times \vector{e}_\Omega)\, .
\end{equation}
The accretion flow starts at the inner edge of the accretion disc (radius of the magnetosphere) $R = \Rm = 0.5 R_{\rm A}$ , where $R_{\rm A}$ is Alfv\'{e}n radius defined as
\begin{equation}\label{eq:Ra}
    R_{\rm A} = \left(  \frac{\mu^2} {2 \dot M \sqrt{2G\Mns}}\right)^{2/7}\, .
\end{equation}
Hereafter we consider a NS with a conventional set of parameters: mass $\Mns = 1.4 M_\odot$, radius $\Rns=10\,\km$, and magnetic moment $\mu=10^{30} \Gcmc$, corresponding to $B=2 \times 10^{12} {\rm G}$ on the poles.

We assume that each column (and the corresponding part of the FF) has the azimuthal extent of $2 \uppi a$, where $a \leq 1$. Accretion in the northern hemisphere occurs in the azimuthal range $\lbrack \phicenter - \uppi a, \phicenter + \uppi a \rbrack$, see Fig. \ref{fig:schema}. 
The azimuth of the column midline, free parameter $\phicenter$, takes into account the azimuthal shift of the flow with respect to $\vector{e}_X$.

This shift may be considerably different from zero, as it was shown, for example, in numerical simulations by \citet{Kulkarni-Romanova2013} (see also \S\ref{s:phase shift}). One may consider this a simplified approach to the twist of the magnetosphere expected when the rotation frequencies of the inner disc and the magnetosphere do not match each other.
As we will show later in Section~\ref{s:dep_on_phi}, \phicenter\ affects the shape of the pulse
and produces a large variety of pulse shapes even if all the other parameters are fixed.
Its role is not equivalent to the phase shift of the pulse, as it described rotation around the magnetic rather than rotation axis.

We assume that the matter enters the magnetosphere at the disc symmetry plane. 
For simplicity, we also assume that all the mass-loaded magnetic field lines 
reach the surface of the star at the same magnetic polar angle $\thetamag = \thetans$. For an inclined magnetic dipole, the magnetic surface with the same $\thetans$ crosses the disc plane at different radii for different azimuths.
This immediately raises the question about the range of radii within the disc that are pierced by the magnetic field lines with the same $\thetans$. 
Obviously, the smallest radius should not be less than the radius of the magnetosphere. 
Constraints on $a$ and $\chi$ are calculated in  \ref{s.about_magnetic_polar_angle}.
Therefore, we consider below mainly two values of the magnetic angle, $\muang = 10^\circ$ and $20^\circ$, see Table~\ref{tab:input params}. 
This allows us  to parametrize the mass-loaded field lines by the maximum extent radius $\Re$ ($R = \Re \sin^2\theta$ along each of the lines)
\begin{equation}\label{eq.Re}
    R_{\rm e} = \frac{\Rns}
    {\sin^2{\thetans}}  \, = \frac{\Rm}
    {\sin^2{\thetadisc}}  \, ,
\end{equation}
where  $\thetadisc$ is the minimum magnetic polar angle of the magnetospheric flow at its starting point (in the disc plane): 

\begin{equation}\label{eq.thetadisc}
    \thetadisc = 
\displaystyle    \begin{cases}
  \displaystyle  \uppi/2 - \muang, & |\phicenter| \leq \uppi a,\\
  \displaystyle  \min\left\{\uppi / 2 - \arctan{[\tan{\chi} \cos{(\phicenter \pm \uppi a)}]}\right\}\, ,  & \textrm{otherwise\,.} \\
    \end{cases}
\end{equation}
The first case describes the scenario when the zero-azimuth direction $\phimag = 0$ is located within the magnetospheric flow. 
If the magnetospheric flow does not contain the zero-azimuth direction, $\thetadisc$ is equal to the magnetic polar angle of the disc plane at one edge of the column.

\begin{figure}
    \centering
    \includegraphics[width=0.9\linewidth]{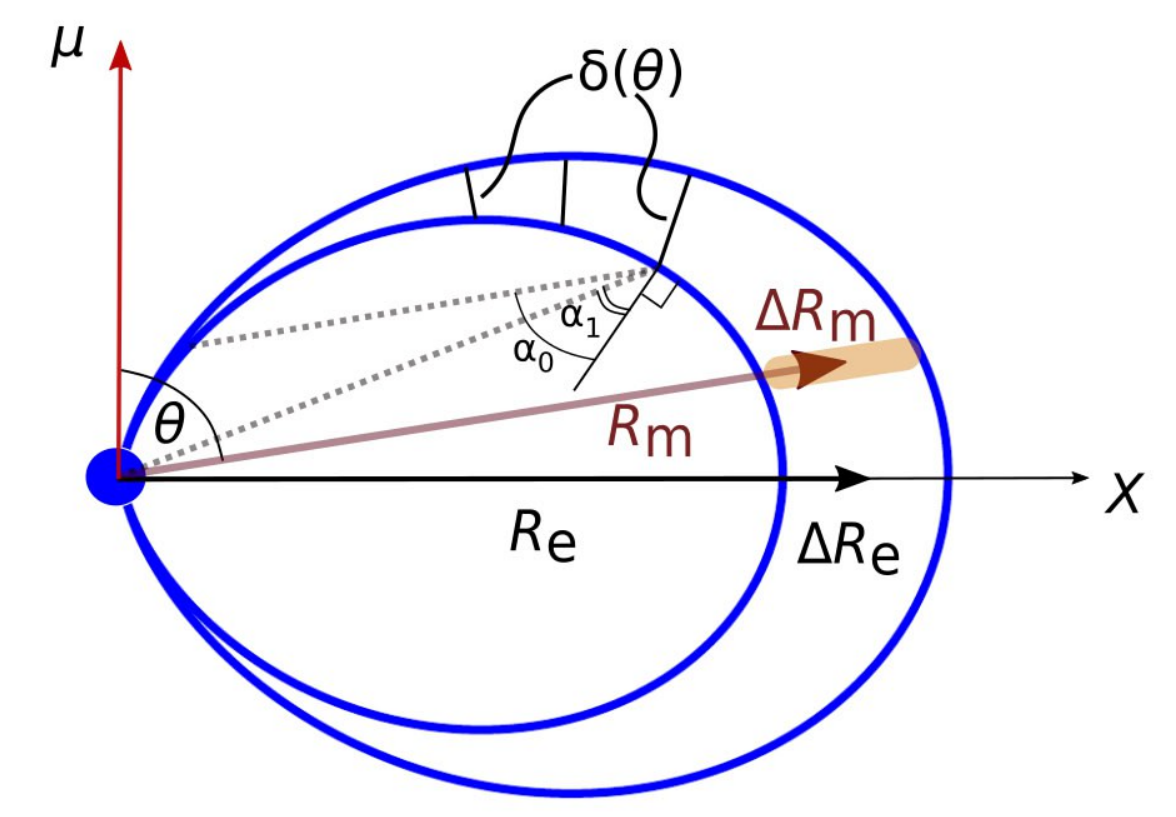}
    \caption{Geometry of the magnetospheric section in the plane $ \phimag = 0$ for the midline azimuth $ \phimag_0 = 0$. 
    The radius-vector of length 
    $\Rm$ connects the origin with the disc, shown as the light brown slab.
    The dotted lines are the rays of light from 
    the accretion column. 
    The transverse thickness of the flow $\delta(\thetamag)$ is given by Eq.~\eqref{eq:delta}. 
    Unlike the transverse thickness $\delta$, the relative thickness of the flow along the radius does not depend on $\thetamag$: $\Delta \Re/\Re = \Delta \Rm /\Rm$, where $\Re$ is the maximum size of the dipole tube. 
    }       
    \label{fig:transverse_delta}
\end{figure}

The range of radii within the disc pierced by mass-loaded field lines is determined by the thickness of the FF.
We define the accretion flow configuration by fixing its footpoints in the disc plane. 
Specifically, we set the funnel flow thickness in the disc plane to be  $\Delta \Rm$, see  Fig.~\ref{fig:transverse_delta}. In the model we fix the value $\Delta \Rm/\Rm=0.25$.

The physical separation between the polar and the equatorial surfaces\footnote{While the polar and equatorial surfaces are physically distinct, in the ray-tracing procedure we assume that they occupy the same geometric locus. Thus, one value of $\Re$ defined by Eqs.~(\ref{eq.Re}) and (\ref{eq.thetadisc}) determines all the four funnel surfaces.} is
$\delta = \delta(\theta)$, and it is calculated from $\Delta \Rm$ ($\Delta \Re$): 
\begin{equation}\label{eq:delta}
    \delta = \frac{R_{\rm e} \sin^3{\thetamag}}{\sqrt{1+3\cos^2\thetamag}} \frac{\Delta R_{\rm e}}{R_{\rm e}}\, , 
\end{equation}

As a result, the mass-loaded magnetic
lines cross the disc  over a range of radii between $\Rm-\Delta \Rm/2$ and $\Rm+\Delta \Rm/2$.
The field lines formally extending towards much larger radii appear if the magnetic angle is large enough, that seems to be an issue likely to be resolved if the magnetic polar angle is allowed to vary within the polar cap.  Simulations by \citet{2024ApJ...960L..12D} indeed show that for large magnetic angles the shape of the polar cap is closer to an arc of the big circle with the magnetic polar angle approaching zero near its centre.

\subsection{Structure of the column}

Once the geometry of the magnetospheric flow is set, it is possible to use the equations from \citetalias{Basko-Sunyaev1976} to calculate the position of the shock $R_{\rm shock}$ (their equation 34), luminosity of the columns $\Lx$ (their equation 37), and the effective temperature distribution $T_{\rm eff}(R)$ (see section 4.5 of \citetalias{Basko-Sunyaev1976}.) 
Applying analytical model of \citetalias{Basko-Sunyaev1976}, we evaluate their solution along the columns' central line. 

The model is initialized with the input parameters listed in Table~\ref{tab:input params}. The dimensionless accretion rate is defined as $\dot m = {\dot{M} c^2}/{L_{\rm Edd}}$, where $L_{\rm Edd}$ is the Eddington luminosity:
\begin{equation}
    L_{\rm Edd} = \frac{4 \uppi G \Mns c}{\varkappa} \approx 2 \times 10^{38} ~\textrm{erg $\rm{s}^{-1}$},
\end{equation}
and $\varkappa = 0.35 \, \rm cm^2 \, g^{-1}$ is the Thomson electron scattering opacity for solar abundances, and $\dot{M}$ is the accretion rate.
The geometry-defining parameters $a, \Delta \Rm$, $\muang$, and $\phicenter$ are converted into the corresponding \citetalias{Basko-Sunyaev1976} model parameters $l_0$ and $d_0$ that characterize the accretion flow geometry at the NS surface.

\begin{figure}[ht]
    \centering  \includegraphics[width=0.9\columnwidth]{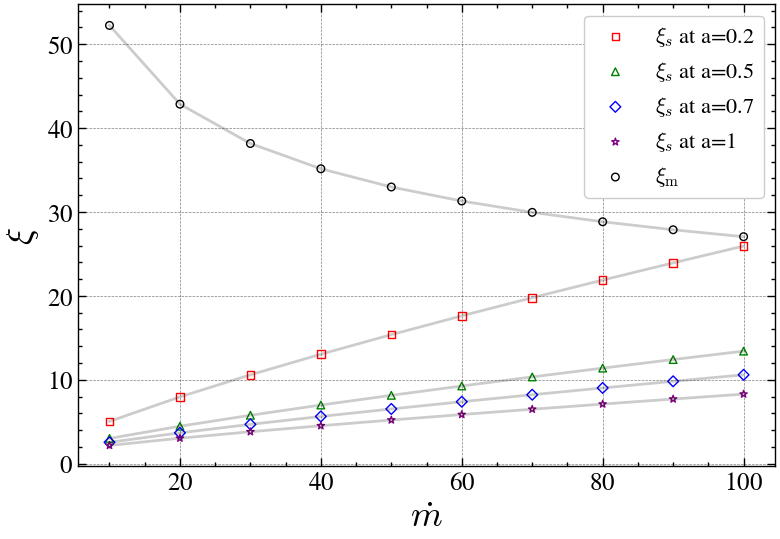}
    \caption{ Dependence of the normalized magnetosphere size $\xi_{\rm m} = R_{\rm m}/\Rns$ and shock wave position $\xishock =  R_{\rm shock} /  \Rns$ on the accretion rate $\dot{m}$. 
    The magnetic angle $\muang = 0\grad$.}
    \label{fig:xi to m}
\end{figure}

\begin{figure}[ht]
    \centering  \includegraphics[width=0.9\columnwidth]{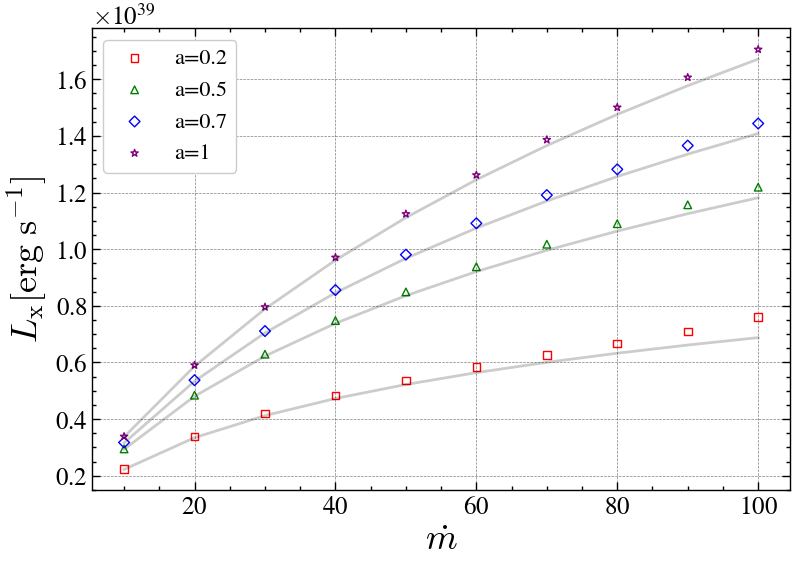}
    \caption{Bolometric luminosity vs. accretion rate $\dot{m}$. The lines show the analytic result of \citetalias{Basko-Sunyaev1976},  and symbols show surface black body flux  integrated over the column surfaces according to  Eq.~\eqref{eq.total_columns_power}. The magnetic angle $\muang = 0\grad$.}
    \label{fig:Lx to m}
\end{figure}

In Fig.~\ref{fig:xi to m}, the shock height $\Rshock$ and disc radius $\Rm$ are plotted as functions of the input parameters $\dot m$ and $a$. For spherical radii, we use a normalized variable $\xi=R/\Rns$. As expected, the height of the column $\Rshock$ increases with the accretion rate and decreases with $a$.
In Fig.~\ref{fig:Lx to m} the total bolometric power $\Lx$ is shown against the same input parameters, $\dot m$ and $a$.

\begin{table*}[htbp]
\caption{
Free parameters of the model. Stellar mass $\Mns = 1.4\Msun$, radius $\Rns = 10\km$, and magnetic moment $\mu = 10^{30}\Gcmc$ are fixed.
}
\centering

\begin{tabular}{c c p{12cm}}
\hline \hline

parameter & values & description \\
\hline
$\dot{m} = {\dot{M} c^2}/{L_{\rm Edd}}$ & $30,60,100$ & dimensionless accretion rate, in units $2.2\times 10^{17}$\,$\rm{g \, s^{-1}}$\\
$a$ & $0.1, 0.2, ..., 1$ 
&  azimuthal covering factor \\


$\muang$ & $10^\circ, 20^\circ$ & angle between the magnetic axis $\bm\mu$ and the rotation axis $\bm\Omega$\\ 


$\cos \thetarot_{\rm{obs}}$ & $0,0.1,...,0.9,1$ 
& cos of angle between the direction towards the observer and the rotation axis $\bm\Omega$ \\

$\phicenter$ & $0\grad, 20\grad, ...,180\grad$
& central azimuth of the column (midline azimuth)  \\
\hline
\end{tabular}
\label{tab:input params}
\end{table*}

\subsection{Processes in the free-falling flow above the shock (FF)}\label{sec:Above shock}

The geometry introduced above suggests complex and variable visibility conditions for all the radiating surfaces. 
In particular, one of the optically thick columns may obscure the other  or a part of itself. 
The FF, if its optical depth is large enough, is also capable of obscuring column emission.
It is probably fully ionized (see \S\ref{s.dif-scattering}), which makes it an efficient reflector in the X-ray range. In turn, the reflected  emission  may be obscured by the accretion flows and  NS, making the pulses at high accretion rates a result of an interplay between various obscuration and reflection processes. 
Reflections may be multiple, especially for $a \gtrsim 0.5$, when a typical photon has a high probability to cross a mass-loaded field line.
Here, we stop after two steps: the column emission is complemented with the emission reflected  by the FF above the shock wave. Visibility of both contributions is altered by the presence of the optically thick matter along the line of sight.
Real columns are not radiating as local blackbodies: effects of radiation transfer in the strong magnetic field and Comptonisation effects need to be considered to make reliable predictions of their spectral properties and beam shapes. 
Here we focus on the visibility effects and bolometric intensities and leave a more elaborate analysis to future studies.

If a beam emitted in the direction of the observer intersects the FF, 
its intensity is reduced by the factor of $\e^{-\tau / \cos{\alpha_0}}$, where $\alpha_0$ is the angle between the beam  and the normal to the surface at the intersection point, and $\tau$ is the scattering optical depth along the normal (see Fig.~\ref{fig:transverse_delta}). Transverse optical depth

\begin{equation}
    \tau = \varkappa \rho \delta = \frac{\varkappa \dot{M} \delta}{\crossec v}\,  , 
\end{equation}
where $v = v(\thetamag)$ is the velocity of the accreting matter along the field line, $\delta = \delta(\thetamag)$ is the accretion funnel's width (Eq.~\ref{eq:delta})
and $A_\perp = A_\perp(\thetamag)$ is the cross-section of the accretion flow (of the two columns)
\begin{equation}\label{eq:S}
    \crossec = 2 \cdot 2\uppi a R_{\rm e} \sin^3{\thetamag} \, \delta\, .
\end{equation}

Assuming the velocity along the field in the FF equal to the free-fall velocity 

\begin{equation}\label{eq:vfree}
    v = \sqrt{\frac{2 G \Mns}{R}}\, ,
\end{equation}
we arrive at
\begin{equation}
    \tau = \frac{\varkappa \dot{M} \sqrt{R_{\rm e} } }{4 \uppi a R \sqrt{2 G \Mns}} \sim 1.44 \frac{\dot{M}_{18} \sqrt{\ReRns}}{a ~ \RRns} \, \sim 0.32  \frac{\dot{m} \sqrt{\ReRns}}{a ~ \RRns} \, ,
\label{eq:tau_1}    
\end{equation}
where $\dot{M}_{18}$ is the accretion rate normalized to $10^{18} \rm g \, s^{-1}$, $\RRns = R/\Rns$, $\ReRns = R_\mathrm{e} / \Rns$
(see also equation 3 in \citealt{Mushtukov+2017}).

To calculate the scattered radiation in a simplified manner, we assume a central isotropic source with the luminosity equal to the total luminosity of the two accretion columns $\Lx$, which is the total radiative power of the four surfaces of the two columns. The source illuminates the FF, whose surface elements re-emit coherently part of the incident flux. We assume that the reflected intensity $I_{\rm refl}$ is isotropic:
\begin{equation}\label{eq:Irefl}
   \uppi\, I_{\rm refl}\, = \frac{ \Lx}{4 \uppi R^2} \, \cos \alpha_1\, (1-e^{- \tau_1})\, 
\end{equation}
where 
\begin{equation}
   \tau_1 = \tau / \cos\alpha_1
\end{equation}  
and $\alpha_1$ is the angle between the normal to the surface element and the radius vector (see Fig.~\ref{fig:transverse_delta}). 
$\Lx$ is calculated analytically according to \citetalias{Basko-Sunyaev1976}.
In our approach, only the fraction $1-e^{-\tau_1}$ of the incident intensity is reflected. 
We neglect the photons that pass through the flow after some, one or more, number of scatterings (see further discussion in \S\ref{s.dif-scattering}). 

\subsection{Observed emission}

We simulate the observed emission using the geometry and physics described in the previous sections. 
The observer's position is defined by the polar angle $\thetaobs$ and the spin phase $\Phi = \phiomega/2\uppi$, with $\phiomega$ denoting the azimuthal coordinate with respect to the spin axis of the star.
Both angles $\thetaobs$ and $\phiomega$ are  measured in the spherical coordinate system set by the rotation axis $\bm\Omega$. The phase $\Phi = 0$ corresponds to the configuration where the angle between the observer's line of sight and the magnetic axis is minimal. A 50-point equally spaced grid in $\Phi$ was adopted to sample one full rotation period.

At a given time or spin phase, we calculate the radiation received by the observer. 
To do this, we integrate the intensity from the surfaces of the magnetospheric flow, taking into account that some fraction of the radiation is eclipsed by the accretion columns or NS, and some of the radiation is attenuated by the FF.

For each of the four column 
surfaces, we use a $100\times 100$ equidistant mesh in $\theta$ and $\phi$. 
The same is done for the NS-facing, illuminated surfaces of the FF. 
Each surface element emits intensity independent of the direction within the relevant hemisphere. 
Thus, the total luminosity of all the four radiating surfaces (which depends neither on $\thetaobs$ nor on the phase $\Phi$) equals
\begin{equation}
    L_{\rm x} = 
    4 \int\limits_{\phicenter - \uppi a}^{\phicenter + \uppi a} \int\limits_{\thetans}^{\thetamag_s} \, \sigma T_{\rm eff}^4(\thetamag) \, \tilde S  \, {\rm d}\thetamag \, {\rm d}\phimag\,
    \label{eq.total_columns_power}
\end{equation}
and coincides with the theoretical prediction by \citetalias{Basko-Sunyaev1976} with the accuracy better than 2\%, see Fig.~\ref{fig:Lx to m}, for the cases when the predicted height of the column is less than $R_{\rm mag}$.
The Jacobian of the mapping $\tilde S(\thetamag)$ of the spherical to Cartesian surface element is calculated according to Eq.~\eqref{eq:tilde S}.

The isotropic bolometric luminosity of the two magnetospheric flows received by an observer at a distance of $d$ is 
\begin{equation}
     L_{\rm iso} = 4\uppi\, d^2\, \int \Fgeom \, I \,\mathrm{d} \Omega\, ,
\end{equation} 
where $I$ is the bolometric intensity, $\diff \Omega$ is the elementary solid angle from the point of view of the observer, and the integrals are taken over the four surfaces of the magnetic tubes.

The multiplier $\Fgeom$ accounts for the eclipses and attenuation:

\begin{equation}
    \Fgeom(\thetamag, \phimag, \thetaobs, \Phi) = f_{\rm eclipse}(\thetamag, \phimag, \thetaobs, \Phi) \cdot f_{\rm attenuation}(\thetamag, \phimag, \thetaobs, \Phi)\, .
\end{equation}
The factors on the right-hand side are calculated using the ray-tracing method described in \ref{appendix:Ray tracing}.
The eclipse function $f_{\rm eclipse}$ is 0 when the ray towards the observer intersects the NS surface or an accretion column, and 1 otherwise. 
The degree of attenuation in the FF depends on the optical depth along the ray as

\begin{equation}
    f_{\rm attenuation}(\thetamag, \phimag, \thetaobs, \Phi) = e^{- \tau / \cos{\alpha_0}}\, ,
\end{equation}
where $\alpha_0$ is the angle between the normal to the surface of the magnetic tube and the incident ray (see Fig.~\ref{fig:transverse_delta}), and the transverse thickness $\tau$ is given by Eq.~\eqref{eq:tau_1}. 

We write below the integral over just one of the four surfaces of the magnetospheric flow. Substituting the solid angle $\mathrm{d}\Omega = ({\tilde S}/{d^2})\cos \psi \,\diff \thetamag \,\diff \phimag$, where $\psi$ is the angle  between the normal to the surface element and the line of sight (see Eq.~\ref{eq:cos_psi}) we obtain
\begin{equation}\label{eq:final L}
    L_{\rm iso} = 4 \uppi \int\limits_{\phicenter - \uppi a}^{\phicenter + \uppi a} \int\limits_{\thetans}^{\thetamag_{\rm disc}}  \Fgeom \,I
    \,  \cos{\psi(\thetamag, \phimag)} \, \tilde S\, {\rm d}\thetamag \, {\rm d}\phimag \, .
\end{equation}

Intensity is calculated differently below the shock and in the FF:
\begin{equation}\label{eq.intensities}
    I =  
    \begin{cases}
        {\sigma\, T_{\rm eff}^4}/{\uppi}, \, & \thetans < \thetamag < \thetamag_{\rm s}\\
         I_{\rm refl}, \, &\thetamag_{\rm s} < \thetamag < \thetamag_{\rm disc}\\
    \end{cases} \, ,
\end{equation}
where $\sigma$ is the Stefan-Boltzmann constant, $\thetamag_{\rm s}$ and $\thetans$ are the polar angles at the shock and at the NS surface, respectively, and the reflected intensity $I_{\rm refl}$ is given by Eq.~\eqref{eq:Irefl}.

In this work, to analyze  the pulse profile morphology, we use isotropic luminosity normalized to its maximum value within each pulse $\tilde{L}_{\rm iso}$:

\begin{equation}
    \tilde{L}_{\rm iso} = \frac{L_{\rm iso}}{\max\limits_\Phi L_{\rm iso}} .
\end{equation}

Additionally, as an integral characteristic of the source's energy output, we will use the phase-averaged luminosity $\Lisoavg$: 

\begin{equation}\label{eq.l_iso_avg}
    \Lisoavg = \int\limits^{1}_{0} L_{\rm iso} {\rm d}\Phi
\end{equation}

\subsection{Symmetry properties of the model and beam pattern}

The model described in the previous subsections allows one to calculate isotropic luminosities as functions of the relative orientation of the observer set by angles $\thetaobs$ and $\varphi$. 
By fixing $\thetaobs$ and considering $\varphi$ as a function of the spin phase $\Phi$, one obtains the pulse profile from the beam pattern. 
This is a commonly used approach both for rotationally and accretion-powered pulsars, valid whenever the beam pattern of the source is stable on the time scales of the spin. The properties of the beam determine the properties of the pulse. For instance, a  compact hot spot on the surface of a slowly rotating NS\footnote{There is a relativistic Doppler boost correction term  important for rapidly rotating pulsars \citep{2000ApJ...531..447B} which makes the profiles asymmetric for spin periods of the order milliseconds.} produces a beam pattern consistent with the RVM.

Broadly viewed, RVM~\citep{Radhakrishnan-Cooke1969} is a beam-pattern model with an infinite-order axis of symmetry, that is identified with the magnetic axis. 
All the pulse profiles in this case are symmetric: mapping $\Phi \to - \Phi$ reproduces the shape of the pulse. 
If this axial symmetry is relaxed, the pulses may become asymmetric, depending on the symmetry properties of the beam.

Our model, and thus its beam pattern, has two independent symmetry elements. 
First is central symmetry, that in terms of coordinate mapping corresponds to invariance under the simultaneous transform $\Phi \to \Phi + 0.5$ and $\thetaobs \to \uppi - \thetaobs$. 
This transform swaps the two polar caps, the corresponding accretion columns and the funnel flows. 
The second symmetry element is mirror reflection with respect to the plane containing the magnetic axis and the middle line of the flow ($\phi \to 2 \phicenter-\phi$).\footnote{Together, these two elements produce the group of symmetry ${\rm C}_{2h}$ consisting of a second-order axis of rotation (orthogonal to the magnetic axis) and a symmetry plane perpendicular to this axis (see for instance \citealt[chapter XII]{landau1981quantum}).} We call this plane the accretion columns' \emph{symmetry plane}.

\section{Results}\label{sec:results}

We have developed a model that calculates the observed pulse profiles for a specific pulsar configuration. 
Each rotational phase corresponds to a specific line-of-sight direction toward the observer and sets the picture plane.
Our model projects the surface elements of the columns and the FF onto the picture plane and integrates their intensity over the solid angle, taking into account visibility conditions, occultation and attenuation.

We also perform a statistical analysis of the entire dataset of the simulation results generated by our model across a grid of input parameters.
The input parameter sets are constructed as the Cartesian product of individual parameter grids. The viewing angle grid, $\thetaobs$, is assumed to be uniform in $\cos \thetaobs$, while the remaining parameters are set by linear uniform grids. The adopted parameter grids may be found in Table~\ref{tab:input params}.

\newcommand{\newParamsValue}[6]{$\mu_{30} = #1$, $\thetarot_{\rm obs} = #2^\circ$, $\muang = #3^\circ$, $\dot{m} = #4$, $a = #5$, $\phicenter = #6^\circ$}

\newcommand{\widthForSubfig}{0.11}

\begin{figure*}[ht]
\captionsetup[subfigure]{labelformat=empty,skip=0pt}
    \centering
    \begin{subfigure}[t]{\widthForSubfig\textwidth}
        \centering
        \includegraphics[width=\linewidth]{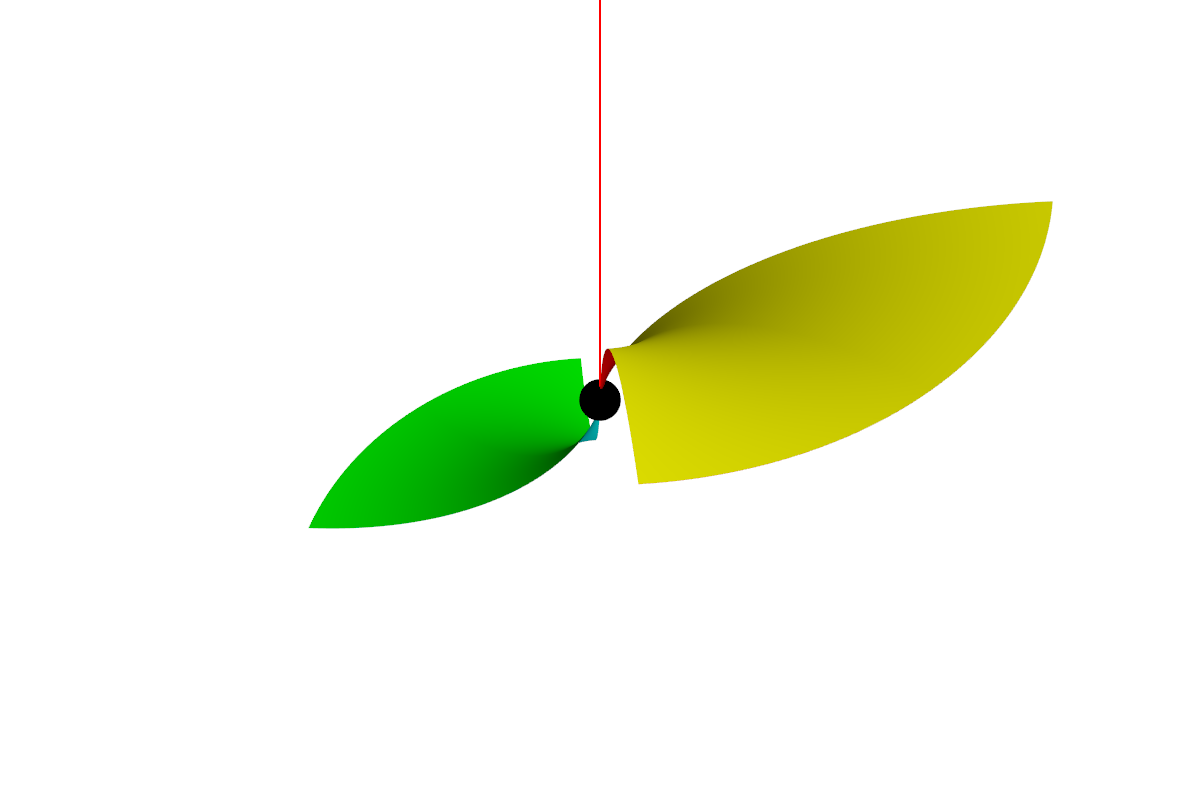}
        \caption{$\Phi$ = 0}
    \end{subfigure}
    ~
    \begin{subfigure}[t]{\widthForSubfig\textwidth}
        \centering
        \includegraphics[width=\linewidth]{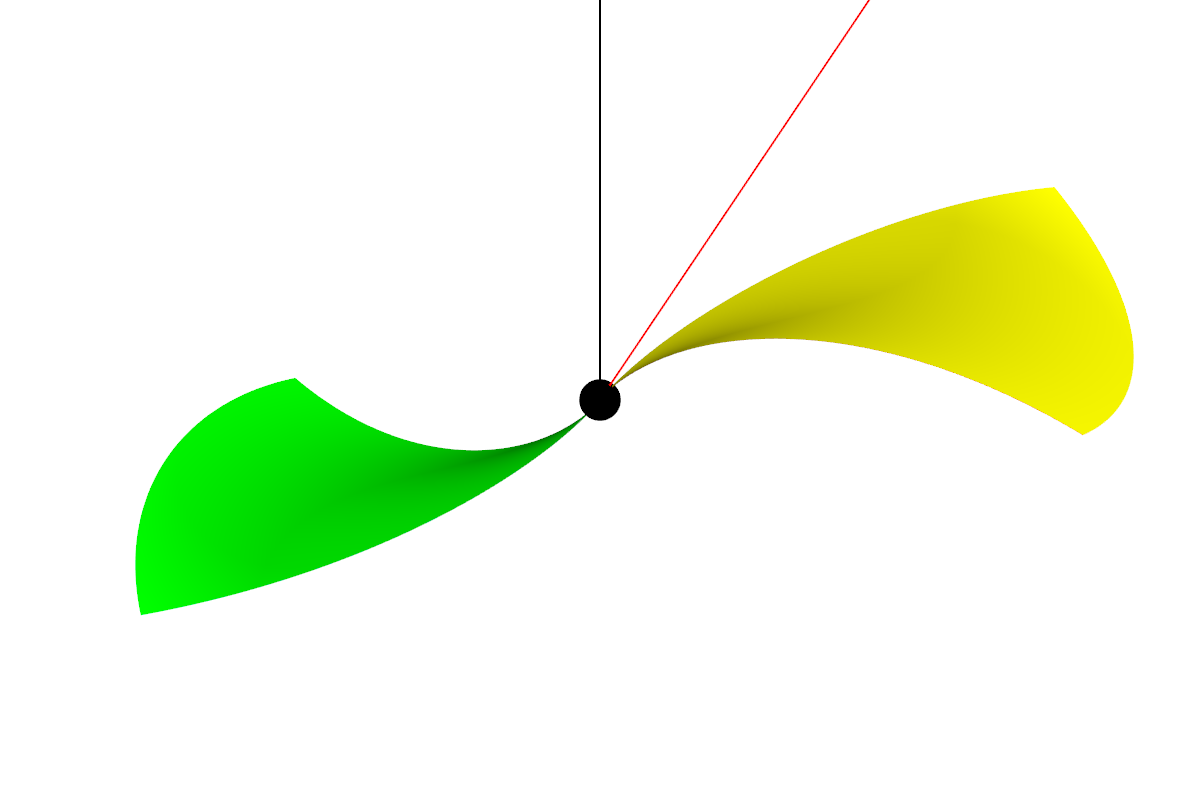}
        \caption{$\Phi$ = 0.12}
    \end{subfigure}
    ~
    \begin{subfigure}[t]{\widthForSubfig\textwidth}
        \centering
        \includegraphics[width=\linewidth]{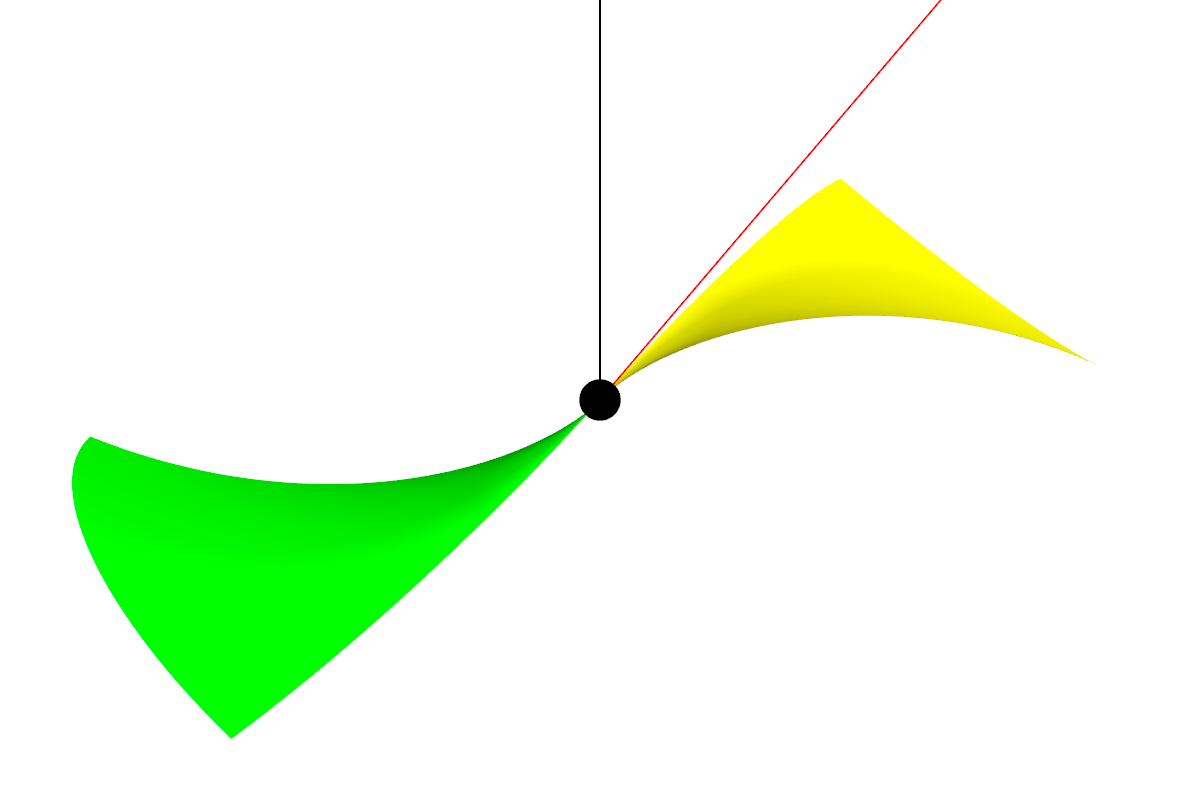}
        \caption{$\Phi$ = 0.25}
    \end{subfigure}
    ~
    \begin{subfigure}[t]{\widthForSubfig\textwidth}
        \centering
        \includegraphics[width=\linewidth]{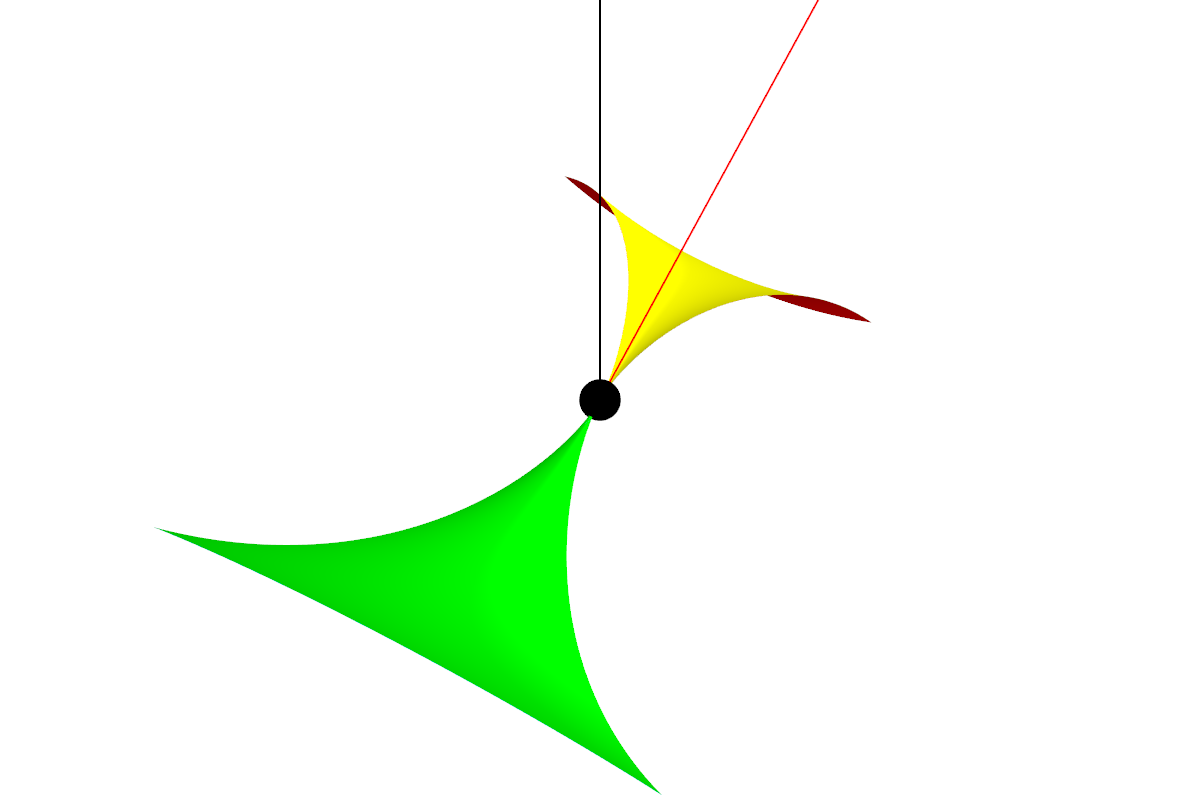}
        \caption{$\Phi$ = 0.38}
    \end{subfigure}
    ~
    \begin{subfigure}[t]{\widthForSubfig\textwidth}
        \centering
        \includegraphics[width=\linewidth]{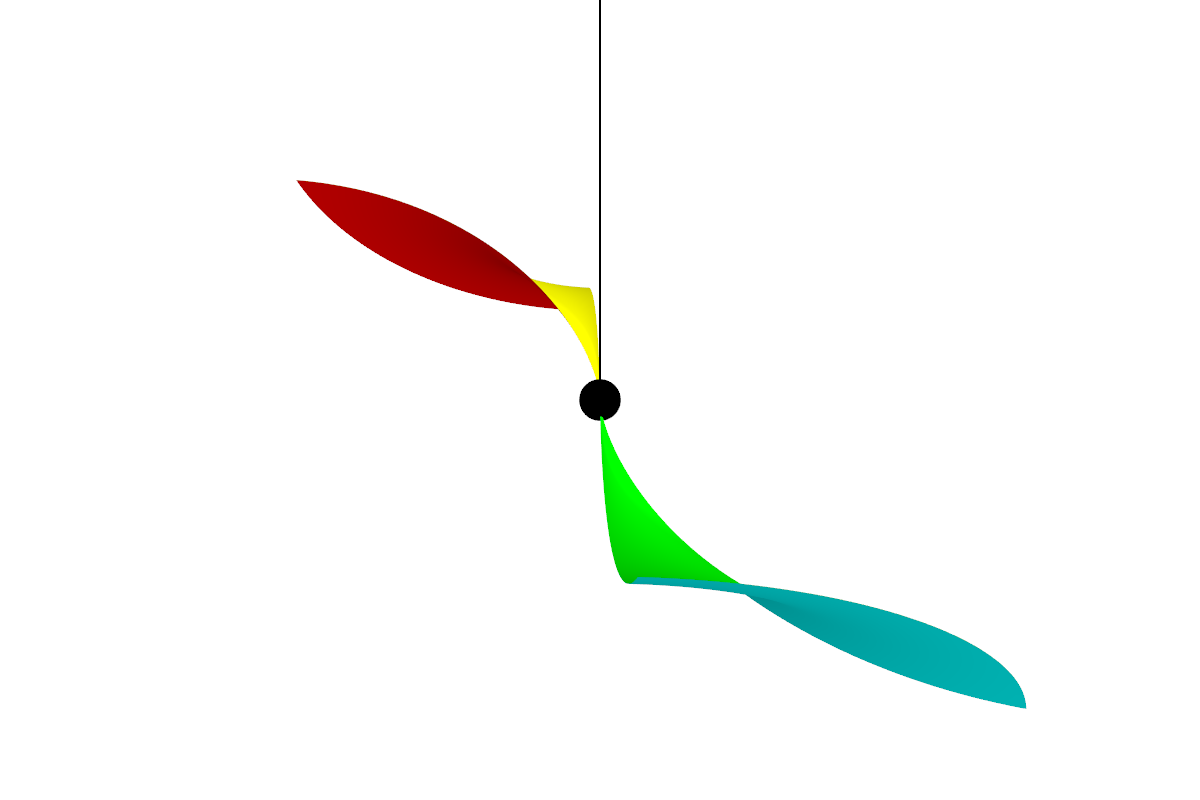}
        \caption{$\Phi$ = 0.5}
    \end{subfigure}
    ~
    \begin{subfigure}[t]{\widthForSubfig\textwidth}
        \centering
        \includegraphics[width=\linewidth]{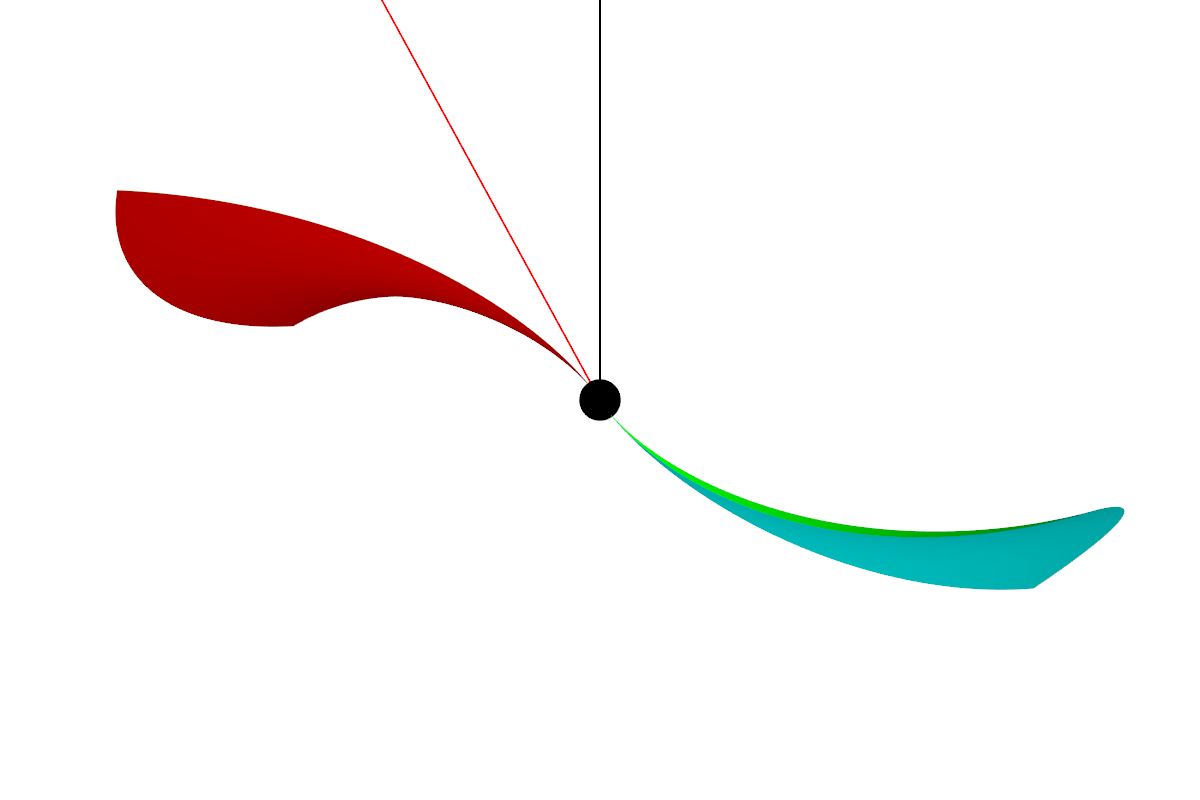}
        \caption{$\Phi$ = 0.62}
    \end{subfigure}
    ~
    \begin{subfigure}[t]{\widthForSubfig\textwidth}
        \centering
        \includegraphics[width=\linewidth]{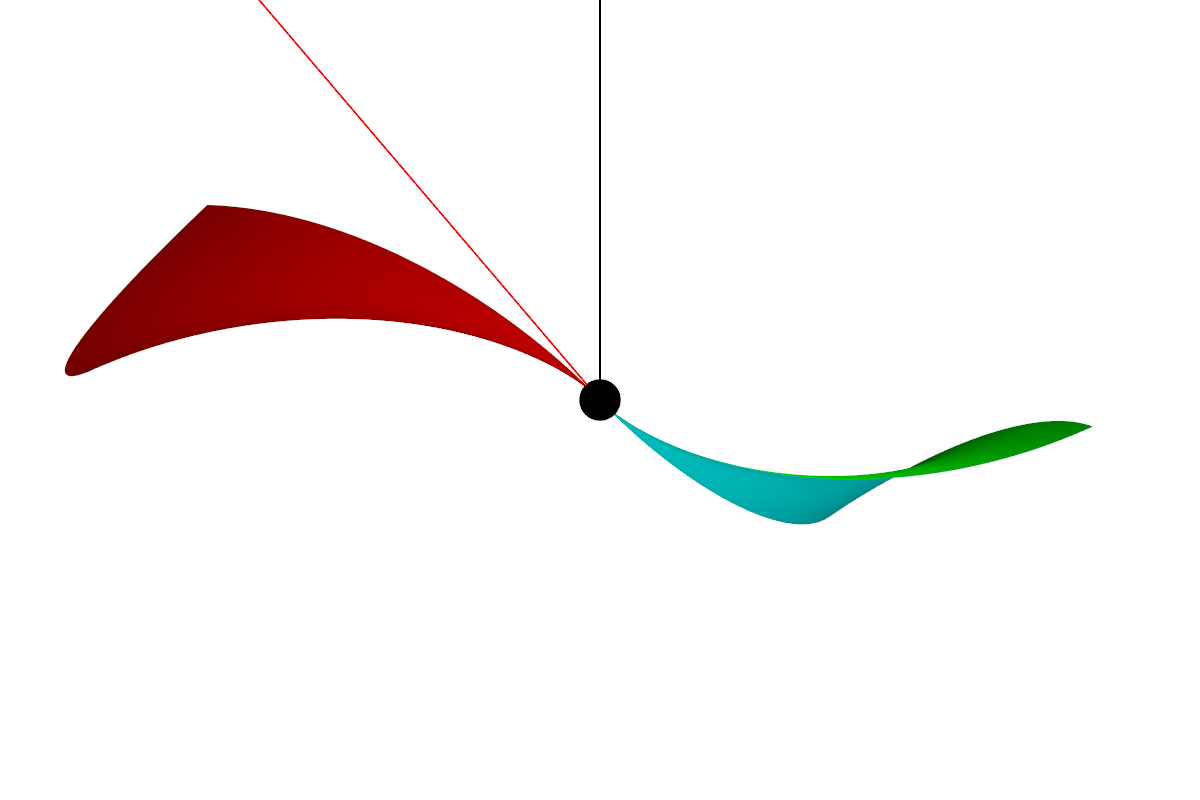}
        \caption{$\Phi$ = 0.75}
    \end{subfigure}
    ~
    \begin{subfigure}[t]{\widthForSubfig\textwidth}
        \centering
        \includegraphics[width=\linewidth]{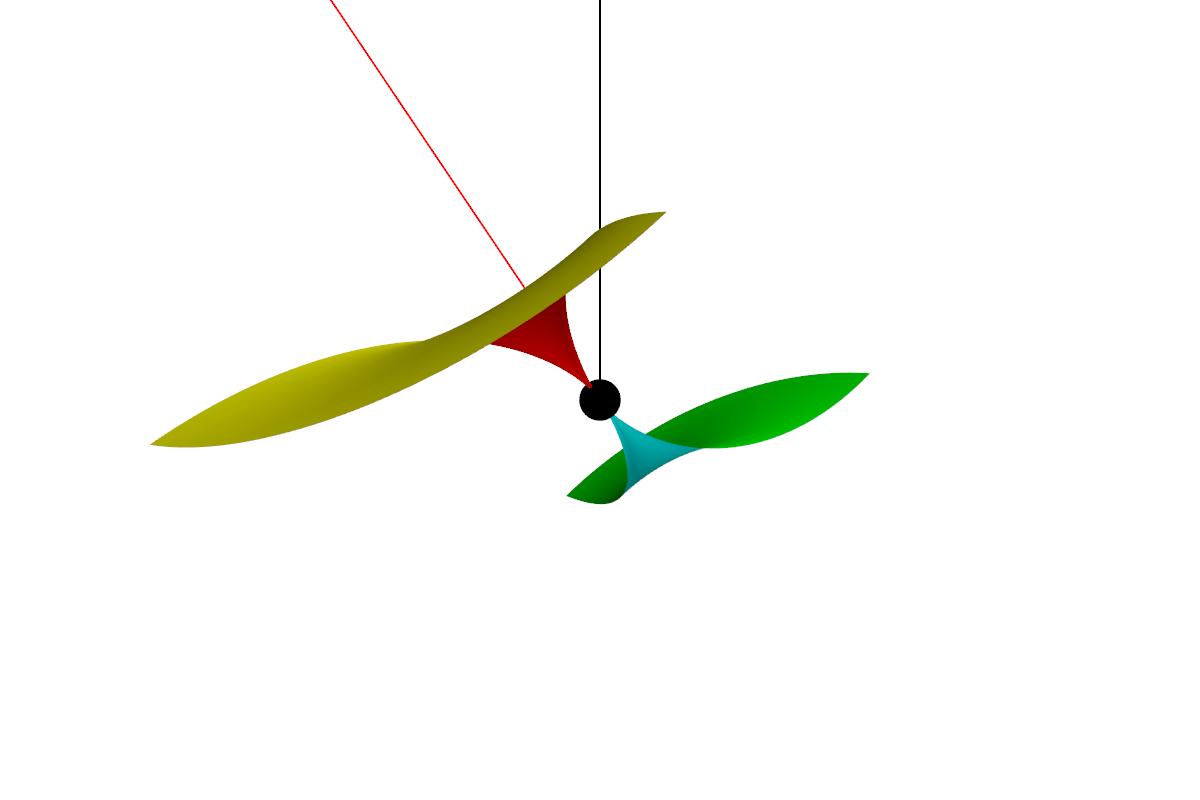}
        \caption{$\Phi$ = 0.88}
    \end{subfigure}
    ~
    ~
    \centering
    \begin{subfigure}[t]{1\textwidth}
        \centering
        \includegraphics[width=1\linewidth]{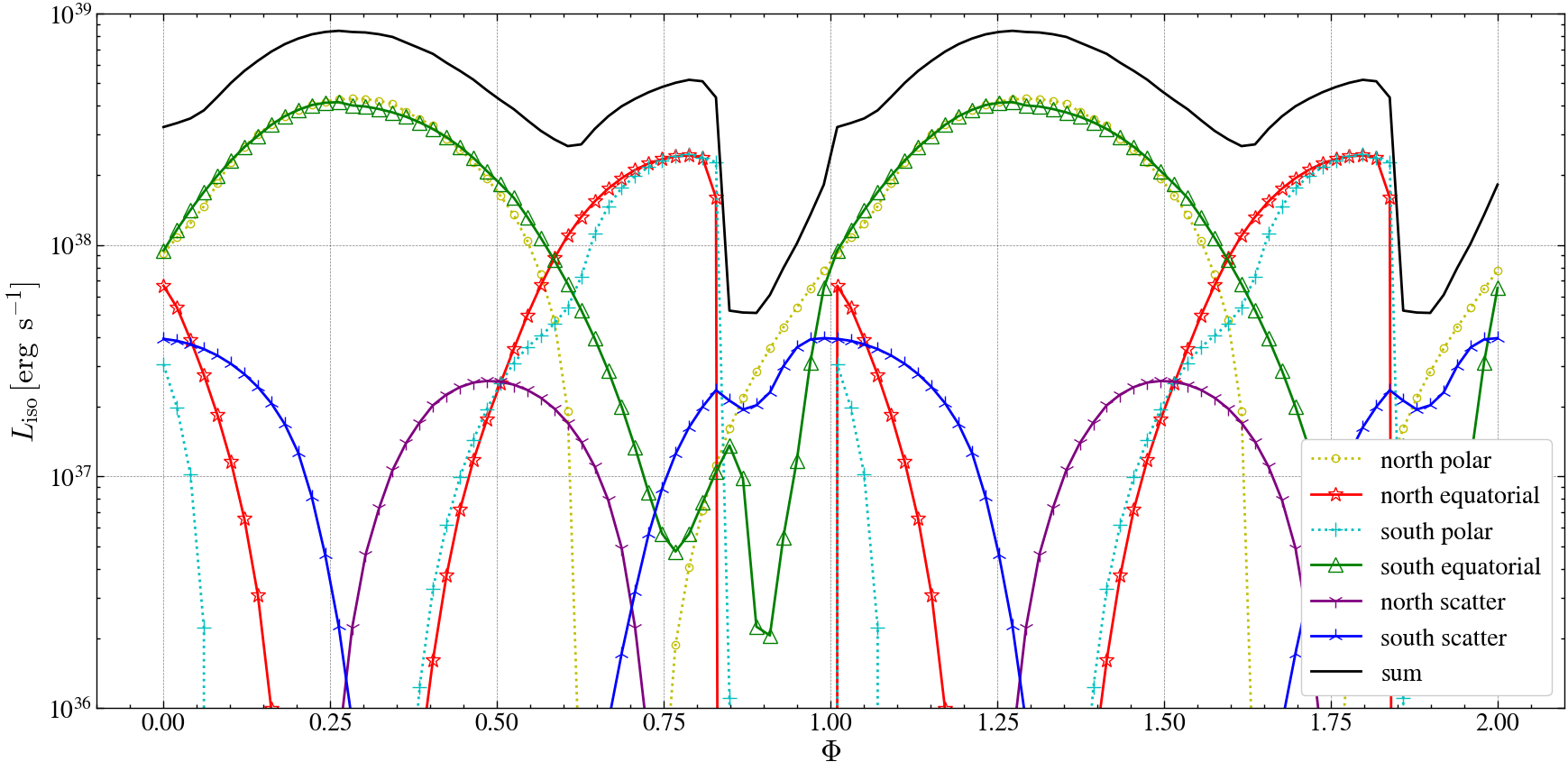}
    \end{subfigure}
    \caption{
        Example of a pulse profile for $\mu_{30} = 1$, $\thetaobs = 80\grad$, $\muang = 40\grad$, 
        $\dot{m} = 100$, $a = 0.2$, $\phicenter = 40\grad$  with the contributions of different radiating surfaces shown separately. The auxiliary upper panels show the 3D renderings of the columns as seen by the observer on different spin phases. The colouring of the surfaces follows the conventions of Fig.~\ref{fig:Geom_config}.
        }
    \label{fig:L_iso phase 2}  
\end{figure*}

\subsection{Pulse profile}\label{sec:Pulses and sky maps}

Figure~\ref{fig:L_iso phase 2} shows an example of a pulse profile with the contributions from the radiating surfaces of the accretion columns and reflecting FF surfaces.
In the auxiliary panels above, we show the 3D-renderings of the columns for an observer at the same $\thetaobs = 80$\textdegree\ for different phases $\Phi$. 

The peaks of the fluxes from specific surfaces are primarily associated with their favourable orientation relative to the line of sight. Effective temperature gradients make the contributions of different parts of the columns unequal.
For example, at $\Phi = 0$, the narrow visible section in the vicinity of the NS of the northern equatorial surface (red) emits a flux comparable in magnitude to that from the much larger visible area  of the northern polar surface (yellow). 

Figure~\ref{fig:L_iso phase 2} clearly demonstrates the influence of attenuation on the pulse shape. 
In particular, the sharp drop of the flux from the north equatorial surface (solid red line)  at $\Phi\sim 0.85$ is due to its self-attenuation. 
At about the same phase, the south polar (cyan dotted line) surface starts to be eclipsed by the NS and northern column. For the configuration we show, several per cent of the observed radiation is reflected by FF, but different configurations may have much larger reflected radiation contributions.
Attenuation and reflection by FF significantly affect the pulse fraction (PF), defined as
\begin{equation}\label{eq:PF}
    {\rm PF} = \frac{\max\limits_\Phi(L_{\rm iso}) - \min\limits_\Phi(L_{\rm iso})}{\max\limits_\Phi(L_{\rm iso}) + \min\limits_\Phi(L_{\rm iso})} \,  ,
\end{equation}
where the maximum and minimum are calculated for one rotation period.
The role of attenuation and reflection will be considered in detail in Section~\ref{sec:impacts}.

\newcommand{\paramsValue}{$\thetarot_{\rm obs} = 60^\circ$, $\muang = 40^\circ$, $\dot{m}=30$, $a = 0.65$, $\phicenter = 0^\circ$}

\renewcommand{\widthForSubfig}{0.32}

\begin{figure*}[!ht]
    \centering
    \begin{subfigure}[t]{\widthForSubfig\textwidth}
        \centering
        \includegraphics[width=\linewidth]{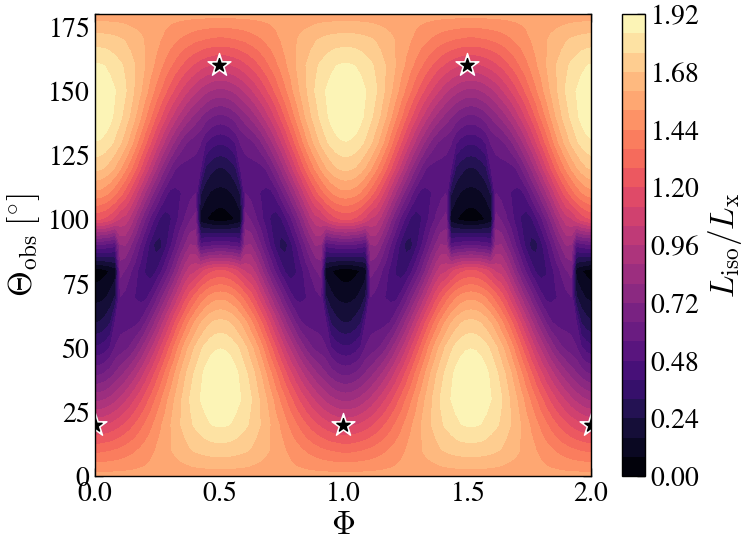}
        \caption{$\muang = 20^\circ$, $\dot{m}=60$, $a = 0.2$, $\phicenter = 0^\circ$}

    \end{subfigure}\hfil
    ~
    \begin{subfigure}[t]{\widthForSubfig\textwidth}
        \centering
        \includegraphics[width=\linewidth]{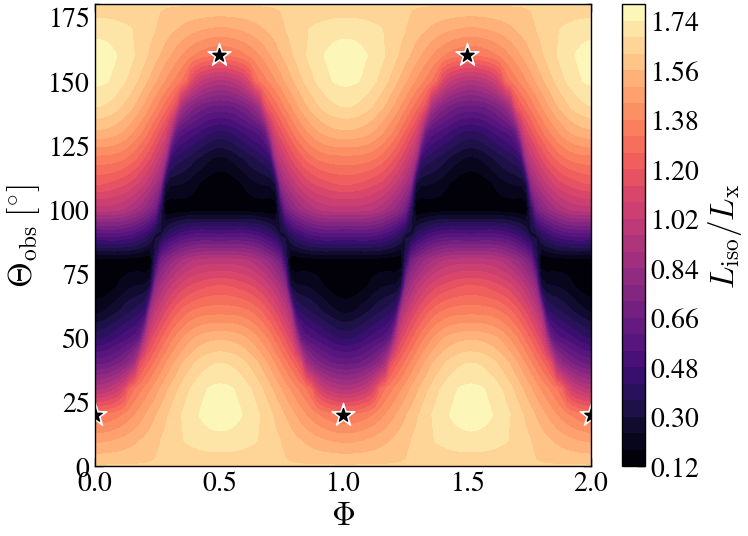}
        \caption{$\muang = 20^\circ$, $\dot{m}=60$, $a = 0.5$, $\phicenter = 0^\circ$}
    \end{subfigure}\hfil
    ~
    \begin{subfigure}[t]{\widthForSubfig\textwidth}
        \centering
        \includegraphics[width=\linewidth]{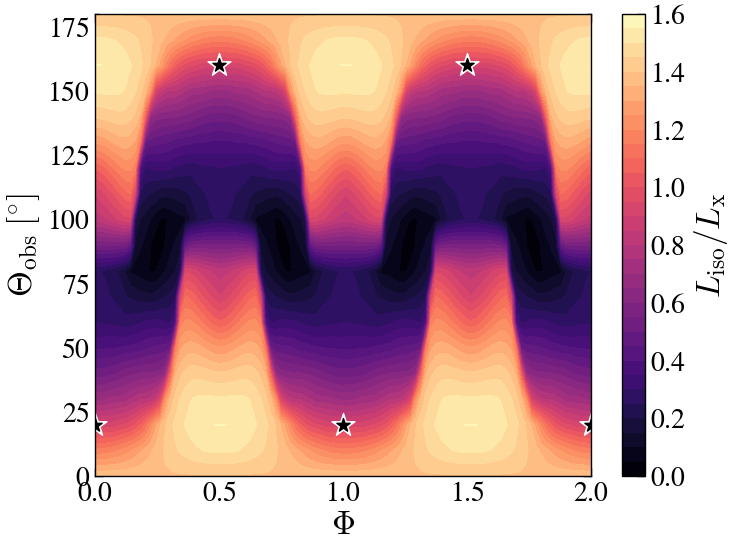}
        \caption{$\muang = 20^\circ$, $\dot{m}=60$, $a = 0.7$, $\phicenter = 0^\circ$}
    \end{subfigure}\hfil

    \begin{subfigure}[t]{\widthForSubfig\textwidth}
        \centering
        \includegraphics[width=\linewidth]{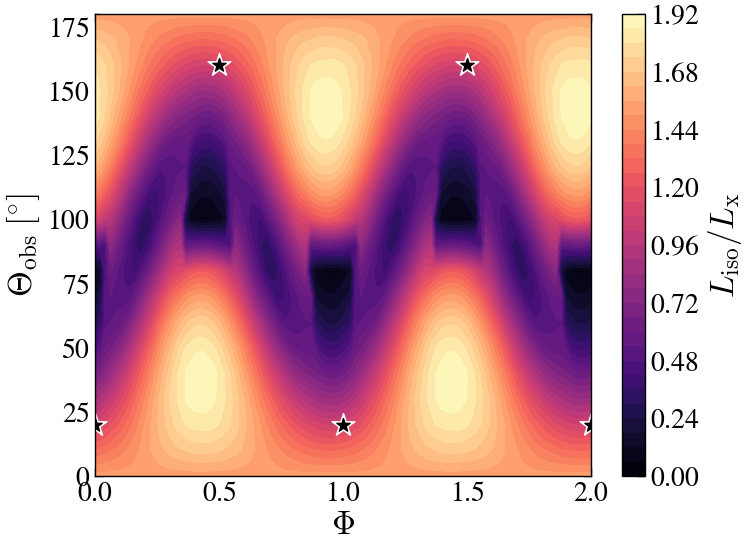}
        \caption{$\muang = 20^\circ$, $\dot{m}=60$, $a = 0.2$, $\phicenter = 20^\circ$}

    \end{subfigure}\hfil
    ~
    \begin{subfigure}[t]{\widthForSubfig\textwidth}
        \centering
        \includegraphics[width=\linewidth]{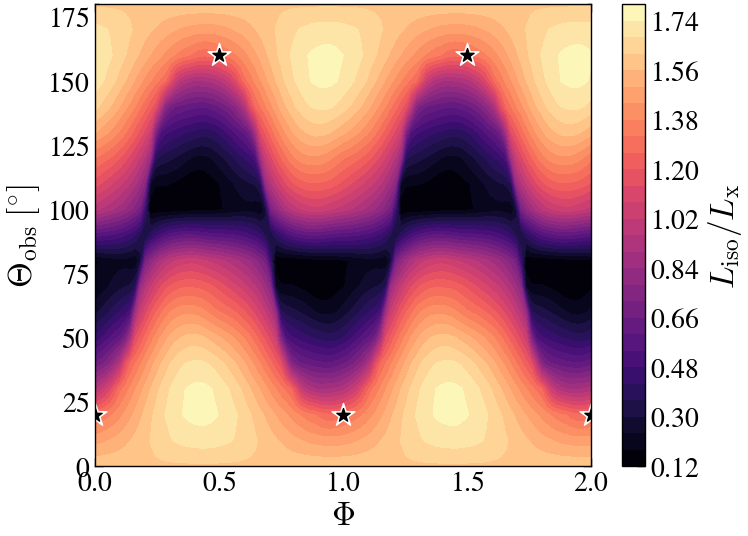}
        \caption{$\muang = 20^\circ$, $\dot{m}=60$, $a = 0.5$, $\phicenter = 20^\circ$}
    \end{subfigure}\hfil
    ~
    \begin{subfigure}[t]{\widthForSubfig\textwidth}
        \centering
        \includegraphics[width=\linewidth]{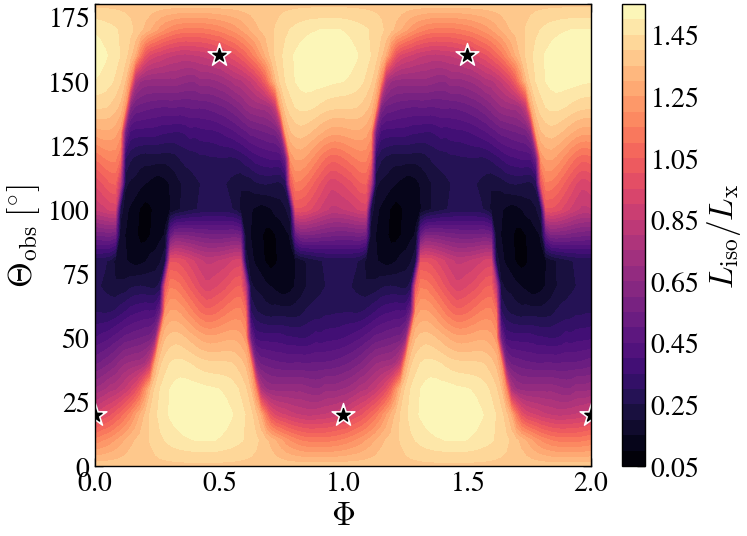}
        \caption{$\muang = 20^\circ$, $\dot{m}=60$, $a = 0.7$, $\phicenter = 20^\circ$}
    \end{subfigure}\hfil

    \begin{subfigure}[t]{\widthForSubfig\textwidth}
        \centering
        \includegraphics[width=\linewidth]{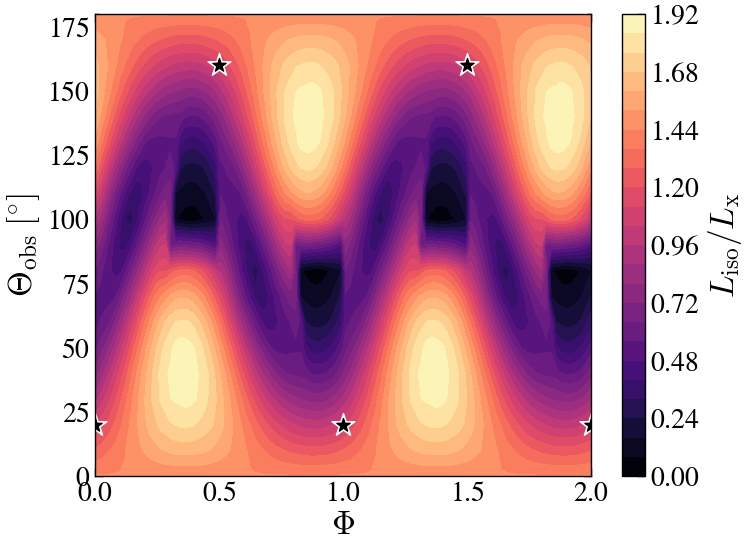}
        \caption{$\muang = 20^\circ$, $\dot{m}=60$, $a = 0.2$, $\phicenter = 40^\circ$}

    \end{subfigure}\hfil
    ~
    \begin{subfigure}[t]{\widthForSubfig\textwidth}
        \centering
        \includegraphics[width=\linewidth]{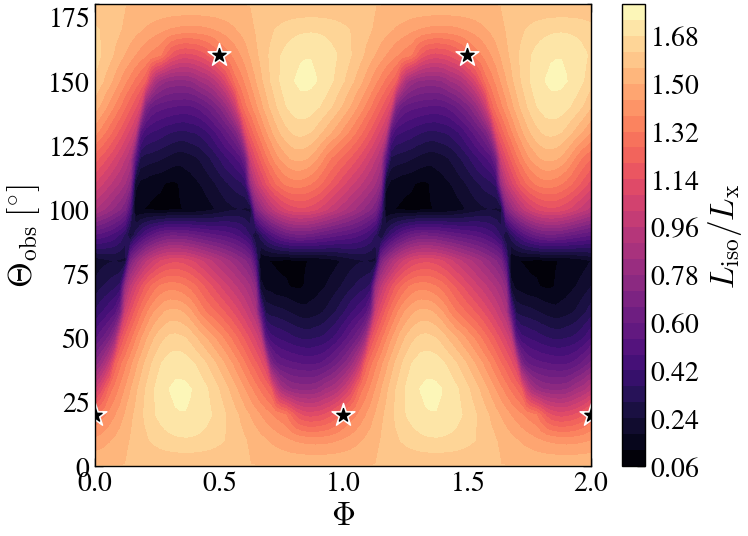}
        \caption{$\muang = 20^\circ$, $\dot{m}=60$, $a = 0.5$, $\phicenter = 40^\circ$}
    \end{subfigure}\hfil
    ~
    \begin{subfigure}[t]{\widthForSubfig\textwidth}
        \centering
        \includegraphics[width=\linewidth]{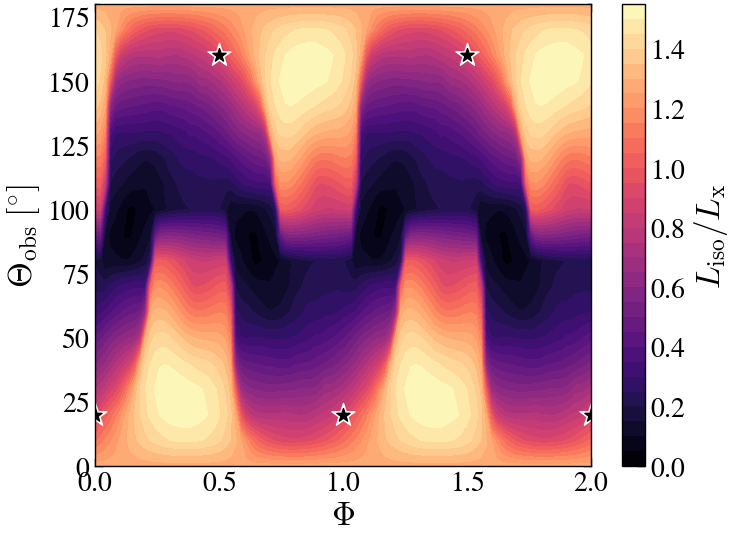}
        \caption{$\muang = 20^\circ$, $\dot{m}=60$, $a = 0.7$, $\phicenter = 40^\circ$}
    \end{subfigure}\hfil

    \begin{subfigure}[t]{\widthForSubfig\textwidth}
        \centering
        \includegraphics[width=\linewidth]{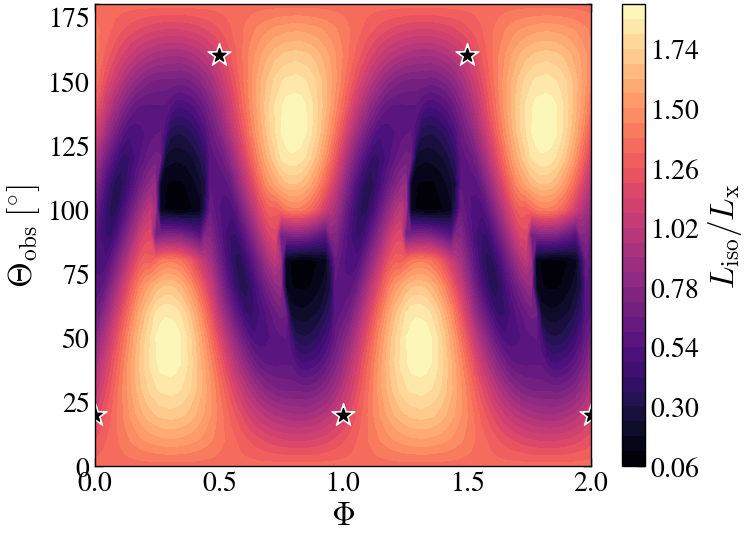}
        \caption{$\muang = 20^\circ$, $\dot{m}=60$, $a = 0.2$, $\phicenter = 60^\circ$}

    \end{subfigure}\hfil
    ~
    \begin{subfigure}[t]{\widthForSubfig\textwidth}
        \centering
        \includegraphics[width=\linewidth]{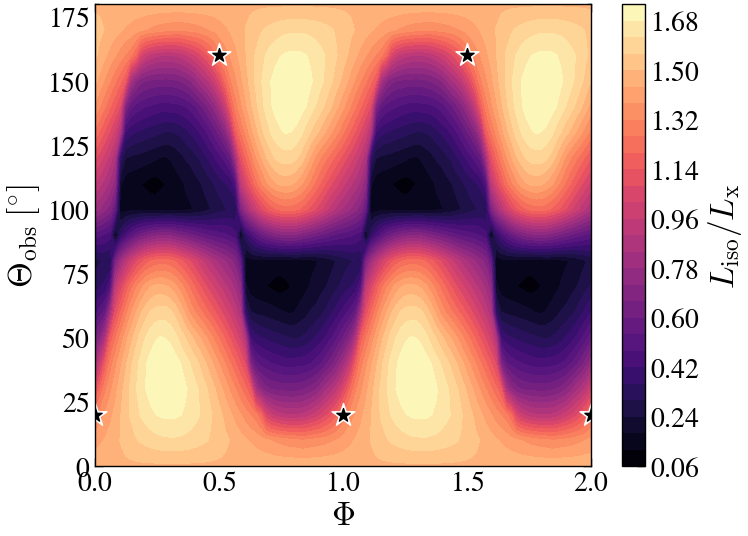}
        \caption{$\muang = 20^\circ$, $\dot{m}=60$, $a = 0.5$, $\phicenter = 60^\circ$}
    \end{subfigure}\hfil
    ~
    \begin{subfigure}[t]{\widthForSubfig\textwidth}
        \centering
        \includegraphics[width=\linewidth]{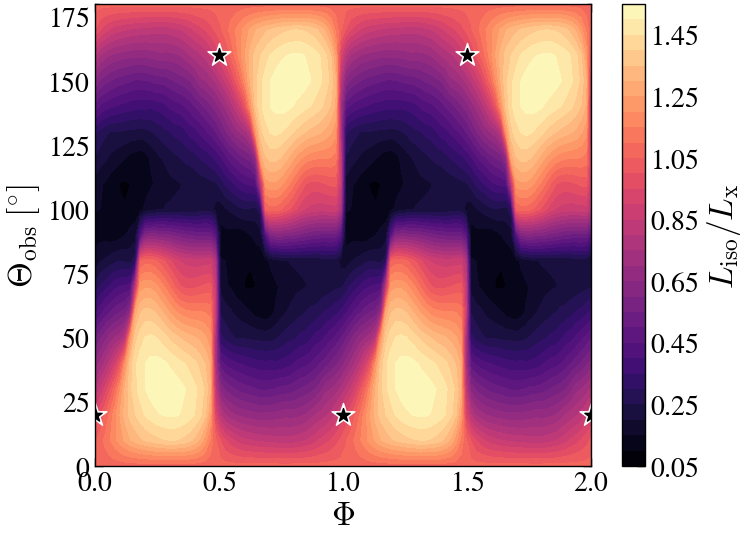}
        \caption{$\muang = 20^\circ$, $\dot{m}=60$, $a = 0.7$, $\phicenter = 60^\circ$}
    \end{subfigure}\hfil

    \caption{Sky maps of the  normalized isotropic luminosity, colour coded.
    The isotropic luminosity is normalized by the columns' power $\Lx$ calculated according to Eq.~\ref{eq.total_columns_power}. 
    Stars indicate magnetic poles.
    }
    \label{fig:sky_map}  
\end{figure*}

\subsection{Sky maps}\label{sec:sky_map}

A useful tool for displaying the variety of pulse profiles is a sky map. 
Sky map is a representation of the beam pattern of the source in the coordinates of the observer's inclination $\thetaobs$ and the spin phase $\Phi$. Each horizontal cross-section of a sky map is the pulse profile measured by the observer at the corresponding $\thetaobs$. Figure~\ref{fig:sky_map}  shows a sample of sky maps with different values of $a$ and $\phicenter$.  The colour-coded bolometric isotropic luminosity  $L_{\rm iso}$ is calculated according to Eq.~\eqref{eq:final L} and normalized by the power of the columns  $L_{\rm x}$.
For all the panels in Fig.~\ref{fig:sky_map}, the accretion rate $\dot{m}$ is the same, but $L_{\rm x}$ is different as it depends on $a$. 
Stars in Fig.~\ref{fig:sky_map} indicate the position of the north and south magnetic poles.

The isotropic luminosity variations in all the sky maps are a combination of two different effects. 
The smooth changes result from gradual changes in the visible radiating areas, while the rapid changes come mostly from eclipses or attenuation in the FF.

In the right row of Fig.~\ref{fig:sky_map} the azimuthal filling $a$ is the largest. 
For large $a$, changes related to FF attenuation affect the largest areas on the sky map. 
Also, the column height is the smallest (see also Fig. \ref{fig:xi to m}), that makes the dark regions of the maps deeper and sharper.

For $\phicenter=0$ and $|\thetaobs - 90\grad| \gtrsim 10\grad$, there is a single maximum at $\Phi\sim 0$ or $0.5$, depending on the observer's hemisphere.
For $\phicenter=0$, the pulse profiles are symmetrical with respect to the phase of the maximum or the minimum.
For positive $\phicenter$, the maximum shifts to smaller phases, its position becomes dependent on $\thetaobs$, and the shape of the pulse becomes asymmetric.
We will consider the asymmetries of the pulse shapes in \S\ref{s.asymmtry} and the role of $\phicenter$ in \S\ref{s:dep_on_phi}.

\renewcommand{\widthForSubfig}{0.32}

\begin{figure*}[!ht]
    \centering
    \begin{subfigure}[t]{\widthForSubfig\textwidth}
        \centering
        \includegraphics[width=\linewidth]{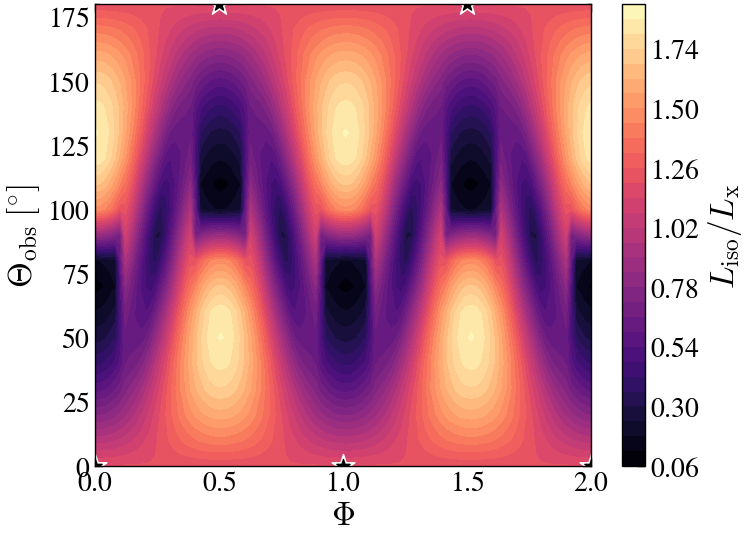}
        \caption{$\muang = 0^\circ$, $\dot{m}=60$, $a = 0.2$, $\phicenter = 0^\circ$}

    \end{subfigure}\hfil
    ~
    \begin{subfigure}[t]{\widthForSubfig\textwidth}
        \centering
        \includegraphics[width=\linewidth]{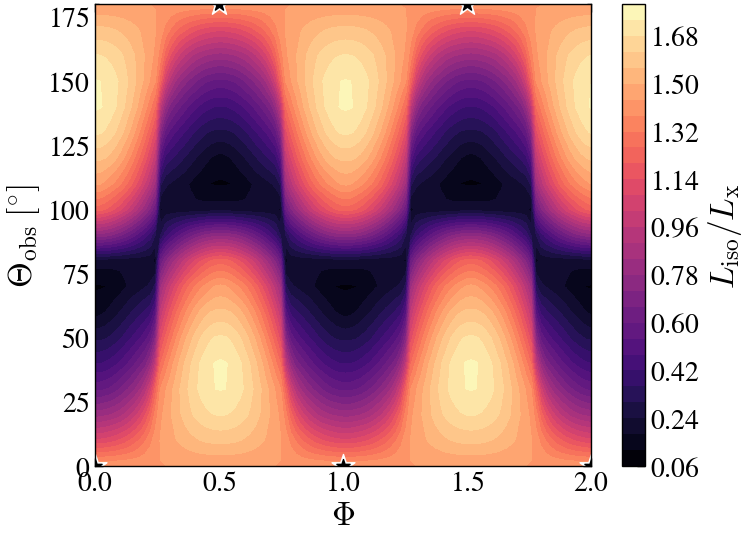}
        \caption{$\muang = 0^\circ$, $\dot{m}=60$, $a = 0.5$, $\phicenter = 0^\circ$}
    \end{subfigure}\hfil
    ~
    \begin{subfigure}[t]{\widthForSubfig\textwidth}
        \centering
        \includegraphics[width=\linewidth]{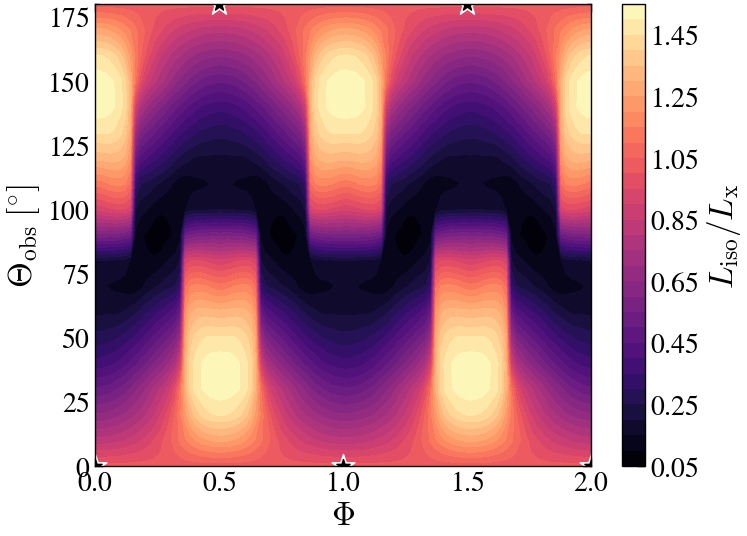}
        \caption{$\muang = 0^\circ$, $\dot{m}=60$, $a = 0.7$, $\phicenter = 0^\circ$}
    \end{subfigure}\hfil

    \caption{Sky maps of the  normalized isotropic luminosity (colour coded) for $\muang = 0$.
    The isotropic luminosity is normalized by the columns' power $\Lx$ calculated according to Eq.~\ref{eq.total_columns_power}.
    Stars indicate magnetic poles.
    Section with $\thetarot_{\rm obs} = $const gives a pulse seen by an observer located at $\thetarot_{\rm obs}$.
    }
    \label{fig:sky_map_mu_0}  
\end{figure*}

In Fig.~\ref{fig:sky_map_mu_0}, we show that even for zero magnetic angle $\muang=0$, the pulses are observed {under the conditions $a<1$ and \hbox{$0 < \thetaobs < 180^\circ$.}}

\subsection{Asymmetry of pulses}
\label{s.asymmtry}

Unless the central azimuth of the column $\phicenter=0$ or $\phicenter=180^\circ$, the pulse shapes produced by our model are in general not symmetric, meaning there is no such value $\Phi_0$ that maps the pulse into itself with the $\Phi \to 2 \Phi_0 - \Phi$ transform. 
This is clearly visible for larger $\phicenter$ in Fig.~\ref{fig:sky_map}. 
To measure the asymmetry of some signal $f(\Phi)$, we propose the following metric
\begin{equation}\label{eq:asym_integral}
    \mathcal{A} = \min_{\Phi_0 \in \textrm{Phase} [0, 1]} \frac{\sqrt{\int\limits^{1}_{0} \left[f(2\Phi_0 - \Phi) - f(\Phi)\right]^2 ~ \diff \Phi}}
    {f_{\rm max}-f_{\rm min}}\, .
\end{equation}
For a uniform discrete phase coverage containing $n$ equal phase bins, 
\begin{equation}\label{eq:asym_nrmse}
    \mathcal{A} = \min_{\Phi_0 \in \{\Phi_i\}} \frac{\sqrt{\frac{1}{N} \sum\limits_{i}^{N} \left[f(2 \Phi_0 - \Phi_i)  - f(\Phi_i)\right]^2}}
    {f_{\rm max}-f_{\rm min}}\, ,
\end{equation}
where $f$ is the pulse profile and the phase $ \Phi_0 \in [0,1]$.
The minimal possible value of $\mathcal{A}$ is zero and is acquired only if the profile is symmetric. 
If a periodical function $f(\Phi)$ linearly depends on $\Phi$ for $\Phi=0..1$ (sawtooth profile), then $\mathcal{A}$ $ = (2\sqrt{3})^{-1} \simeq 0.29$.

The discrete nature of Eq.~\eqref{eq:asym_nrmse} creates a systematic error that makes $\mathcal{A}$ non-zero even for a perfectly symmetric profile.
The irreducible error can be estimated by applying a phase offset to a symmetric reference pulse (we assumed a sinusoidal signal). Our evaluations confirm that this error depends on the number of grid points over the phase $\Phi$ as $\Delta \mathcal{A} \approx 2/N$. For $N=50$ points, $\Delta \mathcal{A} \sim 4.4 \times 10^{-2}$.

\begin{figure}[!htb]
    \centering
    \includegraphics[width=1\linewidth]{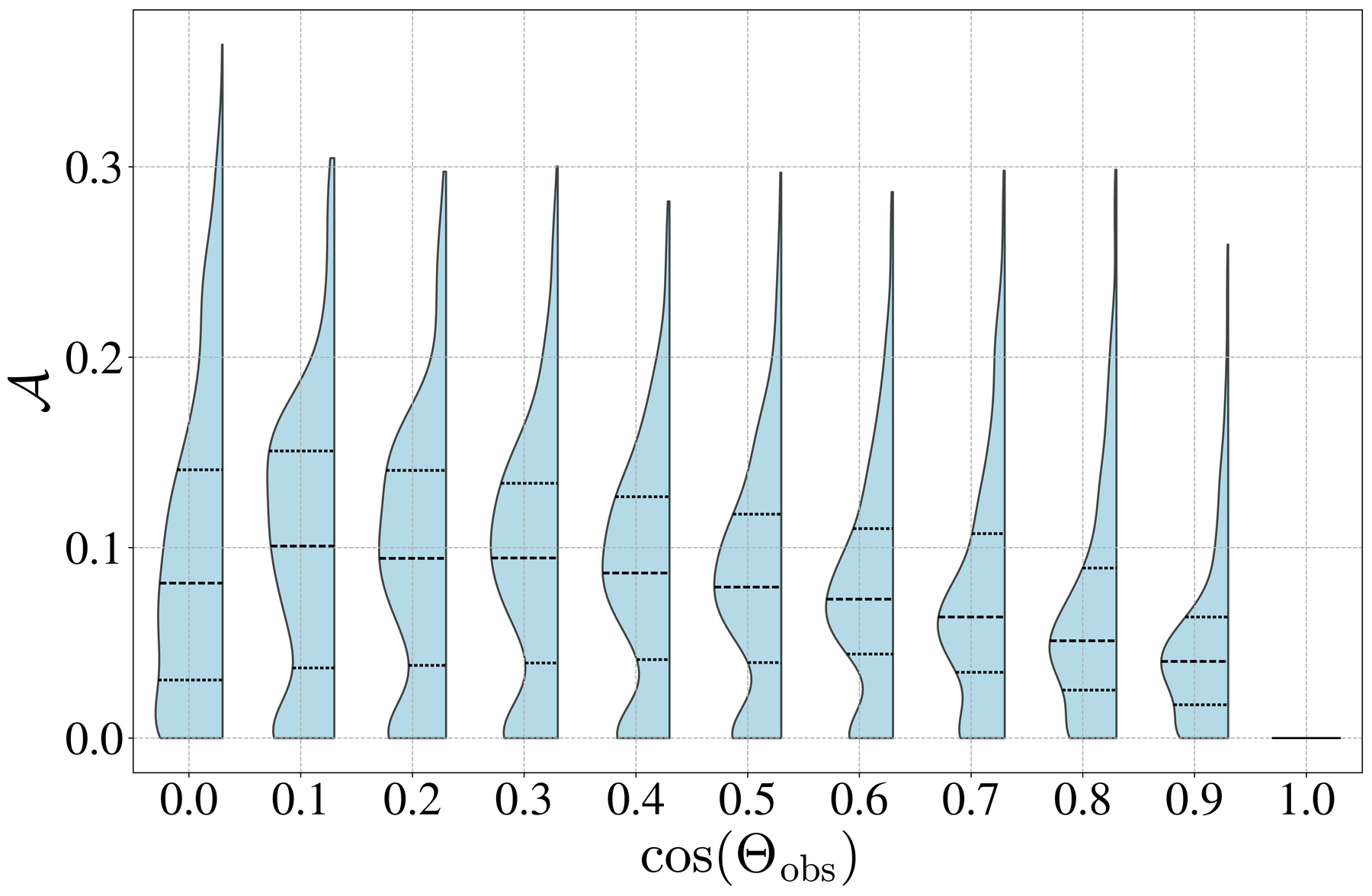}
    \caption{Distributions of asymmetry $\mathcal{A}$ for different $\thetaobs$.
    Individual violin plots show estimated probability density functions of asymmetry $\mathcal{A}$. 
    Dashed and dotted lines show the median and quartiles (25 and 75\%) of the distributions, respectively.
    }
    \label{fig:asymmetry_to_theta_obs_box}
\end{figure}

\begin{figure}[!htb]
    \centering
    \includegraphics[width=1\linewidth]{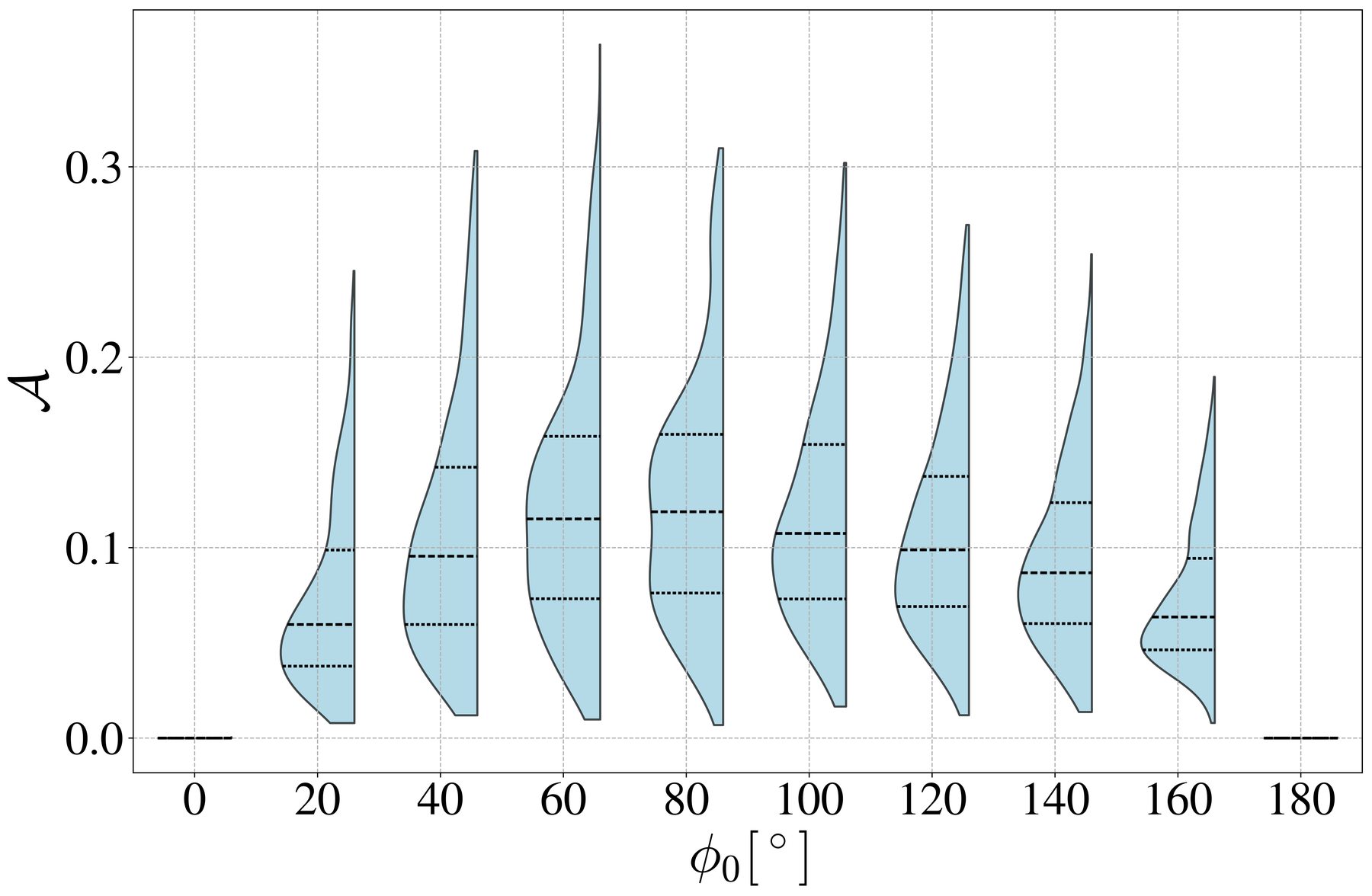}
    \caption{
    Distributions in asymmetry $\mathcal{A}$ for different $\phicenter$. All the notation is the same as in Fig.~\ref{fig:asymmetry_to_theta_obs_box}. 
    }
    \label{fig:asymmetry_phi_0_box}
\end{figure}

\begin{figure}[!htb]
    \centering
    \includegraphics[width=1\linewidth]{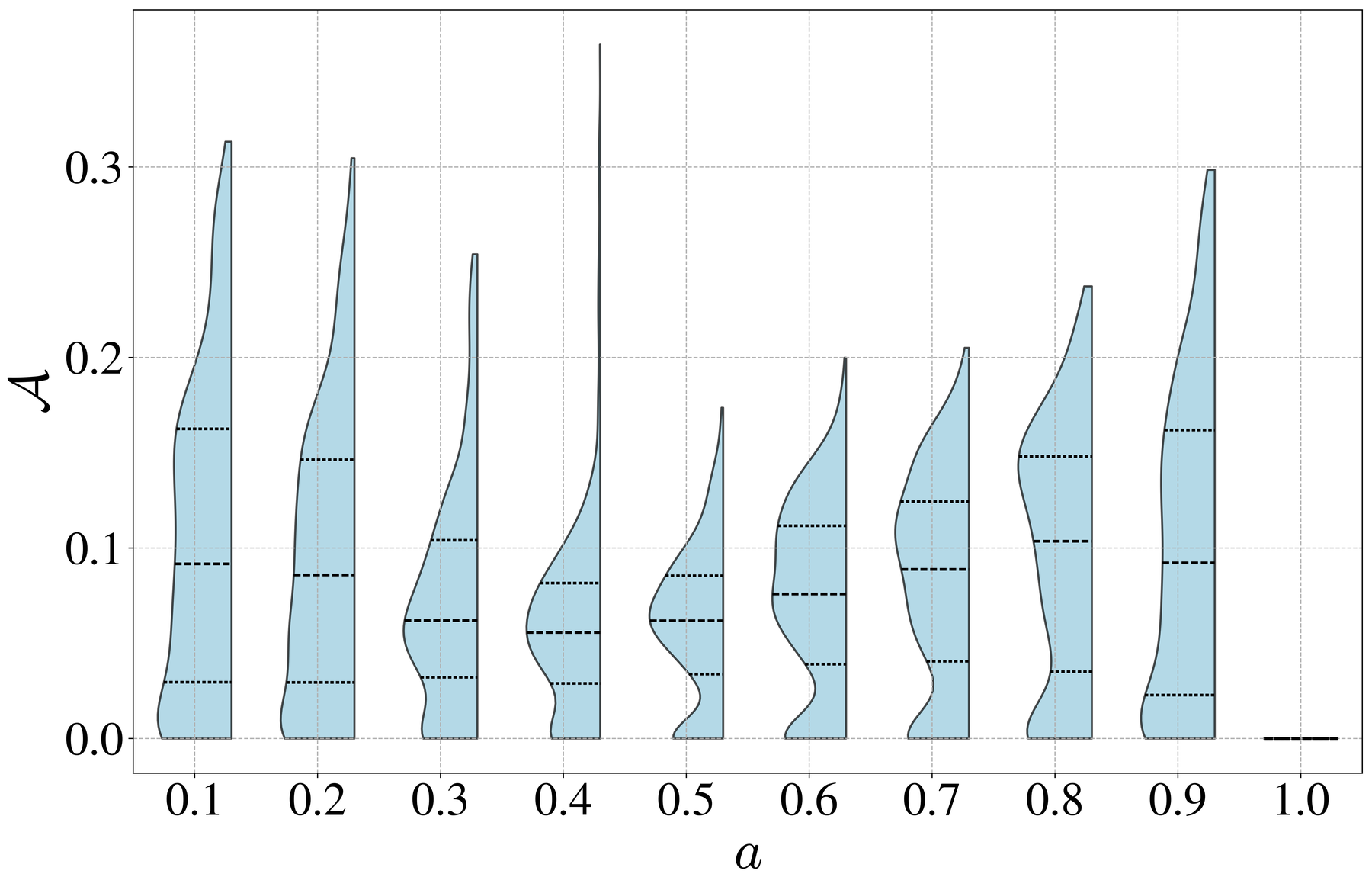}
    \caption{Distributions in asymmetry $\mathcal{A}$ for different $a$. All the notation is the same as in Fig.~\ref{fig:asymmetry_to_theta_obs_box}. 
    }
    \label{fig:asymmetry_a_box}
\end{figure}

When $\thetaobs = 0$, there are no pulsations (see Fig.~\ref{fig:asymmetry_to_theta_obs_box}) and $\mathcal{A}$ is not defined. 
Therefore, we remove this case from the analysis of the symmetry metric of the pulses, and from Figs.~\ref{fig:asymmetry_phi_0_box} and \ref{fig:asymmetry_a_box}, which show the distributions in $\mathcal{A}$ for different $\phicenter$ and $a$. 
For the sample of models described by Table~\ref{tab:input params}, 95\% of metric values lie in the range between 0 and 0.236 (models with $\thetaobs=0$ are excluded).

Pulse profiles are symmetrical ($\mathcal{A}=0$) in the following particular cases:
\begin{itemize}
    \item
    The accretion columns' symmetry plane contains the rotation axis of the NS.
    As demonstrated in Fig.~\ref{fig:asymmetry_phi_0_box}, the metric $\mathcal{A}$ becomes zero 
    when $\phicenter = 0\grad$ or $\phicenter=180\grad$. 
    At intermediate $\phicenter$ values, this alignment is lost, allowing asymmetry to emerge.
    \item Full azimuthal filling. When $a=1$, the parameter $\phicenter$ becomes physically meaningless as the column occupies all azimuthal angles. The system reduces to the $\phicenter=0$ symmetric case, with the symmetry plane containing $\mathbf{\Omega}$, resulting in  zero metric values (see Fig.\ref{fig:asymmetry_a_box}).
    \item Magnetic angle $\muang = 0$.  In this case, because the rotational and magnetic axes coincide, the parameter $\phicenter$ reduces to a pure phase shift. 
    When $a < 1$ and $\muang = 0$, pulsations arise due to the rotation of the column, whose projection onto the observer's plane varies with phase. 
\end{itemize}

\begin{figure}[!htb]
    \centering
    \includegraphics[width=1\linewidth]{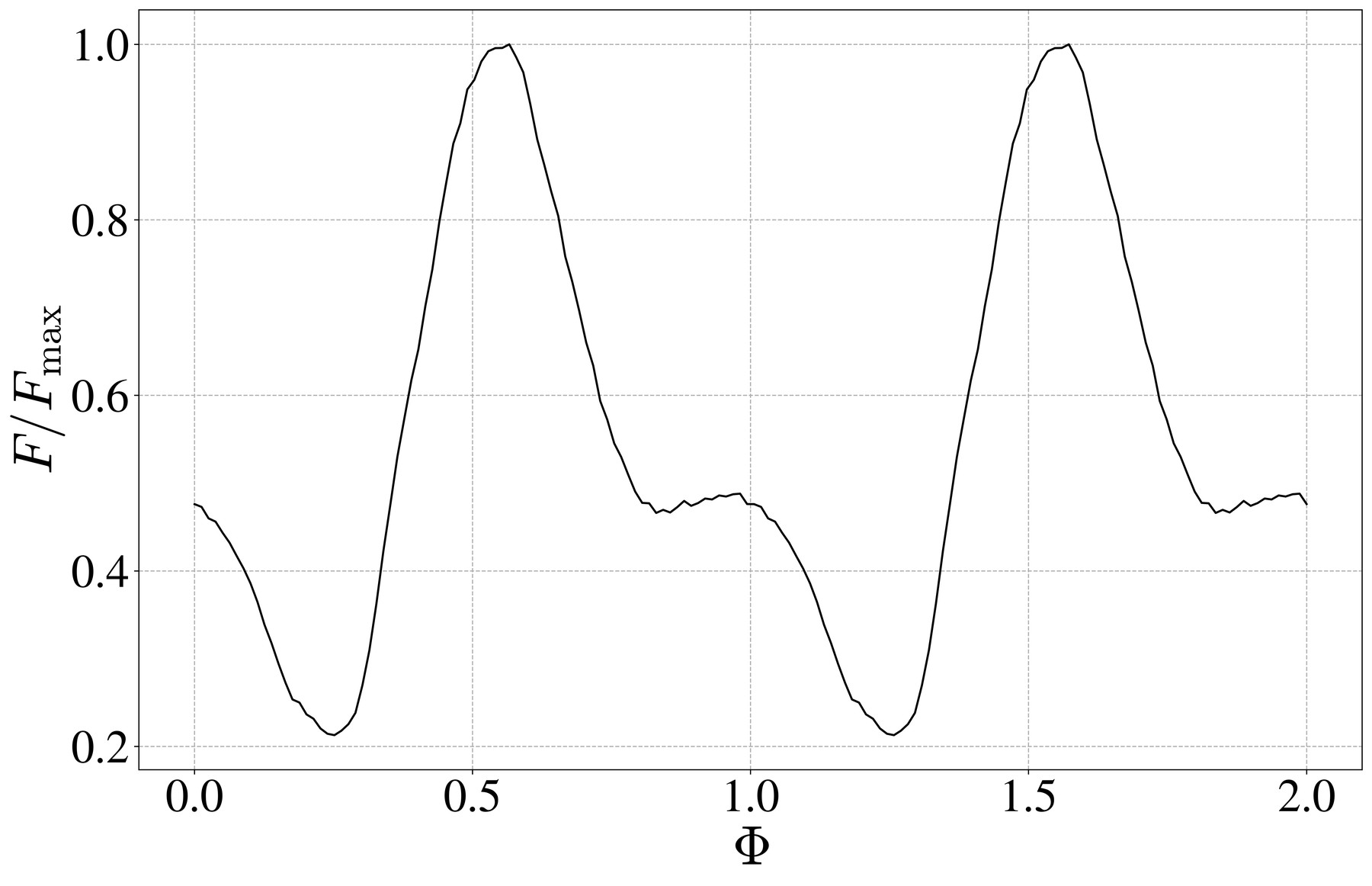}
    \caption{Normalized flux of Cen X-3 at 2-8 keV energy band.}
    \label{fig:cen-x3}
\end{figure}

To illustrate the relation between the metric and the observed profile morphology, we apply it to a real source. 
We took the pulse profilve obtained by \citet{2022ApJ...941L..14T} for \object{Cen~X-3} in the $2-8$keV energy range (shown in Fig.~\ref{fig:cen-x3}) and obtained for it $\mathcal{A} = 0.19 \pm 0.025$.
The metric $\mathcal{A}$ is defined solely by the pulse profile shape and can be applied to any observed profile, regardless of the source's luminosity or energy band.

\subsection{Impact of attenuation and reflection by the optically thick FF}\label{sec:impacts}

\begin{figure*}[!htb]
    \centering    
    \includegraphics[width=1\textwidth]{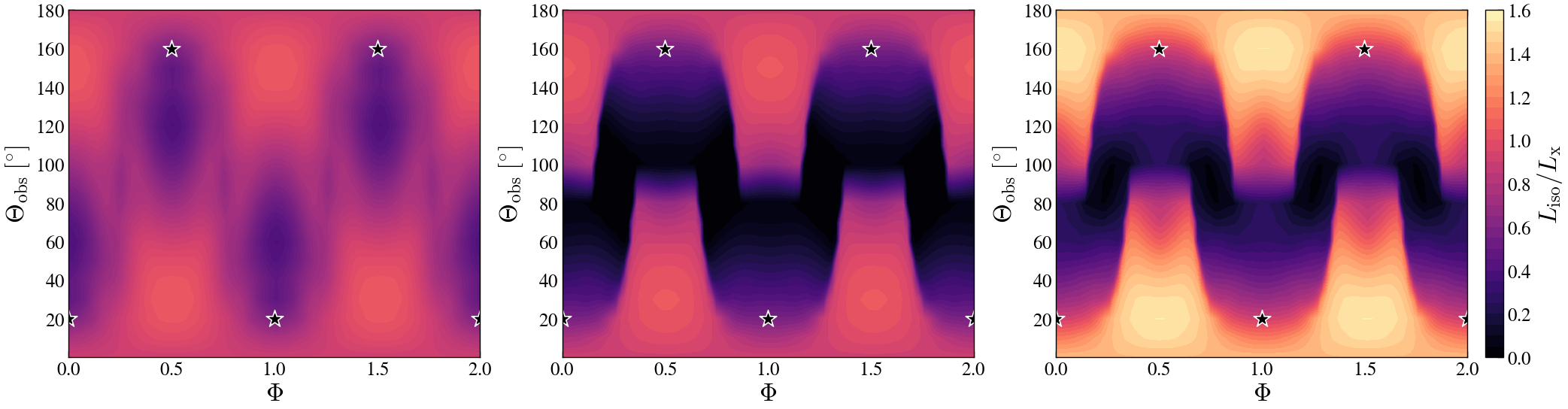}
    {\footnotesize
    \makebox[0.33\textwidth][c]{(a)}%
    \makebox[0.33\textwidth][c]{(b)}%
    \makebox[0.33\textwidth][c]{(c)}%
    }
    \caption{Comparison of the sky maps with different effects taken into account.  
    (a) Emission of the accretion columns eclipsed by the columns and \add{add} star. (b) Attenuation by the free-falling magnetospheric flow is taken into account. (c) Scattering by the free-falling magnetospheric flow is added to the model. Parameters $\muang = 20^\circ$, $\dot{m}=60$, $a = 0.7$, $\phicenter = 0^\circ$. The colour bar is universal for all the panels and shows isotropic luminosity normalized by the column power $L_x$.  
    }
    \label{fig:sky_map_diff}
\end{figure*}

\begin{figure}[!htb]
    \centering    
    \includegraphics[width=0.95\columnwidth]{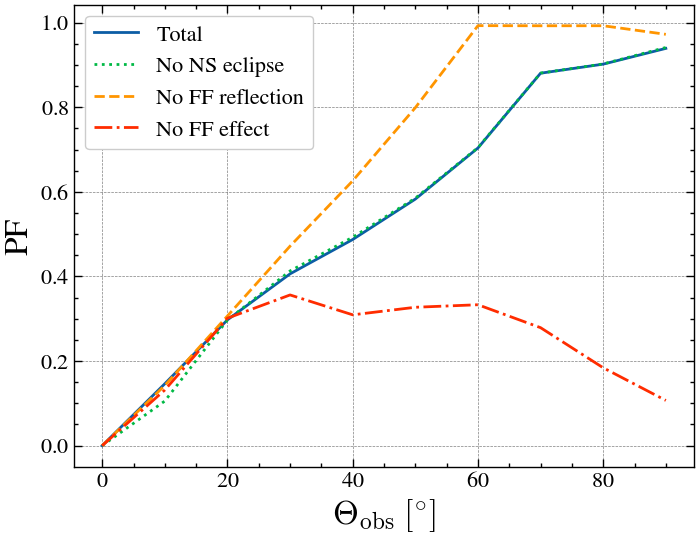}
    \caption{Effects of attenuation by  funnel flow (FF) and reflection from FF on pulse fractions.  The solid blue line shows the model with all effects included, the green dotted (almost {coinciding with the previous one}) shows the model without NS eclipses, the dashed orange line is for the model without reflection from FF, and the dot-dashed red line {ignores all the effects of the FF}. 
    The latter  corresponds to the configuration consisting of only accretion columns, yielding lower PF. Model parameters are $\muang = 20^\circ$, $\dot{m}=60$, $a = 0.7$, and $\phicenter = 0^\circ$.
    }
    \label{fig:Pf_on_off}
\end{figure}

In our calculations, the physical effects
that affect the pulse shape can be enabled separately. 
The sky maps in Fig.~\ref{fig:sky_map_diff}  show  the emission from the columns ignoring the impact of the FF (panel a), with the attenuation by the FF added (b), and taking into account both attenuation by the FF and emission scattered from its surface (c).
Comparing panels (a) and (b), we see that the pulse profile is strongly influenced by the attenuation of emission by the FF. 
Variations of the luminosity with the phase are the strongest in  panel (b), implying that the attenuation drives the PF  to high values (Fig.~\ref{fig:Pf_on_off}). 
Reflection makes the isotropic luminosity significantly higher (c). Compared to (a), the maximum flux is higher by 40\%. The account for FF leads to especially high PFs at high observer angles than without FF.
Reflection from  FF, on the other hand, results in some  decrease of PF, which is also visible in Fig.~\ref{fig:Pf_on_off}.

\begin{figure}[!htb]
    \centering
    {\includegraphics[width=\columnwidth]{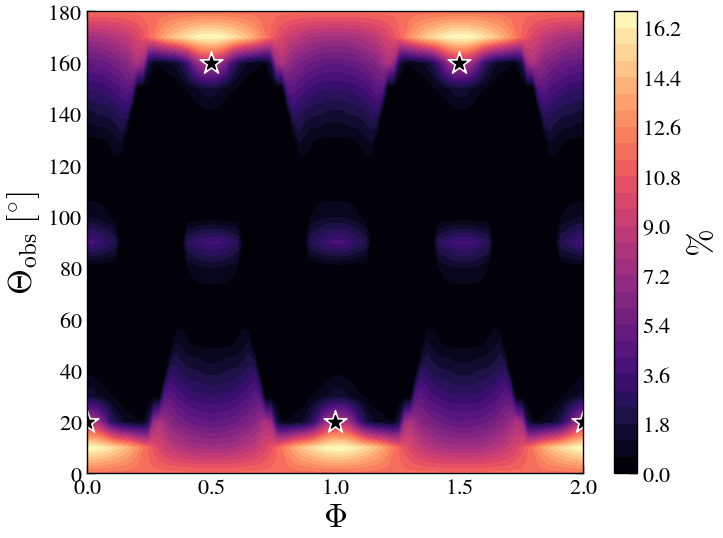}}

    \caption{Difference of the sky maps with  and without eclipses by the NS. Parameters $\muang = 20^\circ$, $\dot{m}=60$, $a = 0.7$, $\phicenter = 0^\circ$. 
    }
\label{fig:NS_diff}    
\end{figure}

If the accretion columns are sufficiently high, the eclipses caused by the NS are weak compared to the other visibility effects. 
Fig. \ref{fig:NS_diff} illustrates the contribution of the NS eclipse that for typical parameter sets considered in this paper is about several per cent. 
When the base of one of the columns is obscured by the NS, this contribution reaches $\sim 16\%$ when the observation angles $\thetaobs$ lie in a narrow interval of values.
This has a minor effect on the final PF (see Fig.~\ref{fig:Pf_on_off}, compare the dotted green and solid blue line).
Thus, the relatively large spatial size of the columns, which is  $\xishock \sim 7$ for the particular configuration, would allow us to neglect, to a certain degree, the occultations by the NS surface. 
The same reasoning may be used to justify our neglect of the relativistic light bending effects (see also \S\ref{sec:Light bending}).

\subsection{\texorpdfstring{Impact of the azimuthal  filling $a$}{Impact of the azimuthal filling a}} \label{s:varying_a}

The azimuthal filling factor $a$ defines the phase ranges affected by eclipses and attenuation.
This is demonstrated in Fig.~\ref{fig:L to a}, which displays how the pulse profile 
varies with changing $a$.
The configurations shown in the figure have values of $\thetaobs=$ $20\grad, 40\grad$, and $60\grad$, and $\phicenter=0\grad, 40\grad$, and $\muang=20\grad$.

For $\phicenter = 0,$ FF attenuates the accretion columns  in the range of azimuths from $-\uppi a $ to $\uppi a$, or, equivalently,  in the range of phases  between $-a /2$ to $a/2$. 
As the value of $a$ grows and approaches unity, there is a growing range of phases where the entire column is attenuated by the FF. 
The pulse-averaged luminosity $\Lisoavg$ Eq.~\eqref{eq.l_iso_avg} gradually varies with $a$ and attains maximum values for $a\sim 0.5-0.6$ (see Fig.\ref{fig:L_iso_to_a_box}).


Configurations with $a=1$ have symmetric pulse profiles (Fig.~\ref{fig:sky_map_rvm}).
The contours of constant angle $\thetamag$ (the polar angle counted from $\bm\mu$) illustrate that  the configuration with the filled accretion columns is a kind of RVM model with the maximal intensity along the magnetic axis.

Distributions of PF hardly depend on $a$, if $a<1$ (see Fig.~\ref{fig:PF_to_a_box}).
For filled magnetospheres ($a=1$), the pulsations are suppressed for polar or equatorial observers (for our values of $\muang = 10\grad$ and $20\grad$). 
Namely, the pulses are  hardly seen for  $|\thetaobs - 90\grad| \lesssim 20\grad$ (when  both columns are seen through FF at all the phases) or  $|\thetaobs - 0\grad| \lesssim 20\grad$ or $|\thetaobs - 180\grad| \lesssim 20\grad$  (when only one column is visible).

Thus, for $a=1$, there could be a situation when PF is so low that the pulsations are missed.
This cannot happen for $a<1$, as the observed pulse fractions always exceed $\sim 0.2$ (see Fig.~\ref{fig:PF_to_a_box}).
Naturally, for $\muang=0\grad$ and $a=1$, pulses are never observed.

\renewcommand{\widthForSubfig}{0.45}

\begin{figure*}[!htb]
\captionsetup[subfigure]{labelformat=empty,skip=0pt}
    \centering
    \begin{subfigure}[t]{\widthForSubfig\textwidth}
        \centering
        \includegraphics[width=\linewidth]{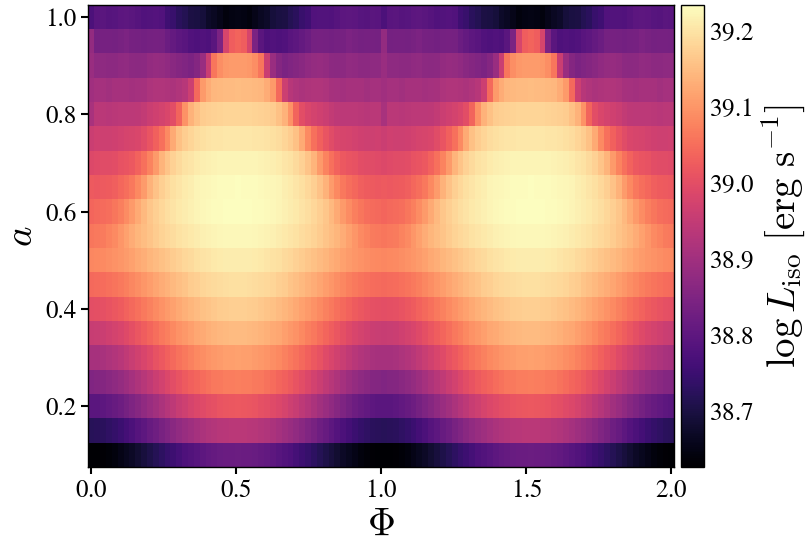}
        \caption{$\thetarot_{\rm obs} = 20^\circ$, $\muang = 20^\circ$, $\dot{m} = 60$, $\phicenter = 0^\circ$}

    \end{subfigure}\hfil
    ~ 
    \begin{subfigure}[t]{\widthForSubfig\textwidth}
        \centering
        \includegraphics[width=\linewidth]{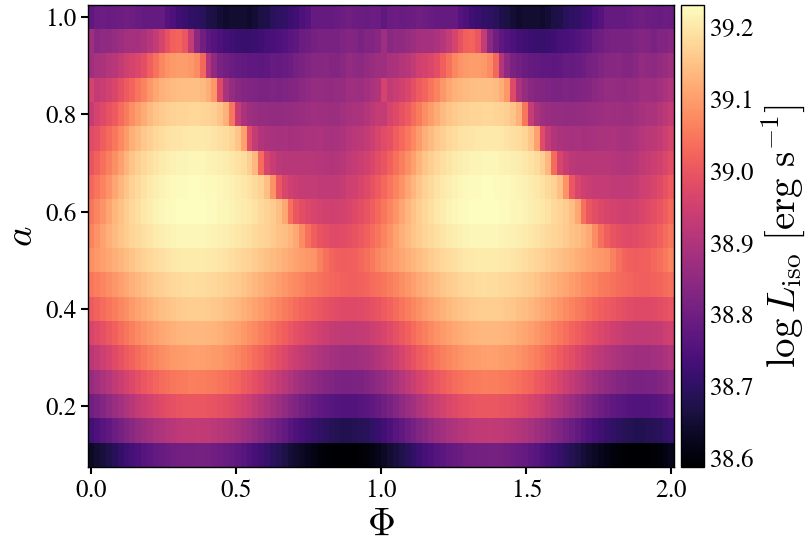}
        \caption{$\thetarot_{\rm obs} = 20^\circ$, $\muang = 20^\circ$, $\dot{m} = 60$, $\phicenter = 40^\circ$}

    \end{subfigure}\hfil

    \begin{subfigure}[t]{\widthForSubfig\textwidth}
        \centering
        \includegraphics[width=\linewidth]{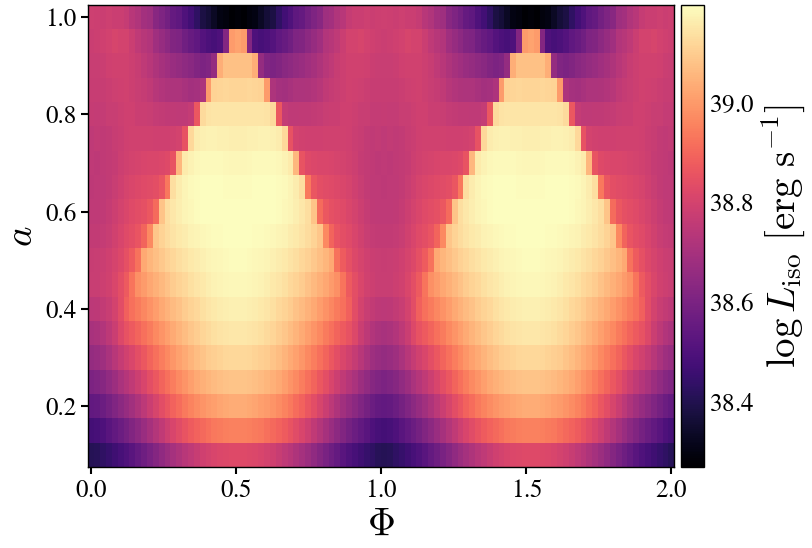}
        \caption{$\thetarot_{\rm obs} = 40^\circ$, $\muang = 20^\circ$, $\dot{m} = 60$, $\phicenter = 0^\circ$}

    \end{subfigure}\hfil
    ~
    \begin{subfigure}[t]{\widthForSubfig\textwidth}
        \centering
        \includegraphics[width=\linewidth]{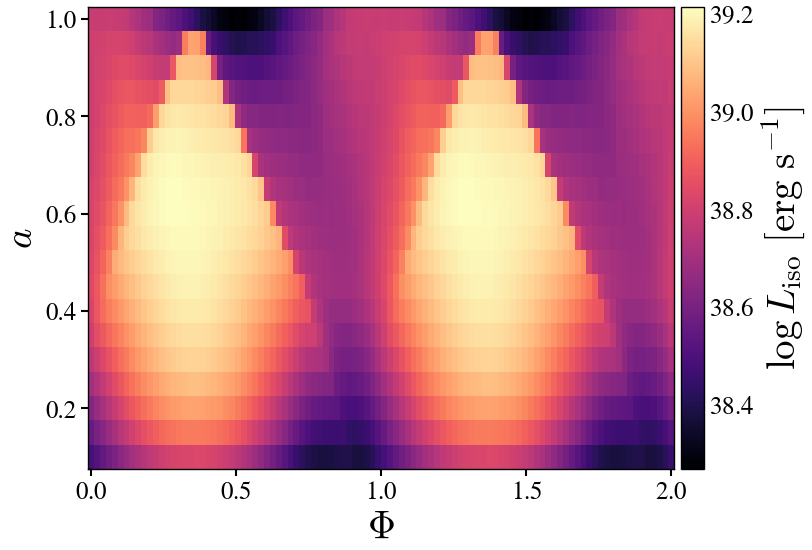}
        \caption{$\thetarot_{\rm obs} = 40^\circ$, $\muang = 20^\circ$, $\dot{m} = 60$, $\phicenter = 40^\circ$}

    \end{subfigure}\hfil

    \begin{subfigure}[t]{\widthForSubfig\textwidth}
        \centering
        \includegraphics[width=\linewidth]{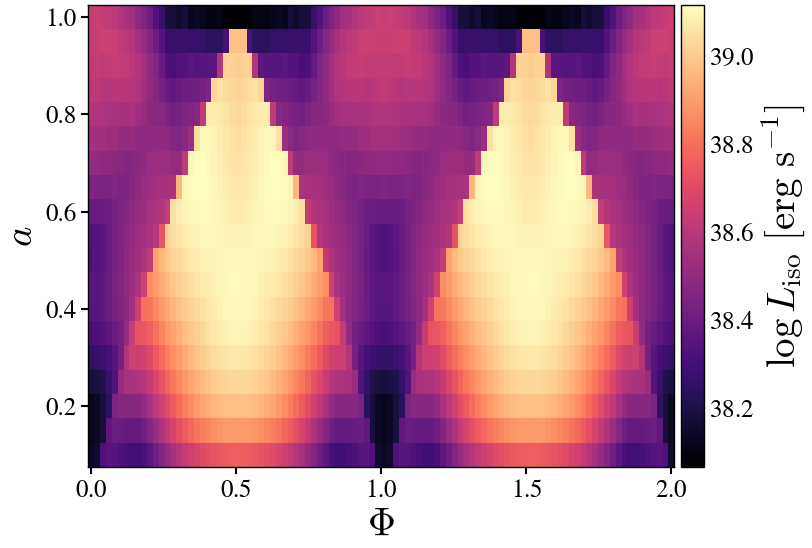}
        \caption{$\thetarot_{\rm obs} = 60^\circ$, $\muang = 20^\circ$, $\dot{m} = 60$, $\phicenter = 0^\circ$}

    \end{subfigure}\hfil
    ~
    \begin{subfigure}[t]{\widthForSubfig\textwidth}
        \centering
        \includegraphics[width=\linewidth]{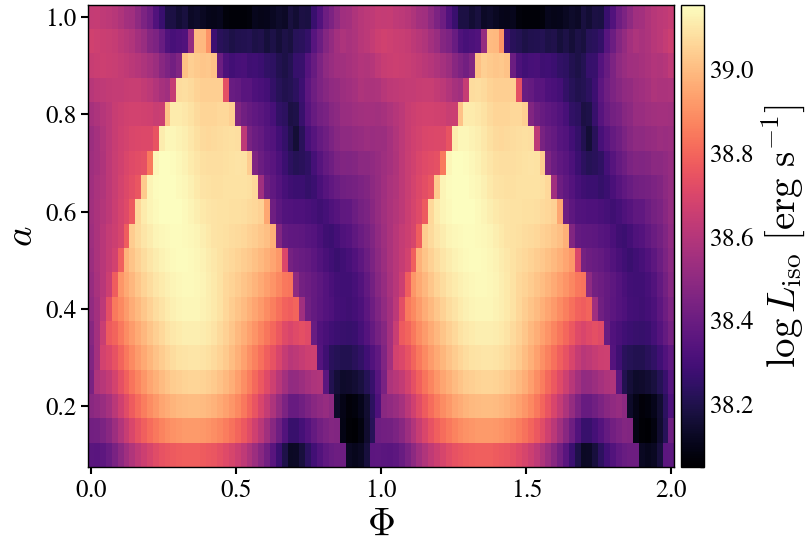}
        \caption{$\thetarot_{\rm obs} = 60^\circ$, $\muang = 20^\circ$, $\dot{m} = 60$, $\phicenter = 40^\circ$}

    \end{subfigure}\hfil
    \caption{Dependence of luminosity $L_{\rm iso}$ on phase $\Phi$ and parameter $a$. For low inclinations of the observer and thin accretion funnels (small $a$), the {attenuation affects only a narrow range of phases}.  For higher $a$, 
    accretion columns are lower, and the attenuation effect makes pulsations more prominent. }
    \label{fig:L to a}
\end{figure*}

\begin{figure}[!htb]
    \centering
    \includegraphics[width=1\linewidth]{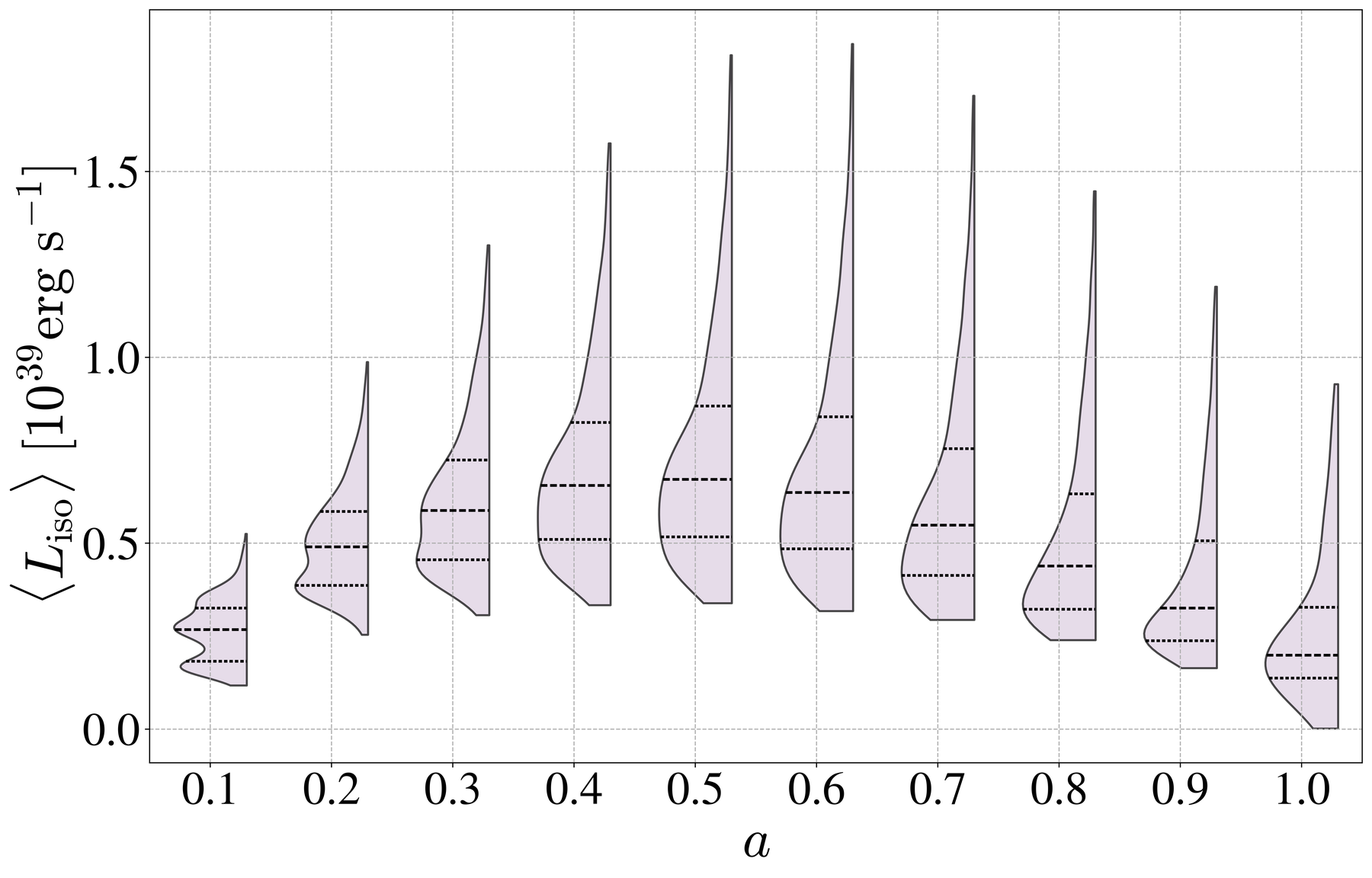}
    \caption{ 
    Distributions in phase-averaged $\Lisoavg$ for different $a$. All the notation is the same as in Fig.~\ref{fig:asymmetry_to_theta_obs_box}.  Models with $\thetaobs$ = 0 are excluded.
    }
    \label{fig:L_iso_to_a_box}
\end{figure}

\begin{figure}[!htb]
    \centering
    \includegraphics[width=1\linewidth]{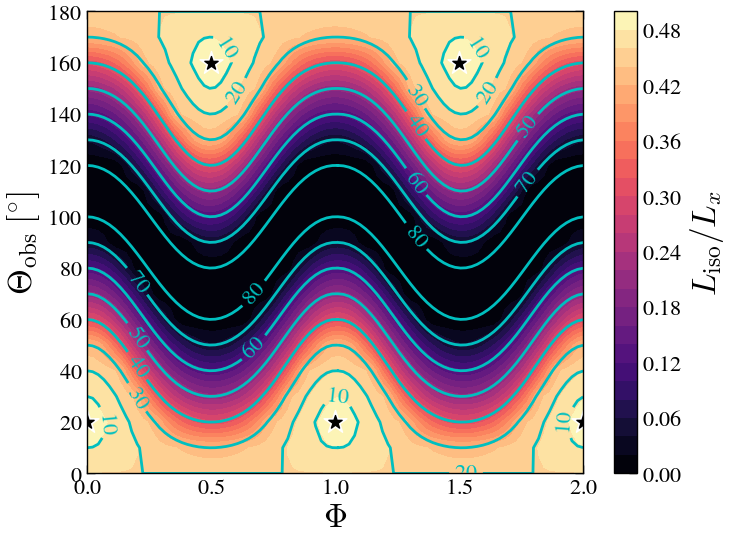}
    \caption{Sky map for the full-azimuthal accretion flow with parameters $\muang = 20^\circ$, $\dot{m}=100$, $a = 1$, $\phicenter = 0^\circ$. Contours indicate the angular distance $\thetamag$ (in degrees) to the nearest of the two magnetic poles. It can be seen that the observed flux depends only on $\thetamag$.}
    \label{fig:sky_map_rvm}
\end{figure}

\begin{figure}[!htb]
    \centering
    \includegraphics[width=1\linewidth]{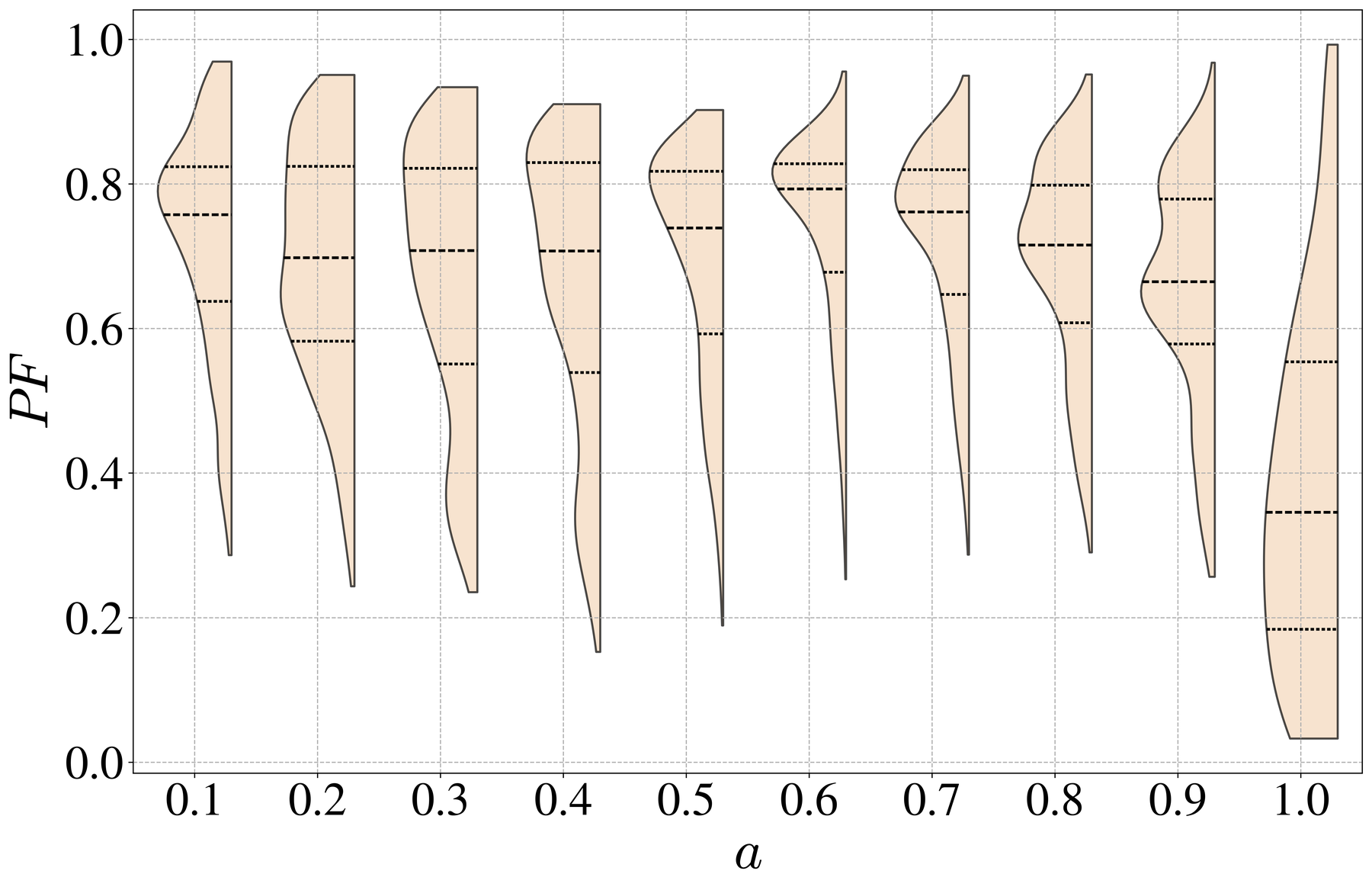}
    \caption{
    Distributions in $PF$ for different $a$. All the notation is the same as in Fig.~\ref{fig:asymmetry_to_theta_obs_box}. Models with $\thetaobs$ = 0 are excluded.}
    \label{fig:PF_to_a_box}
\end{figure}

\subsection{\texorpdfstring{Impact of $\phicenter$}{Impact of phi0}}
\label{s:dep_on_phi}

\renewcommand{\widthForSubfig}{0.32}


\begin{figure}[!htb]
\centering
\captionsetup[subfigure]{labelformat=empty,skip=0pt,font=footnotesize}
\begin{subfigure}[t]{0.5\textwidth}  
    \centering
    \includegraphics[width=\linewidth]{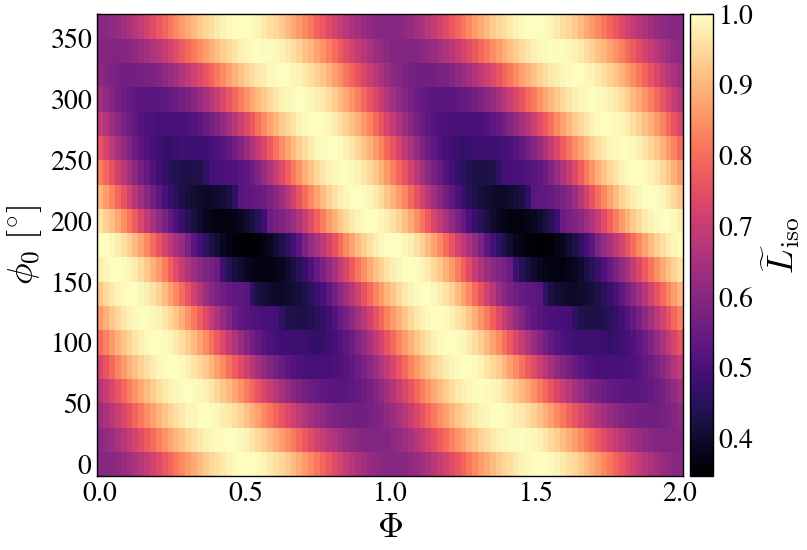}
    \caption{$\theta_{\rm obs}=20^\circ$, $\mu_{\rm ang}=20^\circ$, $\dot{m}=60$, $a=0.2$}

\end{subfigure}
\vspace{0.5\baselineskip}  

\begin{subfigure}[t]{0.5\textwidth}
    \centering
    \includegraphics[width=\linewidth]{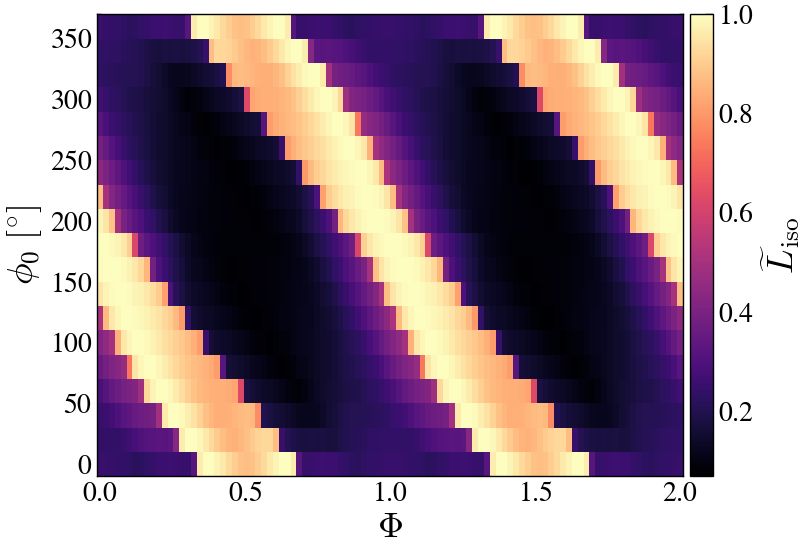}
    \caption{$\theta_{\rm obs}=60^\circ$, $\mu_{\rm ang}=20^\circ$, $\dot{m}=60$, $a=0.7$}

\end{subfigure}
\caption{Dependence of luminosity on the value of the {midline azimuth } $\phicenter$ and phase $\Phi$.   It can be seen that the change in the value of $\phicenter$ is not equal to the phase shift. The pulse shapes for different $\phicenter$ differ significantly. The  sharp edges (steps) in the figure are the rendering artifacts due to the limited grid resolution.}
\label{fig:L_to_phi_0}
\end{figure}

\renewcommand{\widthForSubfig}{0.3}

\begin{figure*}[htb]

\captionsetup[subfigure]{labelformat=empty,skip=0pt}
    \centering
    \begin{subfigure}[t]{\widthForSubfig\textwidth}
        \centering
        \includegraphics[width=\linewidth]{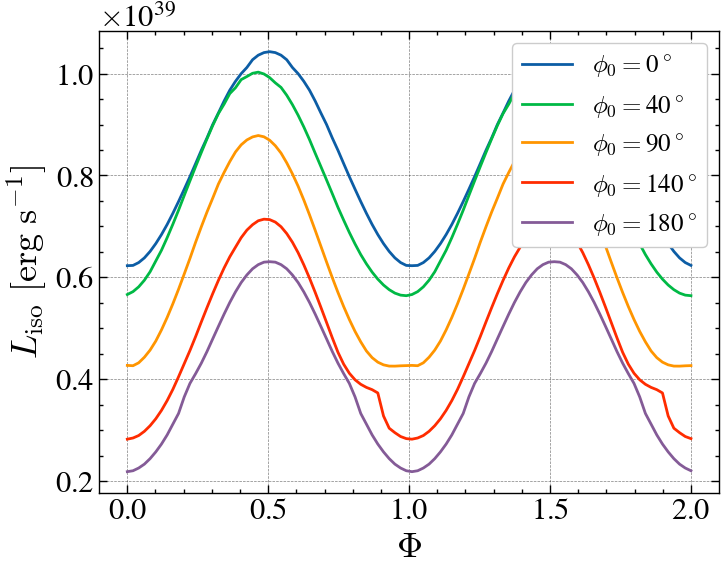}

    \end{subfigure}\hfil
    ~
    \begin{subfigure}[t]{\widthForSubfig\textwidth}
        \centering
        \includegraphics[width=\linewidth]{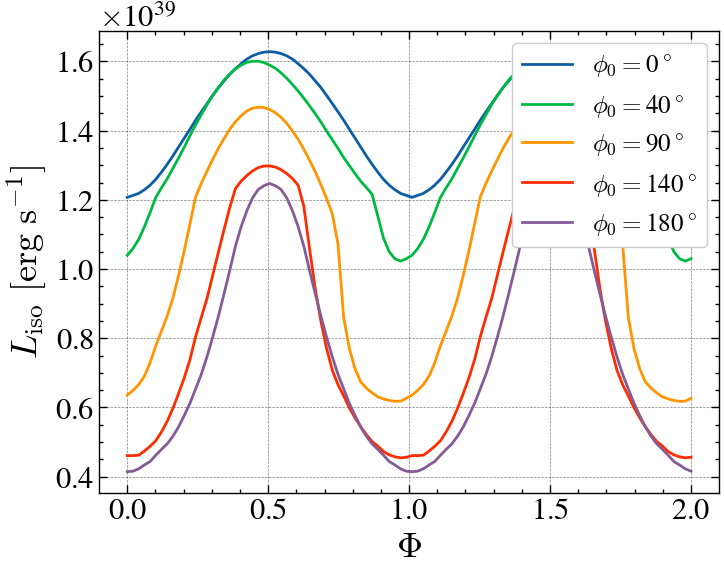}
    \end{subfigure}\hfil
    ~
    \begin{subfigure}[t]{\widthForSubfig\textwidth}
        \centering
        \includegraphics[width=\linewidth]{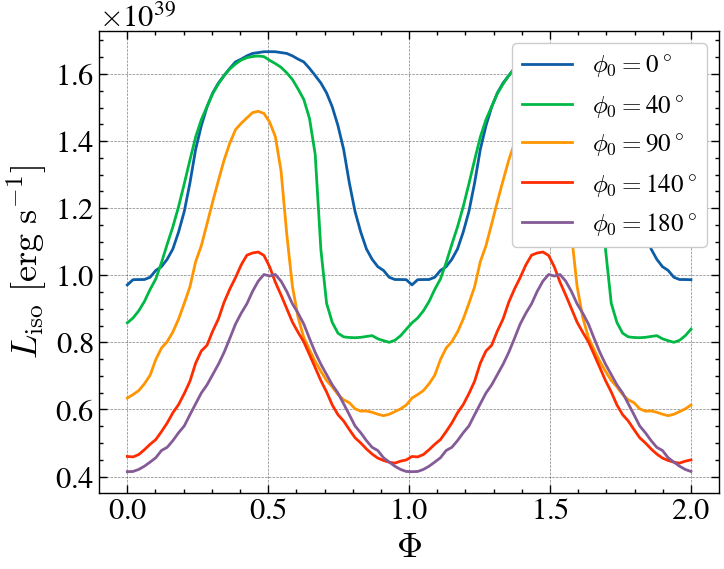}
    \end{subfigure}\hfil
    \caption{Pulse profiles for different midline azimuths $\phicenter$. Values of $\phicenter$ vary from $0^\circ$ to $180^\circ$, from the top curve to the bottom one, and azimuthal filling $a$ is 0.2, 0.5, and 0.7, from left to right. Each pulse is also moved horizontally by the phase value equal to $-\phicenter/2\uppi$, so that the maxima do not move.  
    Other parameters are fixed: $\dot m  =60$, $\muang=20^\circ$, and $\thetaobs=20^\circ$.
    }
    \label{fig.specific_profiles}
\end{figure*}

\begin{figure*}[htb]

\captionsetup[subfigure]{labelformat=empty,skip=0pt}
    \centering
    \begin{subfigure}[t]{\widthForSubfig\textwidth}
        \centering
        \includegraphics[width=\linewidth]{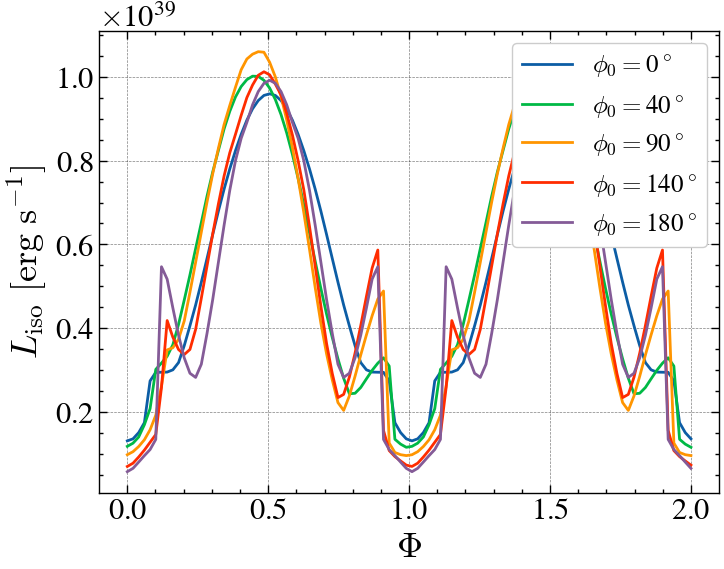}

    \end{subfigure}\hfil
    ~
    \begin{subfigure}[t]{\widthForSubfig\textwidth}
        \centering
        \includegraphics[width=\linewidth]{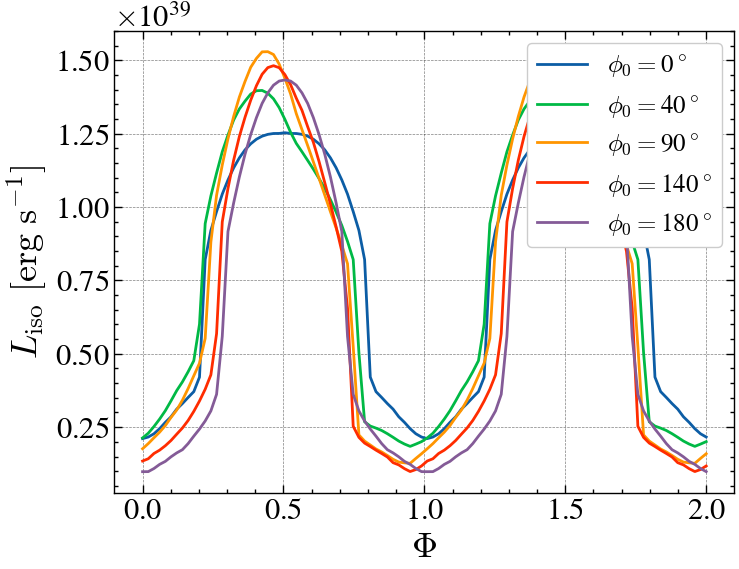}
    \end{subfigure}\hfil
    ~
    \begin{subfigure}[t]{\widthForSubfig\textwidth}
        \centering
        \includegraphics[width=\linewidth]{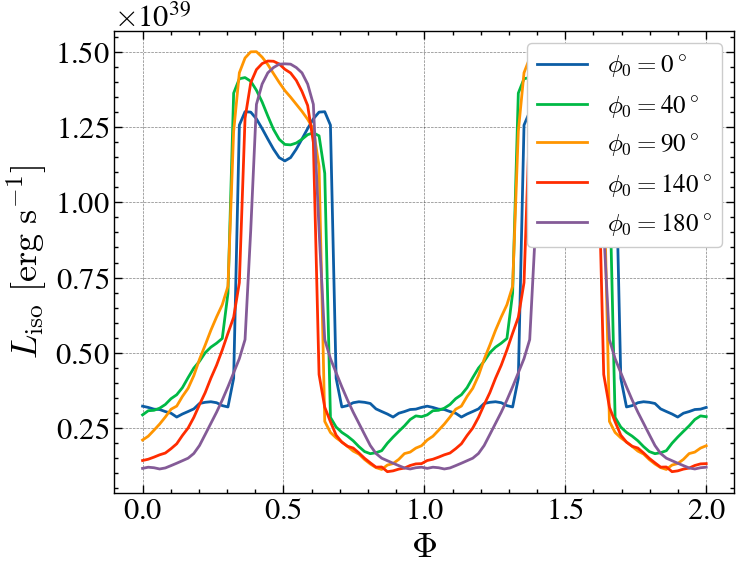}
    \end{subfigure}\hfil
    \caption{Pulse profiles for different midline azimuths $\phicenter$. Values of $\phicenter$ vary from $0^\circ$ to $180^\circ$, from the top curve to the bottom one, and azimuthal filling $a$ is 0.2, 0.5, and 0.7, from left to right. Each pulse is also moved horizontally by the phase value equal to $-\phicenter/2\uppi$, so that the maxima do not move. 
    Other parameters are fixed: $\dot m  =60$, $\muang=20^\circ$, and $\thetaobs=60^\circ$.
    }
    \label{fig.specific_profiles_add}
\end{figure*}

The parameter $\phicenter$ is a key parameter responsible for the geometry and, consequently, for the observational properties of the object. 
Figure~\ref{fig:L_to_phi_0} shows the evolution of the pulse profile as a function of $\phicenter$, evaluated for two different values of $a$ and $\thetaobs$.

In general, at least for small $\muang$ as in our simulations, the pulse profile experiences a negative phase shift $\Delta \Phi \sim  -\phicenter/(2\uppi)$ with an increase in $\phicenter$. 
To illustrate this and how the pulse shapes change with $\phicenter$, in Figs.~\ref{fig.specific_profiles} and~\ref{fig.specific_profiles_add} we plot the pulse profiles phase-shifted to compensate for the rotation of the beam pattern.
The two figures differ in the observation angle $\thetaobs$.
For observers at inclinations about $\thetaobs = 60^\circ$ (Fig.~\ref{fig.specific_profiles_add}), the shape of the pulse acquires additional features, such as additional maxima.
Such secondary maxima are observed in real objects and will be discussed in \S\ref{sec:obs_anticor}.

Both the phase-averaged luminosity $\Lisoavg$ and the pulse shape are affected by $\phicenter$. 
 PF is influenced as well. 
In Fig. \ref{fig:PF_to_phi0}, we show 
how PF varies with $L_{\rm iso}$ as $\phicenter$ changes. 
In general, the lower the value of $\phicenter$ for $0 < \phicenter < 180\grad$, the lower the value of PF 
and the higher the value of $\Lisoavg$.

\newcommand{\captionForPFLnu}[2]{$\thetarot_{\rm obs} = #1^\circ$, $\muang = #2^\circ$}

\renewcommand{\widthForSubfig}{0.32}

\begin{figure*}[!htb]

\captionsetup[subfigure]{labelformat=empty,skip=0pt}
    \centering
    \begin{subfigure}[t]{\widthForSubfig\textwidth}
        \centering
        \includegraphics[width=\linewidth]{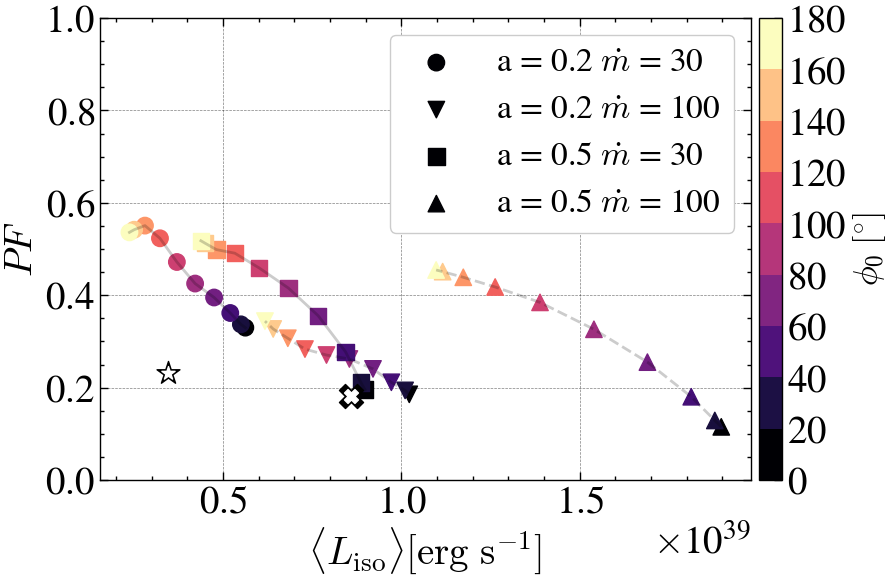}
        \caption{\captionForPFLnu{20}{20}}

    \end{subfigure}\hfil
    ~ 
    \begin{subfigure}[t]{\widthForSubfig\textwidth}
        \centering
        \includegraphics[width=\linewidth]{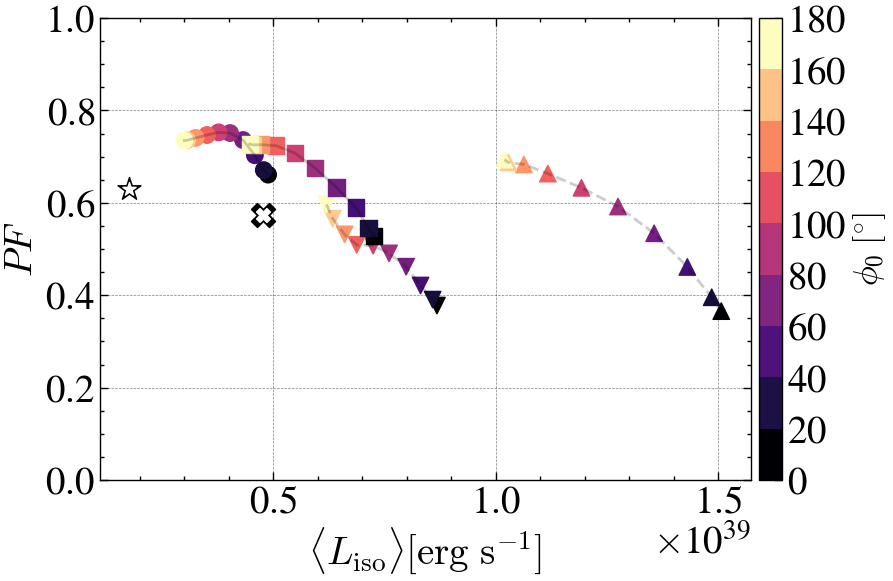}
        \caption{\captionForPFLnu{40}{20}}

    \end{subfigure}\hfil
     ~ 
    \begin{subfigure}[t]{\widthForSubfig\textwidth}
        \centering
        \includegraphics[width=\linewidth]{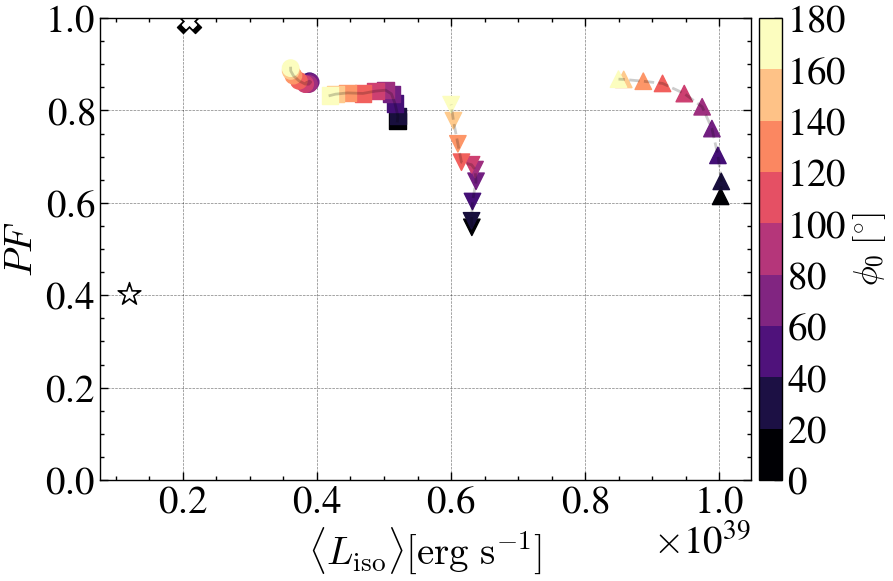}
        \caption{\captionForPFLnu{60}{20}}

    \end{subfigure}\hfil

    \caption{Variations of PF and $L_{\rm iso}$ with $\phicenter$. Panels differ by observation angles $\thetaobs = 20\grad, 40\grad, 60\grad$. Each of the four symbols corresponds to a unique ($a, \dot{m}$) combination (see legend at first panel) and represents a sequence of points at different $\phicenter$ encoded by colour. The single open asterisk and cross in each panel shows the location corresponding to $a=1$ and $\dot m = 30$ and $100$, respectively.
    }
    \label{fig:PF_to_phi0}     
\end{figure*}

\renewcommand{\captionForPFLnu}[2]{$\thetarot_{\rm obs} = #1^\circ$, $\muang = #2^\circ$}

\subsection{\texorpdfstring{Impact of $\dot{m}$}{Impact of m-dot}}\label{s:dep_on_m}

\begin{figure}[!h]
    \centering    
    \includegraphics[width=1\columnwidth]{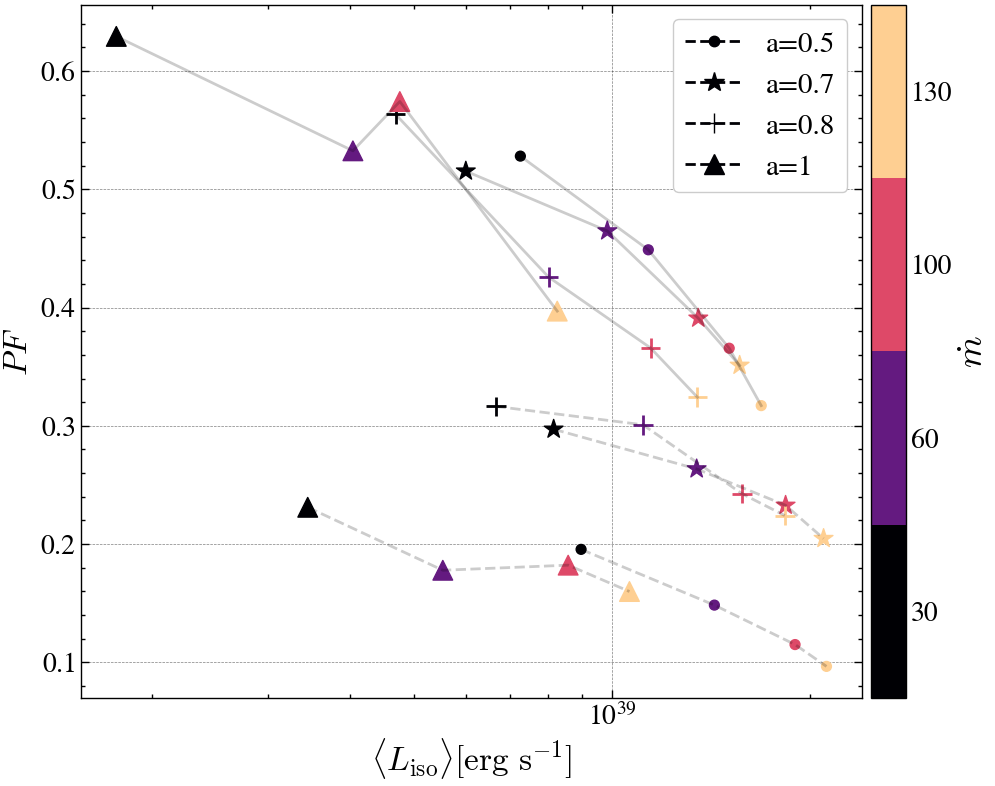}
    \caption{Dependence of PF on phase-averaged isotropic luminosity for two inclinations of observer and magnetic inclination $\muang = 20^\circ$. The middle azimuth of the column is $\phi_0 = 0\grad$. Dashed lines show results for $\thetaobs=20\grad$ and solid,  for $40\grad$. Symbols encode azimuthal filling factor $a$. Colours of the symbol encode different  $\dot{m}$, shown in the colour bar.}\label{fig:PF_incr_m_dot}
\end{figure}

The influence of the mass accretion rate $\dot{m}$ is generally straightforward: luminosity $L_{\rm iso}$ increases with $\dot{m}$ due to {increasing} column size ($\xishock$) and temperature.

Moreover, the optical depth $\tau$ in the FF increases as well. This results in more attenuation of radiation at certain phases. 
In particular, for configurations with $a=1$ 
and large inclinations $\thetaobs$, complete attenuation of the column radiation occurs at some phases for higher $\dot{m}$. As a result, in Fig.~\ref{fig:PF_to_phi0}, PF increases from 0.4 to 1 while the mass accretion rate increases from $\dot{m} = 30$ to $100$.

Figure~\ref{fig:PF_incr_m_dot} illustrates the anticorrelation between PF and $\Lisoavg$. This behaviour arises from the dependence of the column height on $\dot{m}$. As the accretion rate increases, the column grows and bends along the dipole field, exposing additional radiating surfaces at all phases, which reduces the PF. 
Observational evidence for such anticorrelation in real objects is discussed further in \S\ref{sec:obs_anticor}.

\subsection{\texorpdfstring{Impact of $\thetaobs$}{Impact of Theta-obs}}\label{s.thetaobs}

\begin{figure}[!htb]
    \centering
    \includegraphics[width=1\linewidth]{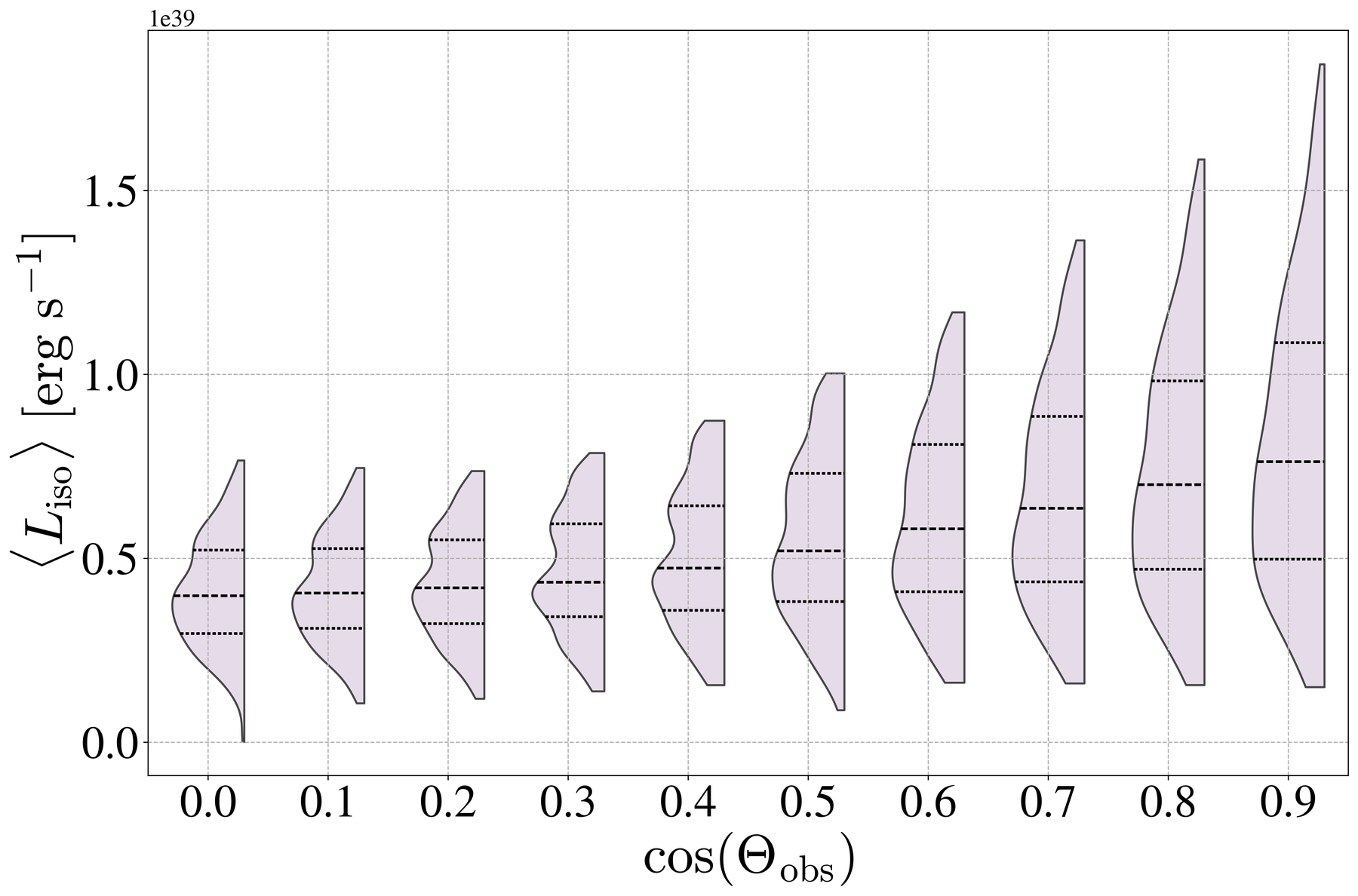}
    \caption{
    Distributions in $\Lisoavg$ for different $\thetaobs$. All the notation is the same as in Fig.~\ref{fig:asymmetry_to_theta_obs_box}.
    }
    \label{fig:L_iso_L_x_to_theta_box}
\end{figure}

\begin{figure}[!htb]
    \centering
    \includegraphics[width=1\linewidth]{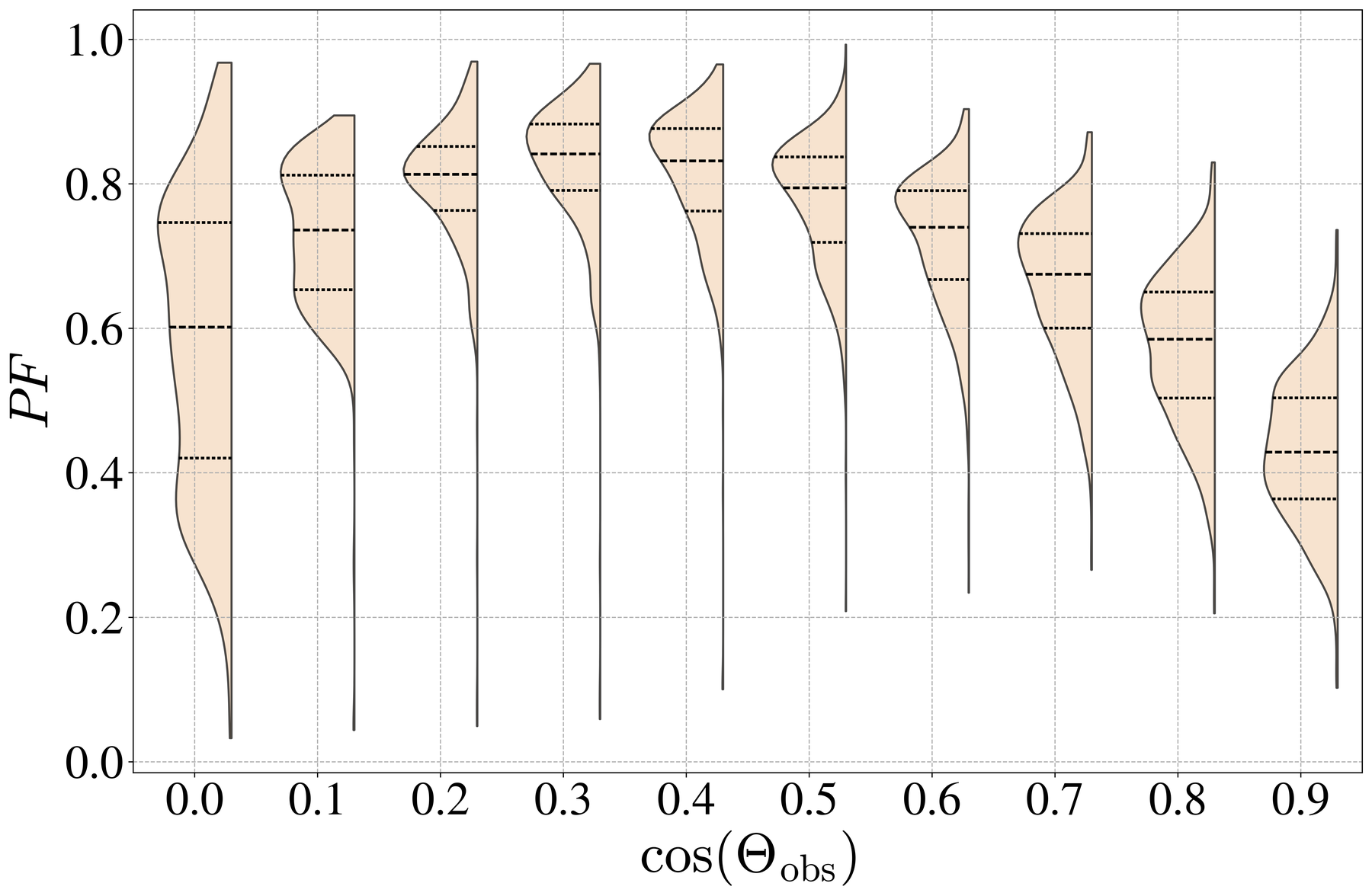}
    \caption{
    Distributions in PF for different $\thetaobs$. All the notation is the same as in Fig.~\ref{fig:asymmetry_to_theta_obs_box}.
    }
    \label{fig:PF_to_theta_obs_box}
\end{figure}

Figure~\ref{fig:L_iso_L_x_to_theta_box} illustrates the gradual decrease in $\Lisoavg$ with the inclination of the observer. 
Face-on observers on average see a brighter and less-variable source.
Figure \ref{fig:PF_to_theta_obs_box}, showing the distributions for PF for different \thetaobs, demonstrates that smaller viewing angles $\thetaobs$ lead to smaller PF for $\cos \thetaobs \gtrsim 0.3$ ($\thetaobs \lesssim 70\grad$).
Furthermore, a face-on observer detects lower values of the asymmetry metric (see Fig.~\ref{fig:asymmetry_to_theta_obs_box}).

At higher $\thetaobs$, the attenuation by FF plays an important role and drives PF to consistently high values. However, this trend is not universal. For instance, in configurations with $a=1$, the accretion columns can remain permanently screened by the optically thick FF layer throughout the whole period. In such cases, the difference between the minimum and maximum fluxes is strongly suppressed, and PF approaches zero.

A distinct regime occurs at $\thetaobs = 90\grad$, i.e., for an observer located exactly in the equatorial plane. Owing to the central symmetry of the system, all obscuration and visibility effects originating from one hemisphere are 
precisely balanced by the opposite effect in the opposite hemisphere. 
In practice, this means that the northern and southern accretion columns become indistinguishable, and the observed pulse profile exhibits a repeated pattern, which effectively reduces the detected period by a factor of two.
Nevertheless, it should be emphasized that the visibility of the system and PF at high angles $\thetaobs$ are to be changed if one includes the disc surface in the problem.


\section{Discussion}\label{sec:discussion}

\subsection{Regimes of accretion and applicability of the model}
\label{sec:Light bending}

Depending on the mass accretion rate on a strongly magnetized NS, the geometry of the radiating regions can change profoundly.
At low accretion rates, when the falling plasma is braked in the NS atmosphere, the emission sites can be considered as 2D structures confined to the surface of the star.
A radiation-supported shock wave forms when $\dot M \gtrsim c A_\perp /\varkappa \delta$~\citepalias{Basko-Sunyaev1976}.


As the mass accretion rate increases, the radiative shock shifts upwards. Then the total luminosity of the column is approximately 
$L^{**} \sim 10^{39}{\rm erg\, s^{-1}}$, which is the  so-called  limiting luminosity  introduced by \citetalias{Basko-Sunyaev1976}. We limit accretion rates by $\dot m=100$, when some of the models, with $a \lesssim 0.2 $, reach the disc radius. 
In this \emph{tall-column} regime, the radiation of the column is relatively easy to calculate, as some of the effects (emission from the shock 
and relativistic light bending) become relatively unimportant.
Our models provide  isotropic luminosities in the range $L_{\rm iso} \sim 10^{38} - 10^{39}$ erg s$^{-1}$, see  Fig.\ref{fig:L_iso_to_a_box}.

\subsection{Light bending}

When the column is not very high, light bending effects are significant.
For a tall accretion column, the impact of the light bending may be estimated in Beloborodov's approximation \citep{2002ApJ...566L..85B}, {in which}
the deflection angle
\begin{equation}
    \delta_{\rm b} \simeq \frac{u}{\sin\alpha (1-\cos\alpha)},
\end{equation}
where $\alpha$ is the angle between the radius vector and the {tangent to the photon trajectory}, and $u = {G \Mns}/({Rc^2})$.
Strong light bending also affects the visibility conditions of the source, especially near the surface of the NS (see, for example, \citealt{falkner}). 

Corrections to the solid angles and thus to the fluxes are of the order of maximal $u$ along the trajectory. 
Thus, for tall columns of height $\xishock$, flux variations due to light bending are $\sim 0.2/\xishock$, excluding the cases when considerable part of the light rays are grazing, as in the case of extreme lensing combined with an eclipse by the NS shown in \citet[figure 4.4]{falkner}. 
If the normalized transverse size of the column $\xishock \sin \thetamag_{\rm s} \gg 1$, both the effects of NS occultations and light bending become negligible. 
Here, $\thetamag_{\rm s}$ is the polar coordinate of the shock wave. 
For dipolar geometry, 
$\sin\thetamag_{\rm s} = \sqrt{\xishock/\ReRns}$, that allows to write the condition for occultation inefficiency as
\begin{equation}
    \xishock \gtrsim \ReRns^{1/3}.
\end{equation}
Thus, when the column is high enough, due to high accretion rate and/or small $A_\perp$, this relation is fulfilled, and the light bending effects may be ignored. 
For $\ReRns = 100$, the requirement for the column height is $\xishock \gtrsim 3$~(see Fig.~\ref{fig:xi to m}). 
All the models in our grid satisfy this condition. 
For a more general case, taking into account light bending is crucial, and we are planning to include it in the subsequent versions of the code. 

\subsection{Observed pulse profile variations at high accretion rate}\label{sec:obs_anticor}

There are several cases when XRPs were thoroughly studied in the luminosity range covered by our model. In particular, \citet{Hou+2022} made timing and spectral analyses of an outburst of the Ultraluminous X-ray Pulsar (ULXP) 
\object{RX~J0209.6-7427} in the Small Magellanic Cloud that reached luminosity of 
$\sim 10^{39}$~erg~s$^{-1}$. 
During the flare, PF behaved non-monotonically, and the pulse shape evolved 
considerably with the bolometric luminosity. 

The PF decreased up to isotropic luminosity of $\sim 2 \times 10^{38}$~erg\,s$^{-1}$, then rose, and then decreased again, after $\sim 6 \times 10^{38}~$erg\,s$^{-1}$. 
The anticorrelation between $L$ and PF at the highest luminosities ~\citep[see figure 7 of][]{Hou+2022} can be explained in our modelas a consequence of the column height dependence on the accretion rate. 
When the column is not too high, much of its radiation is attenuated by FF for a non-polar observer. As we mentioned in Sect.~\ref{sec:impacts}, the attenuation by the FF drives the PF  to high values
(see also Fig.~\ref{fig:Pf_on_off}).
As the accretion rate increases, the column grows in height and bends along the dipole field lines. 
Thus, the observer starts seeing larger and larger fractions of unattenuated column surfaces at all the phases, which decreases the PF and additionally increases the luminosity. 
This behaviour is reproduced in a range of models, see Fig.~\ref{fig:PF_incr_m_dot}.

\begin{figure}[!htb]
    \centering
    \includegraphics[width=\linewidth]{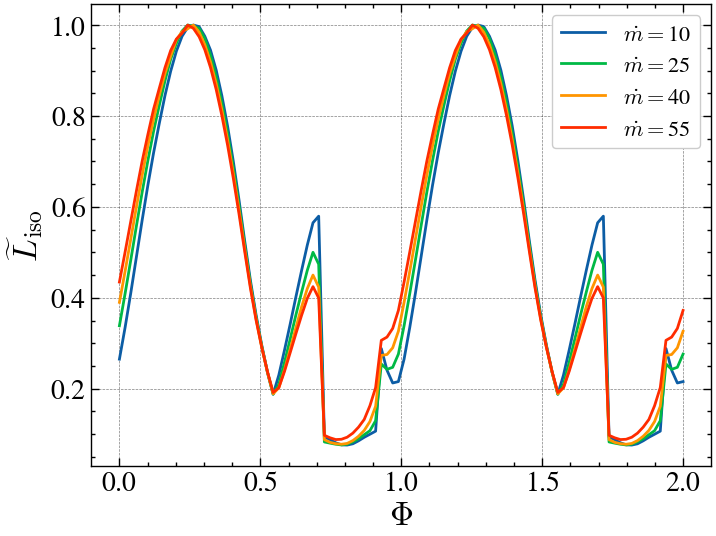}
    \caption{Pulse profile evolution with increasing $\dot{m}$. The secondary maximum becomes fainter relative to the primary one while the shock wave raises
    $\thetaobs=60\grad, \muang=20\grad, a=0.2, \phicenter=80\grad$, 
    }
    \label{fig:L_to_mass}
\end{figure}

Another observed feature was the relative dimming of the second pulse maximum with increasing bolometric luminosity 
(see figure 5 of \citealt{Hou+2022}).  This phenomenon can be explained in our model, when the specific geometry is assumed. In Fig.~\ref{fig:L_to_mass}, we examine the variation of the normalized pulse profile $\tilde{L}_{\rm iso} = L_{\rm iso} / \max\limits_\Phi L_{\rm iso}$ with shock height  for  the parameters  $\thetaobs=60^\circ$, $\muang=20^\circ$, $a=0.2$ and $\phi_0=80^\circ$ . 
The amplitude ratio between the primary (at $\Phi=0.25$) and secondary (at $\Phi=0.7$) maxima depends on the column height.

In terms of Fig.~\ref{fig:Geom_config}, the primary maximum of the pulse is produced by the visibility conditions of the two surfaces (northern polar and southern equatorial, yellow and green in the left panel) which are on average more favourably oriented with respect to the observer. They are responsible for the maximum at $\Phi \simeq 0.25$. The visibility of the other two surfaces is affected by the absorption in FF. For certain values of \phicenter, attenuation by FF is present only when the columns are tall enough and specifically damps the secondary maximum of the pulse. 

At even higher accretion rates, when the Eddington limit is exceeded in the disc, optically thick outflows may be launched from outside the magnetosphere. Their presence restricts the visibility of the source. If the source remains visible, radiation scattered by the wind likely decreases the PF~\citep[e.g.,][]{2023MNRAS.518.5457M}.


\subsection{Phase shift of the maxima and twist of the magnetosphere}\label{s:phase shift}


Some of the phase-resolved observations of the well-studied XRP flares~\citep{2018ApJ...863....9W, 2020MNRAS.493.5680J, Doroshenko+2020} 
show  shifts of the pulse maximum phase by 0.5.
This  shift is a result of gradual  changes in the pulse profile and of the relative contributions of the two maxima. 

In our model, secondary maxima exist and their relative amplitude depends on $\dot{m}$, but they never become stronger than the primary. 
Hence, a phase shift by 0.5 is not possible.
This can be explained by the fact that in our present modelling, the magnetic angle is limited to small values being either $10^\circ$ or $20^\circ$.
For large magnetic angles, we expect that the footprint of the accretion column would have a large spread along the magnetic latitudes. This situation  would require a change in the
model setup, as described in~\ref{s.about_magnetic_polar_angle}.

On the other hand, the position of the pulse maximum is also affected by other parameters, such as the azimuthal filling factor $a$ and the midline shift parameter $\phicenter$, which might vary with the mass accretion rate for different reasons.
Such an explanation was proposed by \citet{miller1996} to explain the phase lags observed for bursting pulsar \object{GRO\, J1744$-$28} during the outbursts when the object is expected to develop a radiation-supported shock.

In MHD  simulations of accreting magnetized stars, 
\citet{Kulkarni-Romanova2013} observed an azimuthal shift of the hot spot on the NS surface relative to the azimuth where the accretion flow originates at the disc’s inner radius. 
The shift is an expected outcome of the twist of  
the magnetosphere \citep{miller1996} which occurs because of the angular velocity difference between the star and the disc.

\citet{Kulkarni-Romanova2013}  have found that the twist of the magnetosphere depends on the fastness parameter, defined as~\citep[][]{1977ApJ...215..897E}
\begin{equation}
    \omega = (\Rm / R_{\rm co})^{3/2}\, ,
\end{equation}
where $R_{\rm co} = (G \Mns \Omega_{\star}^{-2})^{1/3}$ is the corotation radius, which is defined as the radius where the angular velocities of the disc and the magnetosphere, $\Omega_{\star}$, are equal. 
As $\Rm$ depends on $\dot{m}$ (see Eq.~\eqref{eq:Ra}), changing the latter leads to variations of $\omega$.
This makes it possible to relate the twist of the magnetosphere to changes in the rotation rate, particularly during strong bursts, when the fastness parameter can reach small values. 

We expect such a twist to be present in XRP magnetospheres. Even more importantly, we expect this twist to change with the mass accretion rate, as the fastness parameter of the source changes.
If the twist occurs mostly in the outer magnetosphere, it effectively leads to non-zero $\phicenter$ in our model. 

As the twist of the magnetosphere depends on the accretion rate, we expect that this could produce subtle but observable effects of the timing properties (see ~\ref{s.app_timing}).

\subsection{Diffuse scattering in the magnetosphere diminish pulse fraction}\label{s.dif-scattering}


In our model we adopt the approximation that all the photons that do not manage to pass directly through FFs are reflected back. 
Furthermore, we ignore emission from the FF plasma. Below we show that the second assumption is justified, while neglecting the radiation diffusing through the flow may lead to PF being overestimated  by as much as  $\sim 30\%$ for low inclinations.

\begin{figure}
    \centering  
    \includegraphics[width=\linewidth]{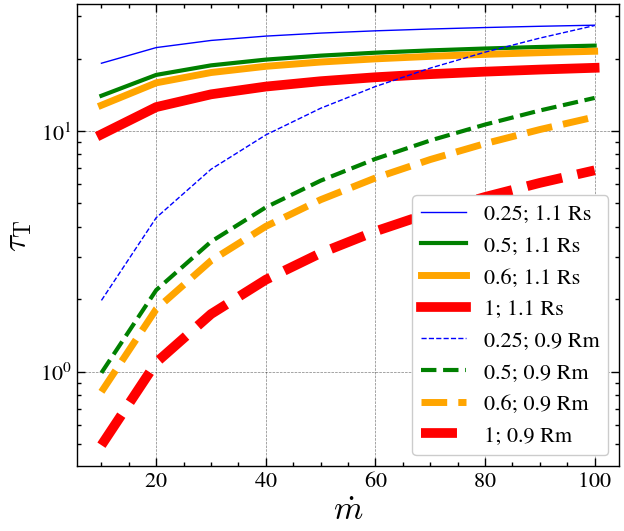}
    \caption{Optical thickness to scattering of the magnetospheric flow just above the shock, at  $R=1.1 \,R_{\rm s}$ (solid lines), and near the disc, $R=0.9\, R_{\rm e}$ (dashed lines). 
    Line thickness encodes the azimuthal parameter, from $a=0.25$ for the thinnest line to $a=1$ for the thickest.
    }
    \label{fig.plasma_properties}
\end{figure}

\begin{figure}[!htb]
    \centering
    \includegraphics[width=\linewidth]{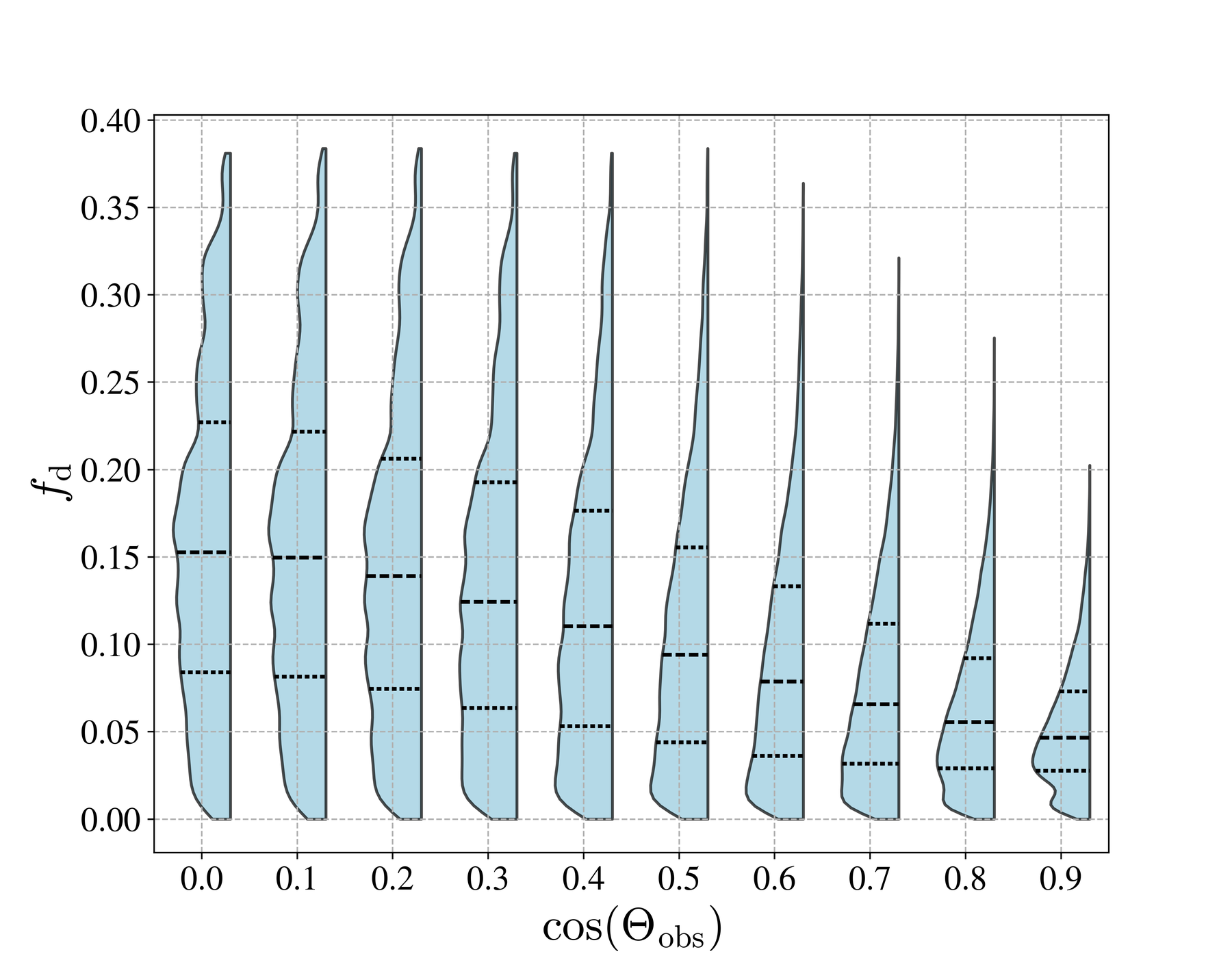}
    \caption{
    Distributions in the diffusion factor $f_{\rm d}$ for different $\thetaobs$. All the notation is the same as in Fig.~\ref{fig:asymmetry_to_theta_obs_box}. 
    }
    \label{fig:f_d_box_new}
\end{figure}

Simple estimates show that the magnetospheric plasma may be considered  completely ionised and purely scattering.  
In a more general case, some fraction of the incident flux is reflected, a small fraction, $\sim \exp(-\tau)$,  passes directly without a single scattering,  and other photons are able to cross the FF after one or several scatterings. 
To estimate the diffuse flux radiated from the outer surface of the FF, we will use the analytic result for the plane scattering slab with $\tau \gg 1$, illuminated by the parallel rays at some angle $\alpha_1$ to the normal~\citep[see ][chapter X, \S1]{Sobolev1963}. The unscattered attenuated component in such a setup may be easily neglected.
The flux  at depth $\tau$ is diminished relative to the incident flux by the following factor\footnote{With accuracy of better than 4\%, Eq.~\eqref{eq.fd}  is  close to equation~(15)  of \citet{2024MNRAS.529.1571F} in the case of $\cos \alpha_1 = 1$. 
}
\begin{equation}
    f_{\rm d} 
    =\frac{ (2+3\cos \alpha_1)}{4+3\tau} \, \cos \alpha_1\, ,
    \label{eq.fd}
\end{equation}
where $\alpha_1$ is the angle between the incident radiation and the normal to the surface. 
The optical thickness to scattering in the FF lies in the interval $1-30$ for our input parameters (see Fig.~\ref{fig.plasma_properties}). 
For all trajectories for all models in our sample, we have calculated the values of the factor $f_{\rm d}$ given by Eq.~\eqref{eq.fd}. 
Resulting distributions for factors $f_{\rm d}$ for our sample of models are shown in Fig.~\ref{fig:f_d_box_new}.

Consequently, account for the diffuse flux  can decrease PF by a factor of $(1+f_{\rm d})$, that is, by 
$\sim 20\%$ for $\sim 75\%$ of the models for $\thetaobs<60^\circ$.
We note that the diffuse flux is more important  for lower accretion columns when PF values are bigger (see Fig.~\ref{fig:PF_incr_m_dot}).
For lower accretion rates and smaller $\tau$, the applicability of our model is also limited by the accretion shock emission, ignored in this study.

The plasma in the FF is highly ionized.
Its ionization parameter can be estimated as follows:
\begin{equation}
    \xi_{\rm ion} \equiv\frac{f_{\rm d}\,L}{n_{\rm H} R^2} \simeq \frac{f_{\rm d}\,G \dot{M} \Mns}{\Rns} \frac{1}{R^2} \frac{m_{\rm H}}{\rho} \, ,
\end{equation}
where  density $\rho = \dot{M}/(\crossec v_{\rm ff})$, and $v_{\rm ff} = \sqrt{2GM/R}$. 
The cross-section of the flow may be estimated, using Eqs~\eqref{eq:S} and \eqref{eq:delta}, as $\crossec \approx 4\uppi a \, R^3\Delta R_{\rm e}/ R_{\rm e}^2$.
Thus,
\begin{equation}
    \xi_{\rm ion} 
    \sim 10^6 \, \frac{\Delta R_{\rm e}}{0.25\,R_{\rm e}}\frac{f_{\rm d}}{0.2}\,\frac{a}{0.25} \frac{\sqrt{\RRns}}{\ReRns} ~  \, \rm erg \, \rm cm ~ s^{-1} \, .
\end{equation}
The FF plasma is cooled by the  bremsstrahlung and  inverse Compton scattering and is heated by the direct Compton scattering in the magnetospheric flow. 
Taking into account diminishing factor $f_{\rm d}$, we approximate the heating rate [erg s$^{-1}$] by the Compton scattering~\citep[c.f. eq. 2.5 in][]{begelman_et1983}, an average electron on the dark side of the column gains energy at a rate
\begin{equation}
 \Gamma = \frac{kT_{\rm IC}} {m_e c^2} \frac{\sigma_{\rm T} f_{\rm d} \Lx} {\uppi R^2}  \, , 
\end{equation}
where  the equilibrium inverse Compton temperature $kT_{\rm IC} = \frac{\int h\nu F_\nu {\rm d} \nu}{4 \int F_\nu {\rm d} \nu}$.
Characteristic time of heating to the  equilibrium temperature $t_{\rm IC} \sim kT_{\rm IC} / \Gamma$ is much less than the characteristic free-fall time:
\begin{equation}
    \frac{v_{\rm ff} t_{\rm IC}}{R} \simeq \frac{m_{\rm e}c^2 \pi \sqrt{G\Mns R}}{\sigma_{\rm T} f_{\rm d}L_{\rm x}} \sim 0.03 \sqrt{\frac{R}{10^8 \,{\rm cm}}} \frac{0.2}{f_{\rm d}} \frac{10^{38} {\rm erg\, s}^{-1}}{L_{\rm x}}. 
\end{equation}
Consequently, the plasma temperature plasma in FF is about  ${T_{\rm IC}} $, whose value depends on the spectrum and is expected to be less than few keV. 


\section{Summary}\label{sec:conclusions}

We have calculated the beam patterns produced by a pair of tall optically-thick accretion columns in an XRP with a dipolar magnetic field. The bolometric luminosity of the columns in the considered regime lies in the range $10^{38}-10^{39}$~erg\,s$^{-1}$. In this case, the expected beam patterns are profoundly different from those produced by hot spots on the NS surface at low accretion rates, as well as  from the conventional fan and pencil diagrams symmetric with respect to the magnetic axis.

A key finding is that even within a dipole magnetic field configuration, taking into account the 3D geometry of the columns makes the pulse profiles complex and diverse.
In particular, we demonstrate that the asymmetric pulse profiles, frequently observed in XRPs, can be produced without any loss of central symmetry in the radiating sites. 
Pulsations in our setup are produced even for magnetic angle $\muang = 0$. 
Two particularly relevant parameters responsible for the diversity of pulse profiles are the columns' height and their central azimuth position $\phicenter$. The latter controls the column rotation with respect to the (inclined) magnetic axis of the star. Non-trivial values of $\phicenter$ (different from $0\grad$ and $180\grad$) produce asymmetric pulse profiles.

To describe the shapes of the pulse profiles, we introduce a quantitative metric $\mathcal{A}$ for pulse profile asymmetry that can be readily applied to observational data (see \S\ref{s.asymmtry}). If either $a=1$ or $\muang=0$, the pulse is always symmetrical ($\mathcal{A} = 0$). 
In most of our simulations, the calculated asymmetries lie in the range $\sim 0.05-0.25$. 

We show that  for the same inclination $\thetaobs$, magnetic angle $\muang$,  azimuthal fraction $a$, and column height, different pulse profiles may be generated  by varying  the central azimuth of the column $\phicenter$ (\S\ref{s:dep_on_phi}). 
The physical motivation for introducing $\phicenter \neq 0$ is the expectation that real XRP magnetospheres are twisted \citep{miller1996}, which is confirmed by the MHD simulations of \citet{Kulkarni-Romanova2013}. This parameter can be considered as an approximation to a rather complex physical situation, where the magnetosphere is deformed by the rotation rate mismatch between the NS and the accretion disc. This difference, and thus the central azimuth $\phicenter$, is expected to depend on the mass accretion rate which affects the size of the magnetosphere. 

An important factor in the formation of the pulse shapes at large mass accretion rates is the absorption by the funnel flow (FF) above the accretion column. It is included  in our model as a source of attenuation and scattering. 
The presence of the FF on the line of sight decreases considerably the visible isotropic luminosity of the columns at certain spin phases, thus increasing the pulse fraction (PF).
Radiation reflected by the inner, illuminated side of the FF compensates to some degree the strong flux variations caused by the attenuation. In future work, we are planning to include the contribution of the radiation diffusing through the FF and emitted by its dark side, which might reach as high as 30\%  for the considered parameter range.

We have investigated the distributions in PF for different inclinations $\thetaobs$. 
For $\thetaobs \leq 20^\circ$, half of the models yield $\mbox{PF}\sim 0.3-0.5$ (see Fig.~\ref{fig:PF_to_theta_obs_box}). At higher $\thetaobs$, PF can be as high as $0.8 - 0.9$, but strongly inclined sources are unlikely to be observed due to attenuation by the disc and disc outflows. 
For magnetospheric flows with full azimuthal filling $a=1$, the values of PF are generally lower than for the rest of the models (see Fig.~\ref{fig:PF_to_a_box}). 

Increase in mass accretion rate onto a NS increases the column height and thereby affects the observed pulse profiles. Two observational effects can be attributed to this. 
First, for luminous XRPs, there is observational evidence for anti-correlation between PF and luminosity in the relevant luminosity range \citep{2021A&A...651A..75F,Hou+2022}. 
Within our model we find a physical explanation for this anti-correlation: increase in $\dot{M}$ elevates the column, which makes a larger part of it visible to the observer throughout the whole spin period.
Such a change in visibility conditions increases the minimal flux observed during the spin period, and thus decreases the PF.

Another observational effect potentially explained by our model is the peculiar pulse shape evolution observed by \citet{Hou+2022} during a bright flare of \object{RX~J0209.6-7427}. As the source reached its peak luminosity of $\gtrsim 10^{39}\ergl$, its pulse shape gradually lost a secondary peak and evolved towards a nearly sinusoidal one. We found similar behaviour in our simulations with $a\sim 0.2$ and $\phicenter \simeq 80\grad$ for the inclination of $\thetaobs\sim 60\grad$.

\section*{Acknowledgements}
We acknowledge the collaboration with Galina Lipunova at the early stage of the work. 
The ray tracing technique, as well as the part of the model handling calculations of processes in the free-falling flow above the shock (FF), were developed by BM with support from the Russian Science Foundation grant 25-12-00012.
PA, whose contribution is the fundamental design of the geometrical model and supervision of the work done by BM, was supported by Simons Foundation grant 00001470 and the International Space Science Institute (ISSI) in Bern, through ISSI International Team project \#495 (Feeding the spinning top). We thank Sergey  Tsygankov for providing the Cen X-3 pulse profile data.


\newpage
\appendix

\section{Derivation of viewing angle and area of a surface element of the magnetic tube}
\label{s.app_dl_dS}

Unit vectors in the spherical coordinate system ($R, \thetamag, \phimag$), anchored with the magnetic dipole vector $\vector{\mu}$, are:
\begin{equation}
    \vector{e}_r=\begin{pmatrix}
        \sin{\thetamag}\cos{\phimag}\\
        \sin{\thetamag}\sin{\phimag}\\
        \cos{\thetamag}
    \end{pmatrix} ,
    ~
    \vector{e}_\thetamag=\begin{pmatrix}
                \cos{\thetamag}\cos{\phimag}\\
\cos{\thetamag}\sin{\phimag}\\
        -\sin{\thetamag}
    \end{pmatrix} ,
    ~
    \vector{e}_\phi=\begin{pmatrix}
    -\sin{\phimag}\\
    \cos{\phimag}\\
    0
\end{pmatrix} .
\end{equation}

The flow geometry in dipole approximation is given by the formula:
\begin{equation} \label{eq:R dipol}
    R = \Re \sin^2{\thetamag} ,
\end{equation}
where $\Re$ is the radius of a specific magnetic line  in the magnetic-equator plane.
The unit vector $\vector{e}_l$ along the field line is
\begin{equation}\label{eq:e_l}
    \vector{e}_l = \frac{2\cos\thetamag\, \vector{e}_r + \sin\thetamag\, \vector{e}_\thetamag}{\sqrt{3\cos^2\thetamag+1}} \, .
\end{equation}
The unit vector $\vector{e}_n$ that is  perpendicular to the field line is
\begin{equation}
    \vector{e}_n=[\vector{e}_l \times \vector{e}_\phi] = \frac{-2\cos\thetamag\, \vector{e}_\thetamag + \sin\thetamag\, \vector{e}_r}{\sqrt{3\cos^2\thetamag+1}}\, 
\end{equation}
or
\begin{equation}
    \vector{e}_n=\frac{1}{\sqrt{3\cos^2\theta+1}} \begin{pmatrix}
        (1 - 3\cos^2{\thetamag})\cos{\phimag}\\
        (1 - 3\cos^2{\thetamag})\sin{\phimag}\\
        3\sin{\thetamag}\cos{\thetamag}
    \end{pmatrix}\, .
\end{equation}
The unit vector of the direction to an observer is
\begin{equation} \label{eq:e_obs_mu vector}
   \vector{e}_{\rm obs} = \begin{pmatrix}
    \sin{\thetamag_{\rm obs}}\cos{\phimag_{\rm obs}}\\
    \sin{\thetamag_{\rm obs}}\sin{\phimag_{\rm obs}}\\
    \cos{\thetamag_{\rm obs}}
    \end{pmatrix} \, .
\end{equation}

To obtain $\cos{\psi}$, we calculate the scalar product of $\vector{e}_{\rm obs}$ and $\vector{e}_{\rm n}$:
\begin{equation}
\begin{split}
    \cos{\psi} &= (\vector{e}_{\rm obs} \cdot \vector{e}_{\rm n}) = \frac{1}{\sqrt{3\cos^2\thetamag+1}} \Big[(1 - 3\cos^2{\thetamag}) \cos{\phimag} \, \vector{e}_{\rm obs, x} \\
    &+  (1 - 3\cos^2{\thetamag}) \sin{\phimag} \, \vector{e}_{\rm obs, y} + 3\sin{\thetamag}\cos{\thetamag} \, \vector{e}_{\rm obs, z}\Big],
    \end{split}
\end{equation}
Thus, 
\begin{equation}\label{eq:cos_psi}
\begin{split}
    \cos{\psi} &= \frac{1}{\sqrt{3\cos^2\thetamag+1}} \Big[(1 - 3\cos^2{\thetamag}) \sin{\thetamag_{\rm obs}}\, \cos{(\phimag - \phimag_{\rm obs} )}\\
    &+ 3\sin{\thetamag}\cos{\thetamag} \, \cos{\thetamag_{\rm obs}}\Big].
    \end{split}
\end{equation}


Let us also obtain the area of elementary surface  ${\rm d}S $. First we find the distance element ${\rm d} l$ along the field line. From \eqref{eq:e_l}
one has:
\begin{equation}
    \frac{{\rm d} R}{{\rm d} l} = \frac{2\cos{\thetamag}}{\sqrt{3\cos^2\thetamag+1}}
\end{equation}
or
\begin{equation}
    {\rm d}l = R_{\rm e} \sqrt{3\cos^2\thetamag+1} \sin\thetamag \,{\rm d}\thetamag \, .
\end{equation}
The distance element  along the azimuth (perpendicular to the field line) is
\begin{equation}
    {\rm d}l_\phimag = R \sin{\thetamag} {\rm d}\phimag= R_{\rm e} \sin^3{\thetamag} {\rm d}\phimag .
\end{equation}

Thus, the area of a surface element is
\begin{equation}
    {\rm d}S = {\rm d}l \, {\rm d}l_\phimag = \tilde S(\thetamag) \,{\rm d}\thetamag {\rm d}\phimag,
\end{equation}
where 
\begin{equation}\label{eq:tilde S}
  \tilde S(\thetamag) = R_{\rm e}^2 \sqrt{3\cos^2\thetamag+1} \, \sin^4{\thetamag}.
\end{equation}

\section{Ray tracing}\label{appendix:Ray tracing}

Using the ray-tracing method, we  check whether each surface element of the column is visible to an observer.

For each surface element we define a ray towards the observer. The ray does not reach the observer if it meets on its way the NS surface or one of the surfaces of accretion columns, that is, an eclipse occurs. 
It is possible to find the eclipses analytically since, in our model, the surfaces of the NS and magnetic tubes are described by explicit equations (a sphere and a dipole surface).  

Steps required to implement the ray tracing method:
\begin{enumerate}
    \item Set the equation of the ray: specify the origin, i.e., the elemental surface area on the column surface and direction of propagation (direction to observer). 
    \item Set the equations of the surfaces that describe the shape of objects that potentially obscures the source
    \item Find roots at which the ray equation and the surface equation are equal.
    A real positive root implies intersection, i.d., eclipse. A negative real root root   means that the intersection point is behind the source.
\end{enumerate}

1) Let us define the origin of a ray as vector $\vector{R_0}$
and the unit vector from the ray origin to the observer as $\vector{D}$. The latter is equivalent to \eqref{eq:e_obs_mu vector} for an infinite observer.
Then any point $P$ of the ray at  distance $t$ from its origin,  $t \geq 0$, satisfies
\begin{equation}\label{eq:eq of ray}
    \vector{P}(t) \equiv  \vector{R_0} + t \, \vector{D}\, .
\end{equation}
We convert it to the coordinate system where the azimuthal angle of the observer $\phimag_{\rm obs} = 0\grad$:
\begin{equation} \label{eq:eq of dipol ray}
    \vector{P}  =
    R_{0}\,\begin{pmatrix}
        \sin{\thetamag}_0 \cos{\phimag}_0\\
        \sin{\thetamag}_0 \sin{\phimag}_0\\
        \cos{\thetamag}_0
    \end{pmatrix}
+ t\, \begin{pmatrix}
        \sin{\thetamag_{\rm obs}}\\
        0\\
        \cos{\thetamag_{\rm obs}}
    \end{pmatrix} \, .
\end{equation}
Here $R_0 = R_{\rm e} \, \sin^2{\thetamag}_0$.




2) If the eclipsing object is the magnetic tube, then
\begin{equation} \label{eq:R dipol vec}
   R = R_{\rm e} \, \sin^2{\thetamag} .
\end{equation}
Let's multiply both parts  by $R^2$:
\begin{equation} \label{eq:R dipol vec mult R2}
   R^3 = R_{\rm e} \, R^2 \sin^2{\thetamag} .
\end{equation}

3) Intersection condition is
\begin{equation}
\label{eq.intersection_cond}
    \vector{P} = \vector{R}\,  ,
\end{equation}
where $\vector{R}$ is the vector to a  point on the surface of an eclipsing object.
Using the latter to express
\begin{equation}
    \begin{gathered}
        R^3 \equiv (R_x^2 + R_y ^2 + R_z^2)^{3/2} = (P_x^2 + P_y ^2 + P_z^2)^{3/2}\, ,\\
        R^2 \sin^2{\thetamag} \equiv R_x^2 + R_y ^2  =P_x^2+P_y^2\, ,
    \end{gathered}
\end{equation}
we rewrite \eqref{eq:R dipol vec mult R2}:
\begin{equation} \label{eq:eq dipole with R}
    \begin{split}
    (R_0^2 \, + t^2 + 2 R_0 t \, (\sin{\thetamag}_0 \, \cos{\phimag}_0 \, \sin{\thetamag_{\rm obs}} + \cos{\thetamag_0} \, \cos{\thetamag_{\rm obs}}))^{3/2} = \\ = R_{\rm e} (R_0^2 \, \sin^2{\thetamag_0} + t^2 \sin^2{\thetamag_{\rm obs}} + 2 R_0 t \, \sin{\thetamag}_0 \, \cos{\phimag}_0 \, \sin{\thetamag_{\rm obs}}) .
    \end{split}
\end{equation}
Using substitutions
\begin{equation} \label{eq:def replacements}
    \begin{gathered}
    t = R_0 x \, ,\\
    \cos{\alpha} = (\sin{\thetamag_0} \, \cos{\phimag_0} \, \sin{\thetamag_{\rm obs}} + \cos{\thetamag_0} \, \cos{\thetamag_{\rm obs}})\, , \\
    \end{gathered}
\end{equation}
one obtains from 
\begin{equation}
    \begin{gathered}
    (R_0^2 + R_0^2 \, x^2 + 2 R_0^2 \, x \cos{\alpha})^3 = R_{\rm e}^2 \, (R_0^2 \sin^2{\thetamag_0} + \\ + R_0^2 \, x^2 \sin{\thetamag_{\rm obs}} + 2 R_0^2 \, x \sin{\thetamag_0} \cos{\phimag_0} \sin{\thetamag_{\rm obs}})^2 \, ,
    \end{gathered}
\end{equation}
%
Lastly, taking into account $R_0 = R_{\rm e} \sin^2{\thetamag_0}$, we obtain:
\begin{equation}
\label{eq.final_eq_intersection}
    (1 + x^2 + 2x \cos{\alpha} )^3 = (1 + x^2 \eta^2 + 2x \cos{\phimag_0} \, \eta )^2 \, .
\end{equation}
Here $\eta \equiv \sin{\thetamag_{\rm obs}}/\sin{\thetamag_0}$.

We open the brackets and move all the terms to the left side: 
 \begin{equation}\label{eq:eq dipole for intersection}
     \begin{gathered}
     x^6 + 6\cos{\alpha} \, x^5 + (12 \cos^2{\alpha} + 3 - \eta^4) x^4 + \\+ (8 \cos^3{\alpha} + 12 \cos{\alpha} - 4 \eta^3 \cos{\phimag_0})x^3 + \\+ (12 \cos^2{\alpha} + 3 - 4 \cos^2{\phimag_0} \, \, \eta^2 - 2 \eta^2) x^2 +\\ + (6 \cos{\alpha} - 4 \cos{\phimag_0} \, \, \eta ) x = 0 .
     \end{gathered}
 \end{equation}
After numerically computing the  roots of \eqref{eq:eq dipole for intersection}, we find the vector of intersection point using~\eqref{eq:eq of dipol ray}.
We note that Eq.~\eqref{eq:eq dipole for intersection} describes the intersection 
with an entire dipole surface for specific $R_{\rm e}$. 
However, because the accretion columns occupy only a limited range of azimuths (determined by parameters $a$ and $\phicenter$), we must further verify that any found intersection point actually lies within the physical domain of the column. This is done by transforming the intersection coordinates into spherical angles $\thetamag$ and $\phimag$, and checking that a column is present at those angular coordinates.


In a similar manner, we search for intersections with the NS surface.
The intersection condition \eqref{eq.intersection_cond}
is written as
\begin{equation}
    |\vector{R_0} + t \, \vector{D}|^2 = \Rns^2 \,
\end{equation}
or
\begin{equation}
    D^2 t^2 + 2\,\vector{R_0} \cdot \vector{D}\, t + (R_0^2 - \Rns^2) = 0 .
\end{equation}
which is a simple quadratic equation with respect to $t$.

\section{\texorpdfstring{Constraint on the possible maximum azimuthal filling $a$}{Constraint on the possible maximum azimuthal filling a}}\label{s.about_magnetic_polar_angle}



\begin{figure}[!h]
    \centering
    \includegraphics[width=0.45\textwidth]{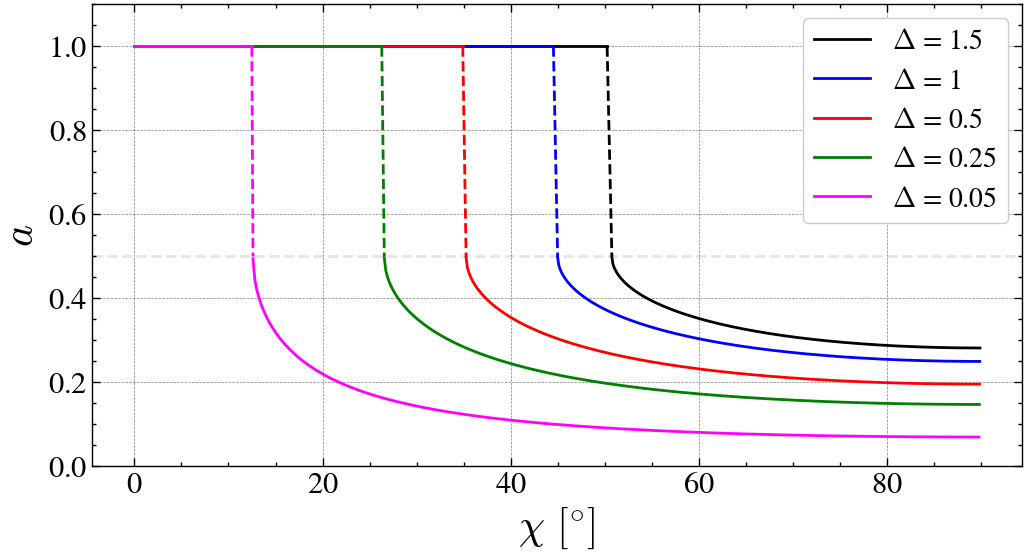}
   \caption{Maximum possible value of $a$ for the model presuming a fixed magnetic latitude for the accretion funnel at the NS surface. 
   }
\label{fig:possible values of a}
\end{figure}

Within our geometry, as the magnetic axis tilts, the parameter $a$ ceases to be free and begins to be constrained. The condition imposed on magnetospheric lines is that the cylyndric coordinate of the dipole line for given $\phimag$ and $\thetamag$ lies in the disc plane in the range from $R_{\rm d}$ to $(1+\Delta) R_{\rm d}$, with the fiducial $\Delta=0.25$. When a dipole line does not satisfy this condition, then matter does not travel along the line.
We have obtained an analytical formula for the allowed angles $\phimag$:
\begin{equation}
    \cos^2 \phimag \geq \frac{1}{\tan^2{\chi}} \frac{1-(1+\Delta)\sin^2{\thetamag_{\rm d}}}{(1+\Delta)\sin^2{\thetamag_{\rm d}}}\, ,
\end{equation}
where $\thetamag_{\rm d} = \min (\uppi/2 - \arctan({\tan{\muang} \cos{\phimag}}))$.
For $\phicenter = 0$ (or $\thetamag_{\rm d} = \uppi/2 - \muang$) this simplifies to
\begin{equation}
    \cos^2 \phimag \geq \frac{1}{(1+\Delta)\sin^2{\chi}} - \frac{1}{\tan^2{\chi}}\, .
    \label{eq.phimax_0}
\end{equation}
Allowed angles $\phimag$ are directly transformed into allowable maximum value of the azimuthal covering factor $a$. It is illustrated by Fig.~\ref{fig:possible values of a}.

As can be seen from the Fig.~\ref{fig:possible values of a}, at small values of $\muang$ the parameter $a\leq 1$ remains free. But at some magnetic angle $\muang$ (which depends on the value of $\Delta$) there is a leap (dashed lines) in the value of the allowed range of values of the parameter $a$, and $a$ ceases to be a free parameter. While the range of values decreases with the increasing inclination magnetic angle, the maximum $a \rightarrow
\arccos ([1+\Delta]^{-1/2}) / \uppi $ at large values of $\muang$.


\section{\texorpdfstring{The influence of the $\dot{m}$-dependent variation of $\phicenter$ on timing properties}{The influence of the m-dot-dependent variation of phi0 on timing properties}}\label{s.app_timing}

During  an outburst, apart from the redistribution of relative amplitudes between the pulse maxima, which can lead to an abrupt shift of the phase of the maximum, our model predicts phase shifts related to changes in column orientation determined by $\phicenter$.
If observed continuously for time $\Delta t$, an XRP with the evolving-azimuth column (varying $\phicenter)$ would experience a period change $\Delta P$,  estimated as a correction to the measured spin period with respect to the real one
\begin{equation}
P_{\rm observed} = P \left( 1+ \frac{P}{2\uppi} \frac{\diff \phicenter}{\diff t}\right),
\end{equation}
or 
\begin{equation}
\Delta P \sim \frac{P^2}{2\uppi} \frac{\diff \phicenter}{\diff t}.
\end{equation}


If the column azimuth changed by $\Delta \phicenter \sim \pi$ in  $\Delta t \sim 10$~days due to considerable variation of $\dot m$, the additional `spin-up effect' would be $\Delta P / P \sim 10^{-6}$, which is a subtle effect likely to be overridden by the changes in the intrinsic spin period.
Variable $\phicenter$ contributes to the observed timing properties of XRPs mainly through the term that depends on $\diff \dot{m} / \diff t$ rather than $\dot{m}$. 
It is most likely to become important during rapid changes in mass accretion rate, including the rising branches of flares and the putative transitions between the propeller and the accretor regimes (see for instance \citealt{2016A&A...593A..16T}).

\bibliographystyle{elsarticle-harv} 
\bibliography{example}






\end{document}